\documentclass[nofootinbib,twocolumn,superscriptaddress,a4paper]{revtex4-1}
\usepackage{graphicx}% Include figure files
\usepackage{bm,bbm}% bold math
\usepackage{xcolor}%colours
\usepackage{tcolorbox}
\usepackage{algorithm}
\usepackage{algpseudocode}
\usepackage{amsmath}
\usepackage{booktabs} 
\usepackage[colorlinks,breaklinks]{hyperref}
\usepackage{lineno}
\usepackage{upgreek}
\usepackage{makecell}
\usepackage{tabularx}
\usepackage{multirow}
\usepackage{minitoc}
\usepackage{enumitem}
\makeatletter
\newcommand\org@hypertarget{}
\let\org@hypertarget\hypertarget
\renewcommand\hypertarget[2]{%
\Hy@raisedlink{\org@hypertarget{#1}{}}#2%
  }
\makeatother
\newcommand{\ket}[1]{\left\vert#1\right\rangle}

\newcommand{\euyso}{$^{151}$Eu$^{3+}$:Y$_2$SiO$_5$}

\newcommand{\um}{$\upmu$m}

\newcommand{\us}{$\upmu$s}

\begin{document}

% \begin{refsection}
 
\title{A Metropolitan-scale Multiplexed Quantum Repeater with Bell Nonlocality}
\author{Tian-Xiang Zhu}
\email{These three authors contributed equally to this work.}
\affiliation{Laboratory of Quantum Information, University of Science and Technology of China, Hefei 230026, China}
\affiliation{Anhui Province Key Laboratory of Quantum Network, University of Science and Technology of China, Hefei 230026, China}
\affiliation{CAS Center for Excellence in Quantum Information and Quantum Physics, University of Science and Technology of China, Hefei 230026, China}
\author{Chao Zhang}
\email{These three authors contributed equally to this work.}
\affiliation{Laboratory of Quantum Information, University of Science and Technology of China, Hefei 230026, China}
\affiliation{Anhui Province Key Laboratory of Quantum Network, University of Science and Technology of China, Hefei 230026, China}
\affiliation{CAS Center for Excellence in Quantum Information and Quantum Physics, University of Science and Technology of China, Hefei 230026, China}
\affiliation{Hefei National Laboratory, University of Science and Technology of China, Hefei 230088, China}
\author{Zhong-Wen Ou}
\email{These three authors contributed equally to this work.}
\author{Xiao Liu}
\affiliation{Laboratory of Quantum Information, University of Science and Technology of China, Hefei 230026, China}
\affiliation{Anhui Province Key Laboratory of Quantum Network, University of Science and Technology of China, Hefei 230026, China}
\affiliation{CAS Center for Excellence in Quantum Information and Quantum Physics, University of Science and Technology of China, Hefei 230026, China}
\author{Peng-Jun Liang}
\affiliation{Hefei National Laboratory, University of Science and Technology of China, Hefei 230088, China}
\author{Xiao-Min Hu}
\author{Yun-Feng Huang}
\email{hyf@ustc.edu.cn}
\author{Zong-Quan Zhou}
\email{zq\_zhou@ustc.edu.cn}
\author{Chuan-Feng Li}
\email{cfli@ustc.edu.cn}
\author{Guang-Can Guo}
\affiliation{Laboratory of Quantum Information, University of Science and Technology of China, Hefei 230026, China}
\affiliation{Anhui Province Key Laboratory of Quantum Network, University of Science and Technology of China, Hefei 230026, China}
\affiliation{CAS Center for Excellence in Quantum Information and Quantum Physics, University of Science and Technology of China, Hefei 230026, China}
\affiliation{Hefei National Laboratory, University of Science and Technology of China, Hefei 230088, China}
%, which is the cornerstone for device-independent security \cite{Acin2007Device-Independent, Pironio2010Random} and foundational tests of quantum mechanics \cite{Hensen2015Loophole}

\begin{abstract}
Quantum repeaters can overcome exponential photon loss in optical fibers, enabling heralded entanglement between distant quantum memories. The definitive benchmark for this entanglement is Bell nonlocality; however, recent metropolitan-scale demonstrations based on single-photon interference (SPI) schemes have been limited to generating low-quality entanglement, falling short of Bell nonlocality certification. Here, we introduce a multiplexed quantum repeater protocol based on time measurements (MQR-TM), successfully combining the high heralding rate of SPI schemes with the phase robustness of two-photon interference (TPI) schemes. This approach achieves heralded entanglement distribution between two solid-state quantum memories over a record 14.5~km separation, generating a Bell state with a fidelity of $78.6 \pm 2.0\%$. We observe a CHSH-Bell inequality violation by 3.7 standard deviations, marking the first certification of Bell nonlocality in metropolitan-scale quantum repeaters. Our architecture supports autonomous quantum node operation without fiber channel phase stabilization, offering a practical framework for scalable quantum-repeater networks.
\end{abstract}

\date{\today}
\maketitle

Heralded entanglement distribution between distant quantum memories (QMs) serves as the elementary link of quantum repeaters (QRs), enabling the construction of large-scale quantum networks through hierarchical entanglement swapping \cite{Briegel1998Quantum, Kimble2008The, Sangouard2011Quantum, Wehner2018Quantum}. Over the past two decades, such heralded entanglement has been demonstrated across various platforms, including atomic ensembles \cite{Chou2007functional, Yu2020Entanglement, Luo2025Entangling}, single neutral atoms \cite{Hofmann2012Heralded,van2022entangling, Zhang2022device-independent}, single ions \cite{Nadlinger2022Experimental, Krutyanskiy2023Entanglement}, single defects in solids \cite{Hensen2015Loophole, Delteil2016Generation, Stolk2024Metropolitan-scale, Knaut2024Entanglement, inc2024distributed}, mechanical oscillators \cite{Riedinger2018Remote} and rare-earth-ion doped crystals \cite{Liu2021Heralded, Lago-Rivera2021telecom, Ruskuc2025Multiplexed}. 

To realize practical quantum networks, laboratory demonstrations must scale to metropolitan distances. The high heralding rate of single-photon interference (SPI) scheme \cite{Zhu2025Remote} has recently enabled heralded entanglement distribution between QMs separated by up to 12.5 km \cite{Liu2024Creation, Stolk2024Metropolitan-scale}. However, the SPI scheme compromises entanglement quality and exhibits high sensitivity to phase fluctuations in fiber channels. In atomic-ensemble implementations, the resulting number-state entanglement must be converted into two-party entanglement, achieving a maximum fidelity of only 0.643(40) \cite{Liu2024Creation}, while single-atom implementations face a fundamental trade-off between signal-to-noise ratio and fidelity, reaching a maximum fidelity of 0.534(15) \cite{Stolk2024Metropolitan-scale}. Neither approach has produced entanglement of sufficient quality to violate a Bell inequality—a prerequisite for many quantum network applications \cite{Acin2007Device-Independent, Zhang2022device-independent, Nadlinger2022Experimental,Pironio2010Random,Adamson2025Parallel,Supic2020self-testing, Sekatski2023Toward}. To date, Bell tests in QR links have been limited to 1.3 km \cite{Hensen2015Loophole}. Although high-fidelity entanglement was heralded using a two-photon detection scheme \cite{Barrett2005Efficient}, their entanglement distribution rate (EDR) remains low ($\sim$0.3 mHz), and would be fundamentally constrained by round-trip communication latency due to single-mode operation \cite{Stolk2024Metropolitan-scale}, hindering scalability to metropolitan distances.
%Phys. Rev. Lett. 91, 110405, Zhao2007Robust, Jiang2007Fast, Bell nonlocality represents the ultimate benchmark of entanglement, enabling device-independent security in quantum key distribution \cite{Acin2007Device-Independent, Zhang2022device-independent, Nadlinger2022Experimental},  certified random number generation \cite{Pironio2010Random}, verifiable blind quantum computation \cite{Adamson2025Parallel}, and self-testing of quantum systems \cite{Supic2020self-testing, Sekatski2023Toward}.

\begin{figure*}[tbph]
\centering
\includegraphics[width=1\textwidth]{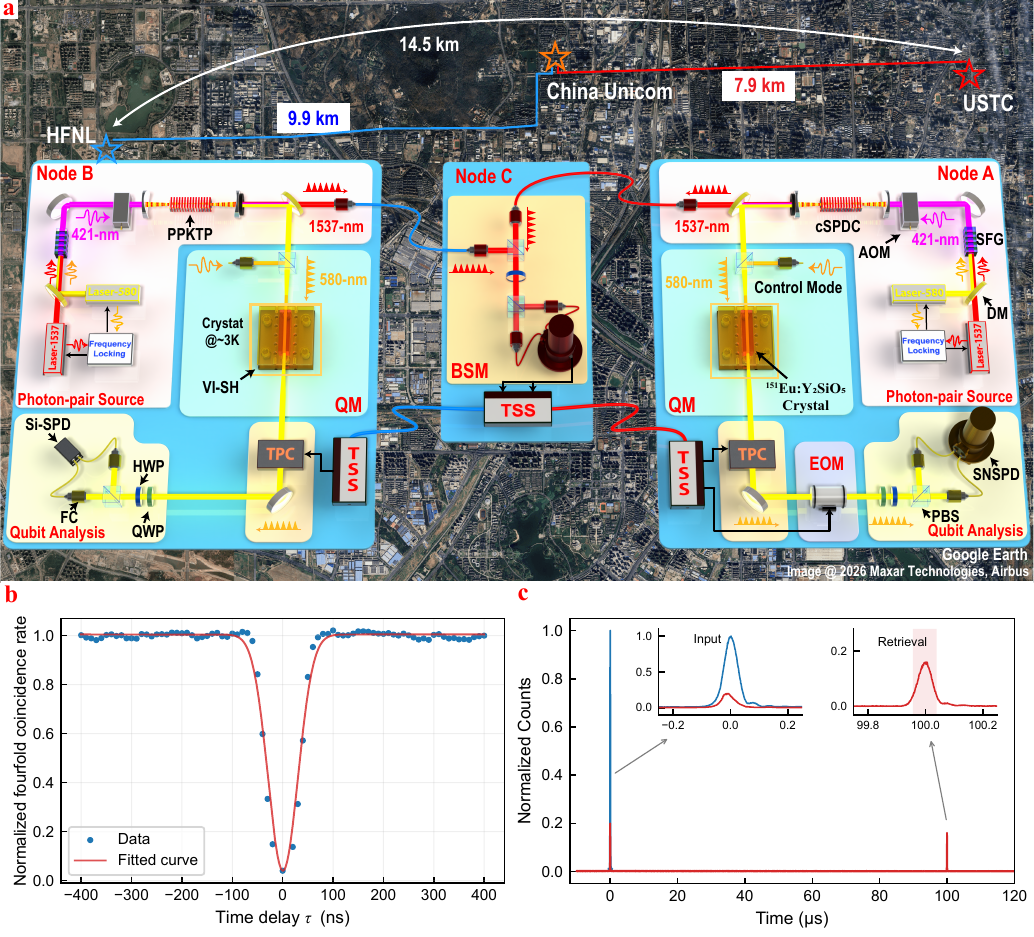} 
\caption{Metropolitan-scale multiplexed quantum repeater (\textit{XingHan 2.0}) deployed in Hefei. \textbf{a}, The network consists of two similar quantum repeater (QR) nodes (node A at USTC and node B at HFNL) and a central Bell-state-measurement (BSM) node (node C, China Unicom). The direct distance between A and B is 14.5 km; the deployed fibers are 7.9 km (A–C) and 9.9 km (B–C). At each QR node, the 1537-nm and 580-nm lasers are independently locked using frequency locking modules, and combined through sum-frequency generation (SFG) to produce a 421-nm laser. Entangled photon pairs are generated via the cavity-enhanced spontaneous parametric down-conversion (cSPDC). The 1537-nm telecom photons are routed through the deployed fiber to node C for BSM, while 580-nm signal photons are stored in \euyso~ crystals cooled to $\sim3$ K on vibration-isolated sample holders (VI-SH). At node C, the two telecom photons are combined and detected.  Successful BSM outcomes herald entanglement between remote quantum memories (QMs), with results fed back to QR nodes via a time-synchronization system (TSS). Then the time-to-polarization conversion (TPC) modules deterministically convert the time-bin encoding to the polarization encoding, while the electro-optic modulator (EOM) at node A performs the corresponding local operation that maps the heralded atomic entanglement into the desired state. PPKTP, periodically poled potassium titanium phosphate; DM, dichroic mirror; AOM, acousto-optic modulator; QWP, quarter-wave plate; HWP, half-wave plate; FC, fiber collimator; SNSPD, superconducting nanowire single-photon detector; Si-SPD, silicon single-photon detector.
City map sourced from Google Earth, with data from Maxar Technologies and Airbus.
\textbf{b}, Normalized four-fold coincidence rate versus relative delay $\tau$. At $\tau = 0$, simultaneous arrival of 1537-nm photons at node C results in a sharp drop in coincidence rate due to photon bunching, yielding a Hong-Ou-Mandel dip with $95.9\pm 0.2\%$ visibility, indicating near-perfect indistinguishability of the independent photon sources. \textbf{c}, Heralded storage of 580-nm photons for 100~\us~at node A. The blue trace shows transmitted photons after a 20-MHz transparency window while the red trace shows the storage process. The inset provides a close-up of the retrieved echo, revealing a single-mode duration of 83 ns and an efficiency of $16.6\pm0.1\%$.}   
\label{fig1}     
\end{figure*}

% This strategy successfully resolves the trade-off between entanglement rate and quality.
Here, we propose a multiplexed QR protocol based on time measurements (MQR-TM) to resolve the trade-off between entanglement rate and quality. The two-photon interference (TPI) based time measurements \cite{Halder2007Entangling} generates high-fidelity two-party entanglement, while a temporal multimode QM simultaneously stores time-bin entanglement with variable delays, lifting the heralding rate to the level of SPI schemes. We distribute heralded entanglement between QMs over a record separation of 14.5 km, observing a CHSH-Bell inequality violation with $S = 2.22 \pm 0.06$ \cite{clauser1970proposed}, exceeding the classical bound of 2 by 3.7 standard deviations. The achieved EDR surpasses recent SPI-based demonstrations \cite{Liu2024Creation,Stolk2024Metropolitan-scale} by two orders of magnitude while maintaining comparable entanglement fidelity. This architecture supports autonomous quantum node operation and eliminates the need for active phase stabilization of fiber channels, enabling seamless integration with existing telecom infrastructure.
%This establishes a robust solid-state platform for scalable quantum networks. This marks the first demonstration of genuine Bell nonlocality in metropolitan-scale QRs.

{\it The network.--} Our network (\textit{XingHan 2.0}) comprises three nodes situated in Hefei, China (Fig. \ref{fig1}a). Two QR nodes—one on the University of Science and Technology of China (USTC, node A) and one at the Hefei National Laboratory (HFNL, node B)- are separated by 14.5 km of direct distance and connected to a middle node (node C) at a China Unicom office via 7.9 km and 9.9 km of deployed fiber, respectively. Each QR node incorporates a multimode QM based on a $\mathrm{^{151}Eu^{3+}}$:$\mathrm{Y_2SiO_5}$ crystal, interfaced with nondegenerate photon-pair source generated through cavity-enhanced spontaneous parametric down-conversion (cSPDC) process. The 1537-nm idler photons travel through deployed fiber to node C, while the 580-nm signal photons are efficiently stored in the local QMs. A successful Bell-state measurement (BSM) at node C heralds the entanglement, with outcomes time-stamped and sent to nodes A and B. Conditioned on this classical information, the local unitary operation performed at node A enables the preparation of the desired entangled state between the remote QMs.

{\it Quantum repeater protocol.--} The source emits time–energy-entangled photon pairs that can be treated as high-dimensional time-bin entangled states. Two detection events at times $t_1$ and $t_2$ at node C, act as a BSM for time-bin qubits, enabling entanglement swapping that projects the signal photons into a two-party entangled state \cite{Halder2007Entangling}. The close temporal spacing of the two detection events eliminates the need for active phase stabilization of the fiber channel. In the standard implementations with linear optics \cite{Halder2007Entangling}, one can only choose a fixed delay between $t_1$ and $t_2$, resulting a low heralding rate, similar to the polarization-based TPI used in \textit{XingHan 1.0} \cite{Liu2021Heralded}. Here we remove this restriction by using two temporally multiplexed QMs to store all entangled signal photons with variable delays $|t_2-t_1|$. Every valid coincidence pairs therefore herald a Bell-nonlocal state of QMs:
\begin{eqnarray}
|\Psi^{\pm}_{AB}\rangle=\frac{1}{\sqrt{2}}\left(|t_1\rangle_A|t_2\rangle_B\pm |t_2\rangle_A|t_1\rangle_B\right),
\end{eqnarray}
where $|t_1\rangle$ and $|t_2\rangle$ label the atomic excitation in the two respective temporal modes and the sign $\pm$ is set by the heralding detectors. The heralding rate  therefore reaches half that of SPI schemes while retaining the phase robustness inherent to TPI.

By employing a spin-wave retrieval strategy with feed-forward control of the delay between two consecutive readouts \cite{Ortu2022Storage, Ma2021Elimination, Liu2025millisecond}, all stored entangled states can be efficiently analyzed, pushing the EDR toward the limit of the heralding rate. Entanglement connection between elementary QR links can be achieved by adjusting the variable delays using random-access QMs or additional spin-wave QMs (see Sec. I in the SI for details). Our MQR-TM protocol—combing TPI based on time measurements \cite{Halder2007Entangling} and temporally multiplexed QMs \cite{Liu2021Heralded}—thus successfully merges the high heralding rate of SPI \cite{Simon2007Quantum,Liu2024Creation,Stolk2024Metropolitan-scale} with the phase robustness of TPI \cite{Halder2007Entangling,Barrett2005Efficient,Liu2021Heralded,Jiang2007Fast,Zhao2007Robust,PhysRevLett.91.110405,Li_2024}, establishing a practical framework for scalable QR networks. 

In this proof-of-principle experiment we convert photonic time-bin entanglement into polarization entanglement with time-to-polarization-conversion (TPC) modules based on unbalanced Mach–Zehnder interferometers whose path difference is fixed at 500 ns (see Sec. VI in the SI for details. 
Consequently, the analysis is confined to entanglement modes satisfying $t_2 - t_1 = 500$~ns, even though the heralding rate remains high. Despite this selective filtering, the resulting EDR already surpasses that of recent metropolitan-scale QRs based on SPI.

%We generate non-degenerate photon pairs through cSPDC process, with the idler photon at 1537 nm to achieve low-loss fiber transmission and the signal photon at 580 nm for efficient quantum storage in $\mathrm{^{151}Eu^{3+}}$:$\mathrm{Y_2SiO_5}$ crystals.
{\it Photon pair sources.--} We employ a high-brightness photon source based on cSPDC, pumped by a continuous-wave 421-nm laser generated via sum-frequency generation of 580-nm and 1537-nm lasers. The source utilizes a Type-0 phase-matched periodically poled potassium titanyl phosphate (PPKTP) crystal within a Fabry-P\'{e}rot cavity. The significant wavelength difference between the signal and idler photons enhances spectral brightness through the clustering effect \cite{Fekete2013Ultranarrow-Band}. The cavity is designed to produce down-converted photons with a linewidth of approximately 10 MHz, matching the bandwidth of QMs. Due to the pump laser's coherence length being significantly longer than that of the down-converted photons, the generated photon pairs are in a time-energy entangled state. Signal and idler photons are collected into single-mode fibers and filtered through two cascaded etalons to ensure a single spatial and spectral mode. This configuration achieves a single-mode pair generation rate exceeding $60$ kHz/mW and a heralding efficiency of $35$\% (with detector efficiency corrected), enabling efficient entanglement swapping with narrowband photons.

% $\eta_A= 19.5\pm0.9 \%$ ($\mathrm{QM_A}$) and $\eta_B= 18.2\pm 0.4\%$ over 300 seconds after QM preparation
{\it Multiplexed quantum memories.--}
With a maximum round-trip latency of 99 \us~between node C and either node A or B, the QM must store the photons for at least this duration to enable the delivery of the desired entangled state \cite{Stolk2024Metropolitan-scale}. We implement the atomic frequency comb (AFC) protocol \cite{de2008solid-state} using 0.01\%-doped $\mathrm{^{151}Eu^{3+}}$:$\mathrm{Y_2SiO_5}$ crystals to achieve multiplexed, long-lived quantum storage. A stabilized laser and a vibration-isolated sample holder (VI-SH) enable preparation of high-quality AFCs with AFC coherence times of approximately 500~\us, approaching the fundamental limits of optical coherence lifetimes (see Sec. VIII in the SI for details). The QR relies on AFCs featuring a $10$ kHz periodicity and a $20$ MHz bandwidth. During the 100-\us~storage time, approximately 1205 temporal modes are stored in parallel, given a single-mode duration of $\sim$83~ns that covers 90\% of the retrieved echo (Fig.~\ref{fig1}\textbf{c}) \cite{Lago-Rivera2021telecom}. The quantum storage efficiencies reach $16.6 \pm 0.1\%$ and $15.7 \pm 0.1\%$ at 100~\us~for nodes A and B, respectively, nearly doubling the efficiency of prior work \cite{Ortu2022Storage}. The reduced doping concentration of 0.01\% $\mathrm{^{151}Eu^{3+}}$ extends the spectral-hole lifetime by a factor of $\sim$10,000 compared to our previous 0.2\% $\mathrm{^{153}Eu^{3+}}$:$\mathrm{Y_2SiO_5}$ crystals \cite{Liu2024Nonlocal}, significantly improving overall storage efficiencies (see Sec. VIII in the SI for details). 
%The photon-counting histogram in Fig.~\ref{fig1}\textbf{c} demonstrates 100-\us~AFC storage of 580-nm signal photons heralded by idler photons, achieving a cross-correlation of $65.5 \pm 1.4$ for retrieved photons.

\begin{figure}[!ht]
\centering
\includegraphics[width=0.48\textwidth]{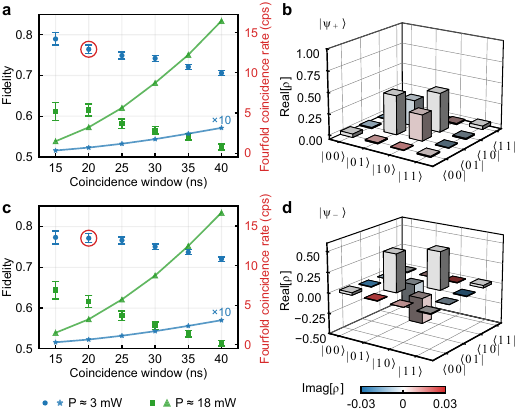}
\caption{Entanglement witness and quantum state tomography for swapped photonic entanglement. \textbf{a},\textbf{c}, Entanglement witness fidelity (blue circles, green squares) and fourfold coincidence rate (blue stars, green triangles) for the $\ket{\psi^{+}}$ (\textbf{a}) and $\ket{\psi^{-}}$ (\textbf{c}) states as a function of coincidence window, with pump powers of $P\approx 3$~mW and $P\approx 18$~mW, respectively. The measurement time for the data is $9$~h ($0.2$~h) for $P\approx 3$~mW (18~mW).
The fourfold coincidence rate at 3 mW is magnified by a factor of 10 for visibility. Data points marked with red circles correspond to those shown in (\textbf{b},\textbf{d}). All error bars in \textbf{a} and \textbf{c} represent one standard deviation. \textbf{b},\textbf{d}, Reconstructed density matrices for $\ket{\psi^{+}}$ (\textbf{b}) and $\ket{\psi^{-}}$ (\textbf{d}), demonstrating fidelities of $\mathcal{F}_+= 76.3 \pm 1.1\% $ and $\mathcal{F}_- = 77.0 \pm 1.2\%$ relative to maximally entangled Bell states. Bar height and color represent the real and imaginary parts of the matrix elements, respectively.}
\label{fig2}
\end{figure}
% 9.48/0.17

\begin{figure*}[!ht]
\centering    
\includegraphics[width=\textwidth]{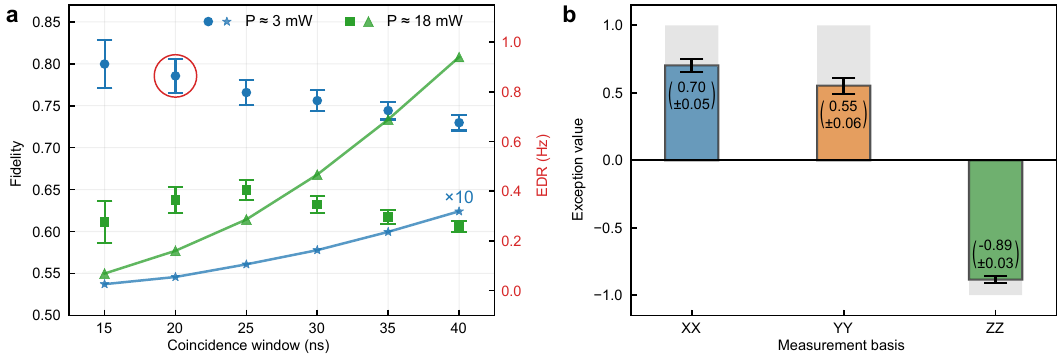}
\caption{Heralded entanglement between remote quantum memories.
\textbf{a}, Fidelity (blue circles, green squares) and entanglement distribution rate (EDR; blue stars, green triangles) of the heralded atomic entanglement versus coincidence window, with pump powers of $P\approx 3$ mW and $P\approx 18$ mW, respectively. The EDR at 3-mW pump power is magnified by a factor of 10 for visibility.
The measurement time for the data is $36$~h ($3$~h) for $P\approx 3$~mW (18~mW).
\textbf{b}, Measurement results in the $XX$, $YY$, and $ZZ$~ bases with a $20$-ns coincidence window and a $3$-mW pump power, corresponding to the data point marked with a red circle in \textbf{a}. All error bars in \textbf{a} and \textbf{c} represent one standard deviation.}\label{fig3}             
\end{figure*}
%Measured fidelity/EDR values are $78.6\pm 2.0\%$/$(5.5\pm 0.2)\times 10^{-3}$~Hz ($P\approx 3$ mW, 20-ns coincidence window) and $60.5\pm 0.7 \%$/$0.94\pm 0.01$~Hz ($P\approx 18$ mW, 40-ns coincidence window).
% 32.07/3.00

{\it Interference between independent photon sources.--}High-visibility TPI is critical for our protocol, requiring excellent indistinguishability of the two 1537-nm photons. Spectral overlap is achieved by independently locking each 1537-nm laser to a local ultra-stable reference cavity, achieving a relative drift of only a few kilohertz per day (see Sec. III in the SI for details), which is much smaller than the $\sim$10~MHz single-photon bandwidth.
Spatial and polarization overlap are enforced using single-mode fibers and polarization beam-splitters (PBSs). A coincidence window shorter than the $\sim$100~ns photon coherence time isolates a unique temporal mode \cite{Halder2007Entangling}. We quantify the indistinguishability of the heralded 1537-nm photons via Hong-Ou-Mandel interference, achieving a visibility of $95.9 \pm 0.2\%$  (Fig.~\ref{fig1}\textbf{b}) with a 20-ns coincidence window.      

{\it Swapped photonic entanglement.--} We first verify swapped entanglement between the two 580-nm photons by preparing a 20-MHz transparency window in the  \euyso~crystals to disable QM functionality. As the storage time is shorter than the round-trip communication delay, the photonic qubits are analyzed in a delayed-choice configuration \cite{Liu2024Creation}. 
At each node, an unbalanced interferometer with a 500-ns imbalance performs the TPC, followed by standard waveplates and PBSs for qubit measurements. Without QMs or classical feed-forward, this conversion succeeds with a 25\% probability.
If two detection events at node C originate from the same detector, the heralded pair is projected onto $|\psi^+\rangle=\left(|01\rangle + |10\rangle\right)/\sqrt{2}$ state; if they come from different detectors, the state is $|\psi^-\rangle=\left(|01\rangle-|10\rangle\right)/\sqrt{2}$, where 0 and 1 denote the photonic time-bin qubit. 
We characterize the photonic entanglement with an entanglement witness \cite{Guhne2009Entanglement}. Fidelity is extracted from the expectation value of the projector $|\psi^\pm\rangle\langle\psi^\pm|=\left(\mathbbm 1-ZZ\pm XX\pm YY\right)/4$, where $X$, $Y$, and $Z$ are Pauli operators. The measured fidelity for $|\psi^+\rangle$ ($|\psi^-\rangle$), along with corresponding fourfold coincidence counts, are obtained across various pump powers and coincidence windows, as shown in Fig. \ref{fig2}\textbf{a} (\textbf{c}). Full quantum-state tomography yields fidelities of $\mathcal{F}_+ = 76.3\pm 1.1\%$ for $|\psi^+\rangle$ (Fig. \ref{fig2}\textbf{b}) and $\mathcal{F}_- = 77.0\pm 1.2\%$ for $|\psi^-\rangle$ (Fig. \ref{fig2}\textbf{d}) with a $20$-ns coincidence window and a $3$-mW pump power, in excellent agreement with the entanglement witness results.

\begin{figure}[!ht]                                    
\centering
\includegraphics[width=0.48\textwidth]{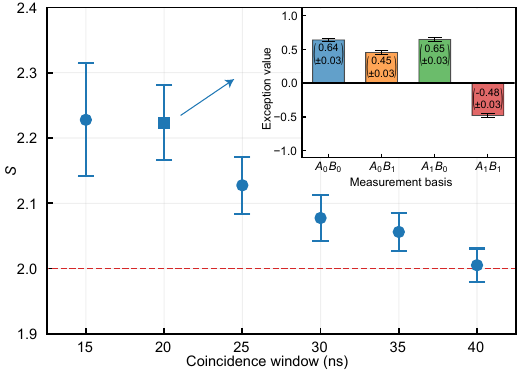}
\caption{CHSH-Bell tests of heralded atomic entanglement. Blue circles depict the measured CHSH values as a function of the coincidence window, with the red dashed line marking the classical bound of 2. The inset presents measured results in the $A_0B_0,~A_0B_1,~A_1B_0$, and $A_1B_1$ bases with a $20$-ns coincidence window, achieving a maximum CHSH value of $S = 2.22 \pm 0.06$. The measurement time for the data is $214$~h. 
All error bars in this figure represent one standard deviation.}
\label{fig4}
\end{figure}

% 214.32

{\it Delivery of desired entanglement.--} Then we activate the QMs by preparing AFCs in the \euyso~crystals, ensuring reliable delivery of the desired entangled state.  Such heralded entanglement is a fundamental resource for quantum networks and can be further used for subsequent entanglement-based quantum-information-processing tasks \cite{Wehner2018Quantum, Zhu2025Remote}. Node C sends the photon-pair detection signal to nodes A and B via classical communication while the 580-nm photons are stored in the QMs. These signals trigger the TPC modules at nodes A and B, flipping the polarization of the early temporal mode of the retrieved photons to convert the time-bin qubit into a polarization qubit with unit probability. When the two detections at node C originate from different detectors, a signal prompts node A to apply a $\pi$ phase shift via the EOM, mapping the heralded state from $|\psi^-\rangle$ to $|\psi^+\rangle$, ensuring the delivered state is consistently $|\psi^+\rangle$. The measured fidelity and EDR for retrieved entangled photons across various pump powers and coincidence windows are shown in Fig. \ref{fig3}\textbf{a}. At a pump power of 3 mW and a 20-ns coincidence window, entanglement-witness measurements on the retrieved atomic excitation yield a fidelity of $78.6\pm2.0\%$ for $|\psi^+\rangle$ (Fig. \ref{fig3}\textbf{b}). We achieve a TPI-based heralding rate of $23.6$ kHz, approaching that of SPI schemes ($50.0$ kHz in our setup), with a 50\%
reduction due to pairing two detection events and a minor reduction due to finite storage time. With our fixed-delay TDC modules, only $107$ Hz of these heralding events being directly analyzed while the remainder could be further unlocked by spin-wave QMs. Due to finite readout efficiency for signal photon pairs, the final EDR is $5.5\pm 0.2$ mHz (See Sec. I in the SI for detailed analysis). 
By increasing the pump power to 18 mW and extending the coincidence window to 40 ns, we achieve an EDR of $0.94 \pm 0.01$ Hz, surpassing recent SPI-based demonstrations \cite{Liu2024Creation, Stolk2024Metropolitan-scale} by more than two orders of magnitude while maintaining comparable entanglement fidelity (see Table S1 in the SI for details). This enhancement of EDR stems from the linear scaling enabled by straightforward multiplexing \cite{Liu2021Heralded}, while much higher EDR could be achieved with spin-wave retrievals.

{\it Bell nonlocality.--} To characterize the high quality and nonlocality of this entanglement, we further measure the CHSH-Bell inequality \cite{clauser1970proposed}, defined as:\begin{eqnarray}
S=\langle A_0 B_0\rangle+\langle A_0 B_1\rangle+\langle A_1 B_0\rangle-\langle A_1 B_1\rangle
\end{eqnarray} where $\langle A_x B_y\rangle$ denotes the expectation value of the product of measurement outcomes at nodes A and B. Each node uses binary inputs (0 or 1) and binary outputs (+1 or -1). Local hidden variable models require $|S| \leq 2$, whereas quantum entanglement can violate this bound up to $2\sqrt{2}$. We measure the CHSH value for the heralded atomic entanglement using the measurement settings $A_0=(Z+X)/\sqrt{2}$, $A_1=(Z-X)/\sqrt{2}$, $B_0=-Z$ and $B_1=X$. As shown in Fig. \ref{fig4}, we observe a maximal violation of $S=2.22\pm0.06$ exceeding the classical bound by 3.7 standard deviations. This high entanglement quality arises from the noise-robust TPI scheme and the exceptional intrinsic fidelity of the solid-state QMs. Notably, the retrieved photon pairs exhibit higher fidelity than the heralded pairs when the memories are bypassed, due to the built-in spectral and temporal filtering of the AFC storage process \cite{Liu2021Heralded}. %\textbf{Future developments in deterministic entangled photon sources \cite{Liu2019solid-state} combined with cavity-enhanced high-efficiency QMs ***Meng Nature Photonics 2026*** could in principle enable loophole-free Bell inequality violations over quantum networks.}
%The certification of Bell nonlocality is pivotal for near-term QR applications, as it provides heralded high-quality atomic entanglement that overcomes the limitations of probabilistic photonic entanglement distribution. Future developments in deterministic entangled photon sources \cite{Liu2019solid-state} combined with cavity-enhanced high-efficiency QMs \cite{PhysRevA.82.022310, PhysRevA.82.022311} could in principle enable loophole-free Bell inequality violations over large-scale networks.

% \addtocontents{toc}{\protect\setcounter{tocdepth}{-1}}
\section*{Discussion}

In summary, we introduce the MQR-TM protocol for high-speed and high-fidelity entanglement distribution and demonstrate heralded entanglement between QMs separated by a distance of 14.5~km. The generated state allows a clear violation of the CHSH-Bell inequality, demonstrating non-locality across a metropolitan-scale QR. By leveraging long-lived quantum storage and feed-forward control, the delivered entanglement is heralded in a target state ready for immediate use or connection with additional elementary links. The achieved EDR of 0.94 Hz over 14.5 km outperforms recent SPI demonstrations \cite{Liu2024Creation, Stolk2024Metropolitan-scale} by two orders of magnitude, driven by the linear speedup from 1205 temporal modes. 

Extending our architecture with spin-wave storage \cite{Ortu2022Storage,Ma2021Elimination,Liu2025millisecond} 
is expected to substantially extend the operational distance, further increase the EDR toward the fundamental limit set by the heralding rate, and enable interconnections between elementary QR links.
The MQR-TM protocol can be readily integrated into any quantum-repeater architecture that employs ensemble-based quantum memories \cite{Sangouard2011Quantum}, removing the requirement for real-time phase stabilization across fibre networks. This guarantees full compatibility with existing telecom infrastructure and facilitates the practical deployment of quantum networks composed of truly independent and geographically separated nodes. 

%Our quantum nodes operate autonomously, with independent laser systems controlling both the photon-pair source and the QMs,

Furthermore, our memory platform is built on the \euyso~crystal, a unique material enabling coherent light storage for up to 1 hour \cite{Ma2021One-hour}. Ongoing research is focused on developing transportable \euyso~memories with minute-to-hour lifetimes. In the near future, the current architecture could enable heralded entanglement distribution involving transportable QMs, facilitating a flexible network architecture where nodes are not constrained by fixed fiber links. Such mobility promises to enable transformative quantum network applications far beyond the capabilities of current fixed-link systems.

\bigskip

\textbf{Data availability}
Data for Fig. 1-4 and Figures in supplementary information are available via Figshare at https://doi.org/10.6084/m9.figshare.31778782 (\cite{Zhu2026Data}). Additional data related to this Article are available from the corresponding authors upon request.
\\

\textbf{Code availability}
The custom codes used to produce the results presented in this paper are available from the corresponding authors upon request.

% % \bibliographystyle{naturemag}
% \bibliography{bibliography}

\bigskip
\textbf{Acknowledgments}
This work is supported by the Quantum Science and Technology-National Science and Technology Major Project (No. 2021ZD0301200 and 2021ZD0301604), the National Natural Science Foundation of China (Nos. 12222411 and 12404572), and the China Postdoctoral Science Foundation (2023M743400). Z.-Q.Z acknowledges the support from the Youth Innovation Promotion Association CAS. The allocation of node C and the deployment of low-loss fiber is supported by China Unicom (Anhui).\\

\textbf{Author contributions}
Z.-Q.Z. conceived the repeater architecture and supervised all aspects of this work; T.-X.Z. and Z.-W.O. constructed the quantum memories; C.Z. constructed the quantum light sources with the help from Y.-F.H., T.-X.Z., Z.-W.O., and X.-M.H; T.-X.Z., Z.-W.O., X.L. and C.Z. constructed the classical communication system; P.-J.L. grew the crystals; T.-X.Z., C.Z. and Z.-W.O. and Z.-Q.Z. wrote the manuscript with the input from others; Z.-Q.Z., Y.-F.H., and C.-F.L. designed the experiment and supervised the project; All authors discussed the experimental procedures and results.\\

\textbf{Competing interests}
The authors declare no competing interests.
% \printbibliography[title={reference}]
% \end{refsection}

% \bibliographystyle{naturemag}
% \bibliography{bibliography}

% \begin{refsection}
% \documentclass[nofootinbib,prl,superscriptaddress,a4paper]{revtex4-1}
% \setcounter{secnumdepth}{1}
% %twocolumn ,linenumbers
% %\usepackage{scicite}
% \usepackage{graphicx}% Include figure files
% %\usepackage{dcolumn}% Align table columns on decimal point
% \usepackage{bm,bbm}% bold math
% \usepackage{xcolor}%colours
% \usepackage{tcolorbox}
% \usepackage{algorithm}
% \usepackage{algpseudocode}
% \usepackage{amsmath}
% \usepackage{float}
% \usepackage{booktabs} 
% \usepackage{upgreek}
% \usepackage{makecell}
% \usepackage{tabularx}
% \usepackage{multirow}

% \usepackage[colorlinks,
%             linkcolor=blue,      % 内部链接（如目录）的颜色
%             citecolor=blue,     % 引用的颜色
%             urlcolor=blue,      % URL 的颜色
%             breaklinks ]{hyperref}
\onecolumngrid
\makeatletter
\let\org@hypertarget\hypertarget
\renewcommand\hypertarget[2]{%
  \Hy@raisedlink{\org@hypertarget{#1}{}}#2%
  }
\setcounter{secnumdepth}{2}
\makeatother

\renewcommand{\figurename}{Fig.}
\renewcommand{\tablename}{Table.}

\setcounter{table}{0}
\renewcommand{\thetable}{S\arabic{table}}
\setcounter{figure}{0}
\renewcommand{\thefigure}{S\arabic{figure}}
\setcounter{equation}{0}
\renewcommand{\theequation}{S\arabic{equation}}

\newpage
% \begin{abstract}
\begin{center}
    {\large\bfseries
    Supplementary Materials for  ``A Metropolitan-scale Multiplexed Quantum Repeater with Bell Nonlocality"} \\[1.5em]
    {Tian-Xiang Zhu, $^{1,2,3,{\color{blue}*}}$ Chao Zhang,\,$^{1,2,3,4,{\color{blue}*}}$ Zhong-Wen Ou,\,$^{1,2,3,{\color{blue}*}}$ Xiao Liu,\,$^{1,2,3}$ Peng-Jun Liang,\,$^{4}$ Xiao-Min Hu,\,$^{1,2,3,4}$ Yun-Feng Huang,\,$^{1,2,3,4,\color{blue}\dagger}$ Zong-Quan Zhou,\,$^{1,2,3,4,\color{blue}\ddagger}$ Chuan-Feng Li,\,$^{1,2,3,4,\color{blue}\S}$  and Guang-Can Guo\,$^{1,2,3,4}$}\\[0.5em]
    % \normalsize
    $^1$\textit{\small Laboratory of Quantum Information, University of Science and Technology of China, Hefei 230026, China}\\
    $^{2}$\textit{\small Anhui Province Key Laboratory of Quantum Network, University of Science and Technology of China, Hefei 230026, China}\\
    $^3$\textit{\small CAS Center For Excellence in Quantum Information and Quantum Physics, University of Science and Technology of China, Hefei 230026, China}\\
    $^{4}$\textit{\small Hefei National Laboratory, University of Science and Technology of China, Hefei 230026, China}
\end{center}
\bigskip\bigskip
\begin{center}
   \large\bfseries CONTENTS 
\end{center}
{\color{red}
\begin{enumerate}[label=\Roman*., left=0pt, align=left, labelsep=0.5em]
    \item Multiplexed quantum-repeater based on time measurements (MQR-TM) \hfill 9
    % 嵌套列表：大写字母编号，缩进1.5em
    \begin{enumerate}[label=\Alph*., left=1.5em, labelsep=0.5em]
        \item Protocol Description \hfill 9
        \item Performance overview of elementary QR links \hfill 11
    \end{enumerate}
    \item Overall experimental setup \hfill 12
    \item Laser systems \hfill 13
    \item cSPDC photon source \hfill 14
    \item Locking system and resonance of the two cSPDC sources \hfill 16
    \item Time-to-polarization conversion \hfill 18
    \item Noise model for the swapped photonic entanglement \hfill 19
    \item Quantum memories \hfill 20
    \item Optical losses \hfill 23
    \item Synchronization and classical communication \hfill 24
    \item Detailed experimental data \hfill 25
\end{enumerate}
}
% \minitoc

% \end{abstract}
% \date{\today}
% \maketitle
\newpage

%\subsection{1. Quantum repeater protocol}
%(张）
% \addtocontents{toc}{\protect\setcounter{tocdepth}{1}
\section{ Multiplexed quantum-repeater based on time measurements (MQR-TM)}
\label{sec.Quantum repeater protocol}

\subsection{Protocol Description}
\label{sec:Protocol Description}

Our MQR-TM protocol integrates a two-photon interference (TPI) scheme based on temporal measurements~\cite{Halder2007Entangling} with temporally multiplexed quantum memories (QMs)~\cite{Lago-Rivera2021telecom,Liu2021Heralded}. 

\begin{figure*}[htbp]
\includegraphics [width=0.75\textwidth]{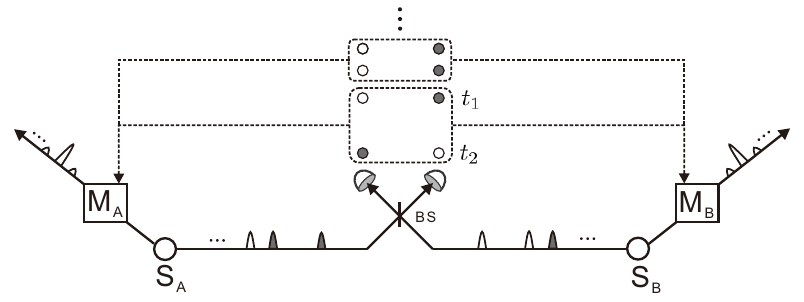}
\caption{The protocol for entanglement distribution between two QMs in an elementary QR link. The two photon sources $S_1$ and $S_2$ are pumped by continuous-wave lasers to generate time-energy entangled photon pairs. The two idler photons are mixed on a beam-splitter and then detected. The solid small circles indicate the response of the detector, while the dashed squares represent the two-detector click events after pairing. Through classical feedback, the QMs could implement two consecutive readouts with a delay corresponding the heralding events, to achieve a equivalent unbalanced MZI for analysis the entanglement with variable delays. }
\label{fig:protcol}
\end{figure*}

As illustrated in Fig.~\ref{fig:protcol}, each quantum repeater (QR) node comprises a photon pair source and a QM. The sources, $S_A$ and $S_B$, are coherently excited with a small probability $p$ to produce a photon pair, yielding the state:
\begin{eqnarray}   
\left[1+\sqrt{p}\left(e^{i\theta_A}a_s^\dagger a_i^\dagger +e^{i\theta_B}b_s^\dagger b_i^\dagger \right)+O(p) \right]|0\rangle,
\end{eqnarray}
where $a_s$ and $a_i$ ($b_s$ and $b_i$) represent the signal and idler modes for $S_A$ ($S_B$) respectively, $\theta_A$ ($\theta_B$) is the phase of the pump laser for  $S_A$ ($S_B$). $O(p)$ represents higher-order emission noise. Signal photons are stored in the QMs, while idler photons are transmitted to the middle station, overlapped on a beam splitter to erase which-path information.
The two output modes after the beam-splitter are $a_i'=\frac{1}{\sqrt{2}}(a_ie^{-i\phi_A}+b_ie^{-i\phi_B})$ and $b_i'=\frac{1}{\sqrt{2}}(a_ie^{-i\phi_A}-b_ie^{-i\phi_B})$, where $\phi_A$ and $\phi_B$ represent the phase accumulated by the photons during transmission. Detecting one photon in mode $a_i'$ or $b_i'$ projects the signal modes into:
\begin{eqnarray}
\frac{1}{\sqrt{2}} \left(a_s^\dagger \pm e^{i\Delta \varphi }b_s^\dagger\right)|0\rangle,
\end{eqnarray} where $\Delta \varphi=\theta_A-\theta_B+\phi_A-\phi_B$. This creates an number-state entanglement between the QMs:
\begin{eqnarray}
\frac{1}{\sqrt{2}}\left(|1\rangle_A|0\rangle_B\pm e^{i\Delta \varphi }|0\rangle_A|1\rangle_B\right),
\end{eqnarray}
where 0 and 1 represent the number of atomic excitations of the two QMs. This single-photon interference (SPI) heralding protocol yields a heralding rate of $R=2R_i\eta_{L_0} \eta_D^i$, where $R_i$ is the the idler rate of the photon source, $\eta_D^i$ is the detector efficiency for idler photon, $\eta_{L_0} = e^{-\alpha L_0/2}$, $\alpha$ is the fiber loss coefficient and $L_0$ is the total distance between nodes. Compared to direct transmission efficiency of $e^{-\alpha L_0}$, SPI halves the channel loss (in dB), enabling high heralding rates~\cite{Liu2024Creation}. However, SPI is sensitive to phase noise from remote lasers and fiber channels. The number-state entanglement further requires a conversion to two-party entanglement to make it useful in quantum information applications~\cite{Liu2024Creation,Simon2007Quantum}.

A key innovation from Ref.~\cite{Halder2007Entangling} involves pairing two adjacent detection events at the middle station for Bell-state measurements (BSM), yielding: $\frac{1}{2}\left(a_{t_1}^\dagger\pm e^{i\Delta \varphi_{t_1}} b_{t_1}^\dagger\right) \left(a_{t_2}^\dagger\pm e^{i\Delta \varphi_{t_2}} b_{t_2}^\dagger \right)|0\rangle$, where subscripts $t_1$ and $t_2$ denote temporal modes. Projecting this state onto a subspace with one photon per location creates a time-bin entangled state:
\begin{equation}
\frac{1}{\sqrt{2}} \left( a_{t_1}^\dagger b_{t_2}^\dagger \pm e^{i(\Delta \varphi_{t_1} - \Delta \varphi_{t_2})} a_{t_2}^\dagger b_{t_1}^\dagger \right) |0\rangle.
\end{equation}
Given the small time difference between two detection events, the phase difference is typically negligible ($\Delta \varphi_{t_1} - \Delta \varphi_{t_2} \approx 0$), enabling a phase-robust TPI heralding of two-party entanglement between the two signal photons. 

Unlike polarization-based TPI implemented in our previous work \cite{Liu2021Heralded}, time-measurement-based TPI allows variable delays between two detection events. However, the original photonic implementations \cite{Halder2007Entangling} were limited to fixed-delay analysis using unbalanced Mach-Zehnder interferometers (uMZIs).

Our MQR-TM protocol interfaces signal photons with temporally multiplexed QMs, storing all variable-delay time-bin entanglement for subsequent use, yielding entanglement between QMs:
\begin{eqnarray}
|\Psi^{\pm}_{AB}\rangle=\frac{1}{\sqrt{2}}\left(|t_1\rangle_A|t_2\rangle_B\pm |t_2\rangle_A|t_1\rangle_B\right),
\end{eqnarray}
where $|t_1\rangle$ and $|t_2\rangle$ label the atomic excitation in the two temporal modes.

%The projection can be realized by converting the atomic excitations back into photons and postselecting there is one photon in each location.

To analyze time-bin entanglement with variable delays, one would require uMZI with active-controlled variable imbalance, which is impossible with linear optics. However, spin-wave QMs can accomplish this task through feed-forward control of the delay between two consecutive readouts \cite{Ortu2022Storage, Ma2021Elimination, Liu2025millisecond}, effectively emulate a variable-imbalance uMZI (Fig.~\ref{fig:protcol}).

Since all entangled states with variable $|t_2 - t_1|$ are stored in our experiments, we can achieve a heralding rate of $R_i\eta_{L_0} \eta_D^i $, approaching SPI rates with 50\% loss due to pairing of two detection events. Spin-wave QMs would be necessary to fully analyze all these entangled states, however, our current experimental setup still relies on
TPC modules incorporating optical uMZIs,  which selectively extract detection pairs with a fixed 500-ns interval. Consequently, the useful heralding rate is given by $R=4R_{i}^2T_w(\eta_{L_0} \eta_D^i)^2$, where $T_w$ represents the coincidence window. The EDR, which is defined as the final fourfold photon counting rate, should further include the detection efficiency of the two retrieved signal photons
$EDR=\frac{1}{2}R(\eta_s\eta_{T_w}\eta_{QM}\eta_T\eta_D^s)^2$, 
where the factor $\frac{1}{2}$ accounts for the postselection probability for one photon at each node  \cite{Halder2007Entangling}, $\eta_{s}$ is the heralding efficiency of the signal photon of the photon source, $\eta_{T_w}$ is the probability that the photon pair of the source is detected within the coincidence window, $\eta_{QM}$ is the storage efficiency of the quantum memory including the photon-memory bandwidth-matching efficiency, $\eta_T$ denotes various transmission losses (see Table~\ref{tab:component efficiencies} for these parameters in our experiments).

Let us examine the scalability of the protocol. Given that all entangled states with variable delays are utilized, entanglement swapping between elementary QR links would require BSM on time-bin entangled states with arbitrary delays. This presents a challenge within linear optics, yet it can be addressed through two potential approaches. The first method employs spin-wave QMs with random-access capability \cite{ou2025multichannel,zhang2024realization} to actively adjust the delay of one time-bin entangled state to match another. The second approach involves deterministically converting time-bin entanglement into polarization entanglement by mapping time bins onto polarization states and compensating the delays between polarization bases using spin-wave QMs. While the practical implementation of spin-wave QMs with both high fidelity and high device remains a technical challenge, both strategies enable efficient connections across elementary quantum repeater links, thereby supporting the development of scalable quantum networks. The MQR-TM protocol further highlights the unique advantages of absorptive ensemble-based QMs \cite{Liu2021Heralded}, since it requires parallel storage of many external photons without prior knowledge, which is challenging to be implemented with single-atom systems even addressing multiple atoms are available techniques.

It's instructive to compare our MQR-TM protocol with the SPI-based multiplexed QR protocol proposed by C. Simon et al~\cite{Simon2007Quantum}. In Ref.~\cite{Simon2007Quantum}, SPI establishes number-state entanglement in elementary QR links, followed by single-photon detection for entanglement swapping to extend number-state entanglement across multiple links. Photons are then recalled from two pairs of end-node QMs to create two-party entanglement, to create a useful two-party entanglement between end nodes. The key distinction between these two protocols lies in the fact that our scheme directly generates two-party entanglement within elementary QR links, followed by entanglement swapping via two-photon BSM. Our method offers enhanced phase robustness in practical implementations, as it requires phase stability of elementary links only during the brief intervals between two detection events within elementary QR links (typically spanning tens of kilometers), which are on the order of microseconds. In contrast, the protocol described in Ref.~\cite{Simon2007Quantum} requires phase stability of the full channel throughout the entire duration needed to establish number-state entanglement over the full channel length—a period that would extend to seconds for practical applications. 

By combining the time-measurements-based TPI scheme with temporally multiplexed QMs, our MQR-TM protocol successfully combines the high heralding rate of SPI schemes with the phase robustness of TPI schemes. This straightforward yet effective extension of Ref.~\cite{Halder2007Entangling} establishes a practical framework for scalable QR networks. Because MQR-TM can be dropped into any ensemble-based repeater architecture, we expect it to find wide application.

\subsection{Performance overview of elementary QR links}
\label{sec:QR links}

\begin{figure*}[tbph]
\includegraphics [width=0.8\textwidth]{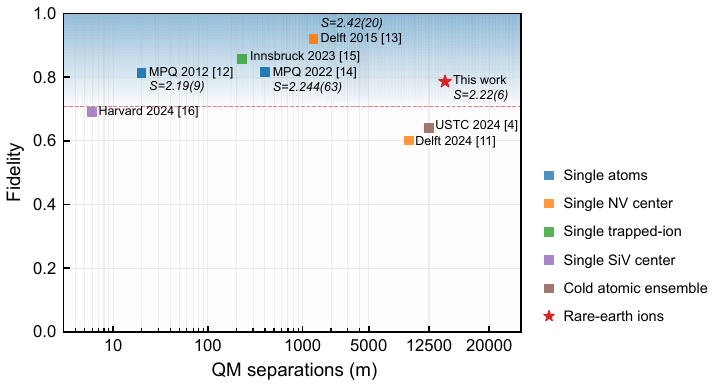}
\caption{Performance overview of heralded entanglement between distant quantum memories, with full data tabulated in Table~\ref{tab:QR_performance}. The dashed line indicates the minimum fidelity of $1/\sqrt{2}$ required for CHSH-Bell inequality violation under dephasing noise (see Eq. S10 for the noise model).}
\label{fig:report_quantum_repeater}
\end{figure*}

In Table~\ref{tab:QR_performance} and Fig.~\ref{fig:report_quantum_repeater}, we present a performance overview of elementary QR links implemented across various physical systems, focusing on those with large QM separations suitable for practical quantum networks. Our work achieves the first demonstration of metropolitan-scale QRs with certified Bell nonlocality. Compared to recent SPI-based metropolitan-scale QRs, our entanglement distribution rate (EDR) of 0.47~Hz surpasses the 0.0037~Hz reported in Ref.~\cite{Liu2024Creation} by a factor of 127, with a comparable fidelity of 0.632(10). Similarly, our EDR of 0.94~Hz exceeds the 0.0083~Hz in Ref.~\cite{stolk2024metropolitan} by a factor of 113, with a similar fidelity of 0.606(7). The achieved EDR of 0.94~Hz also represents the highest among all surveyed demonstrations, highlighting the critical role of multiplexing in enhancing QR performance. Due to the probabilistic nature of SPDC photon sources, the MQR-TM scheme involves a trade-off between entanglement distribution rate and achievable fidelity. Our follow-up study \cite{Liu2026}  will provide an improved strategy with detailed theoretical analysis to assess the optimal operating conditions and scalability of the MQR-TM scheme in global-scale quantum networks.

\begin{table}[htb]
\caption{The reported state fidelity and EDR correspond to fully heralded two-party entanglement, measured after a storage time that exceeds the communication latency. Experiments lacking heralded storage or involving two nearby QMs connected via short-distance fibers are excluded.}
\label{tab:QR_performance}
\centering
\begin{tabular}{|c|c|c|c|c|c|} \hline
Experiment & \makecell QM separations & Entangling protocols & Fidelity / CHSH-Bell Test  & EDR  & Fiber length \\\hline
Single atoms 2012 \cite{Hofmann2012Heralded} & 20 m & TPI & \makecell{F=0.813(28), \textit{S}=2.19(9)} &  0.0094 Hz & 30 m\\\hline
Single NV center 2015 \cite{Hensen2015Loophole} & 1.3 km & \makecell{Two-photon detection} & \makecell{F=0.92(3), \textit{S}=2.42(0.20)} & 0.0003 Hz & 1.7 km \\\hline
\multirow{3}{*}{Single atoms 2022 \cite{van2022entangling}} & \multirow{3}{*}{400 m} & \multirow{3}{*}{TPI} & F=0.816(20),\textit{S}=2.244(63) & 0.053 Hz & 6 km \\\cline{4-6}
& & & F=0.719(12) &  0.012 Hz & 23 km \\\cline{4-6}
& & & F=0.622(15) &  0.0048 Hz & 33 km \\\hline
Single trapped ion 2023 \cite{Krutyanskiy2023Entanglement} & 230 m & TPI & F=0.857(42) & 0.058 Hz & 520 m\\\hline
Single SiV center 2024 \cite{Knaut2024Entanglement}& 6 m & Spin-photon logic & F=0.69(7) & 0.0012 Hz & 35 km\\\hline
\multirow{2}{*}{Single NV center 2024 \cite{stolk2024metropolitan}} & \multirow{2}{*}{10 km} & \multirow{2}{*}{SPI} & F=0.534(15) & 0.022 Hz & \multirow{2}{*}{ 25 km}\\\cline{4-5} & & & F=0.60(2) & 0.0083 Hz & \\\hline
Cold atomic ensemble 2024 \cite{Liu2024Creation} & 12.5 km & SPI & F=0.640(40) & 0.0037 Hz & 19.7 km \\\hline
\multirow{3}{*}{Rare-earth ions, this work 2025} & \multirow{3}{*}{14.5 km}  & \multirow{3}{*}{MQR-TM} & F=0.606(7) & 0.94 Hz & \multirow{3}{*}{17.8 km} \\\cline{4-5}
& & & F=0.632(10) & 0.47 Hz & \\\cline{4-5}
& & & F=0.786(20), S=2.22(6) &0.0055 Hz & \\\hline 
\end{tabular}
\end{table}

\newpage
\section{Overall experimental setup}
\label{Sec:Overall experimental setup}

\begin{figure*}[tbph]
\includegraphics [width=1\textwidth]{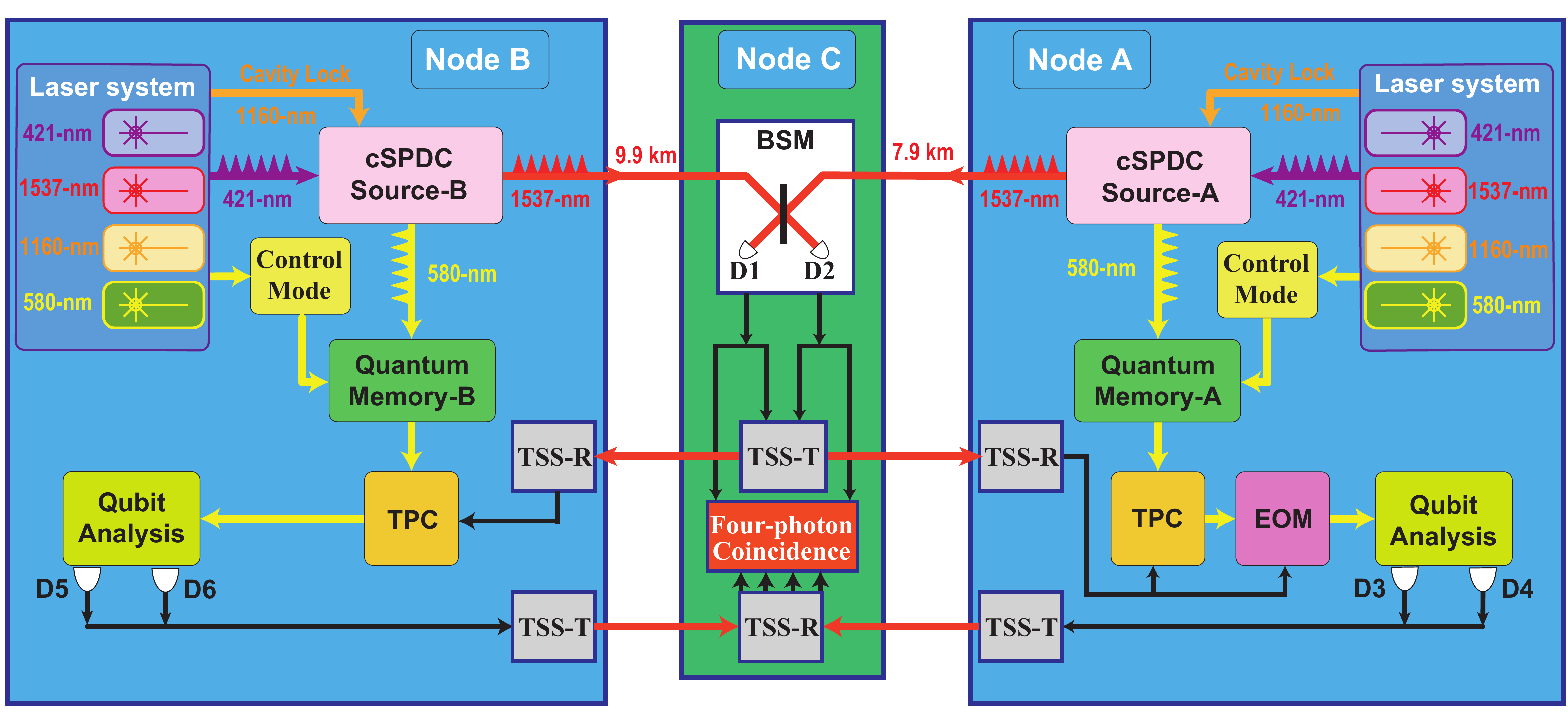}
\caption{The overall experimental setup. Both nodes A and B are equipped with independent-locked laser systems, cavity-enhanced spontaneous-parametric-down-conversion (cSPDC) sources, QMs, time-to-polarization conversion (TPC) modules, qubit analysis modules, and time-synchronization systems (TSSs). Node A additionally incorporates an Electro-Optic Modulator (EOM) to perform the corresponding local operation that maps the heralded atomic entanglement into the desired state.}
\label{fig:totalsm}
\end{figure*}

\begin{figure*}[tbph]
\includegraphics [width=0.45\textwidth]{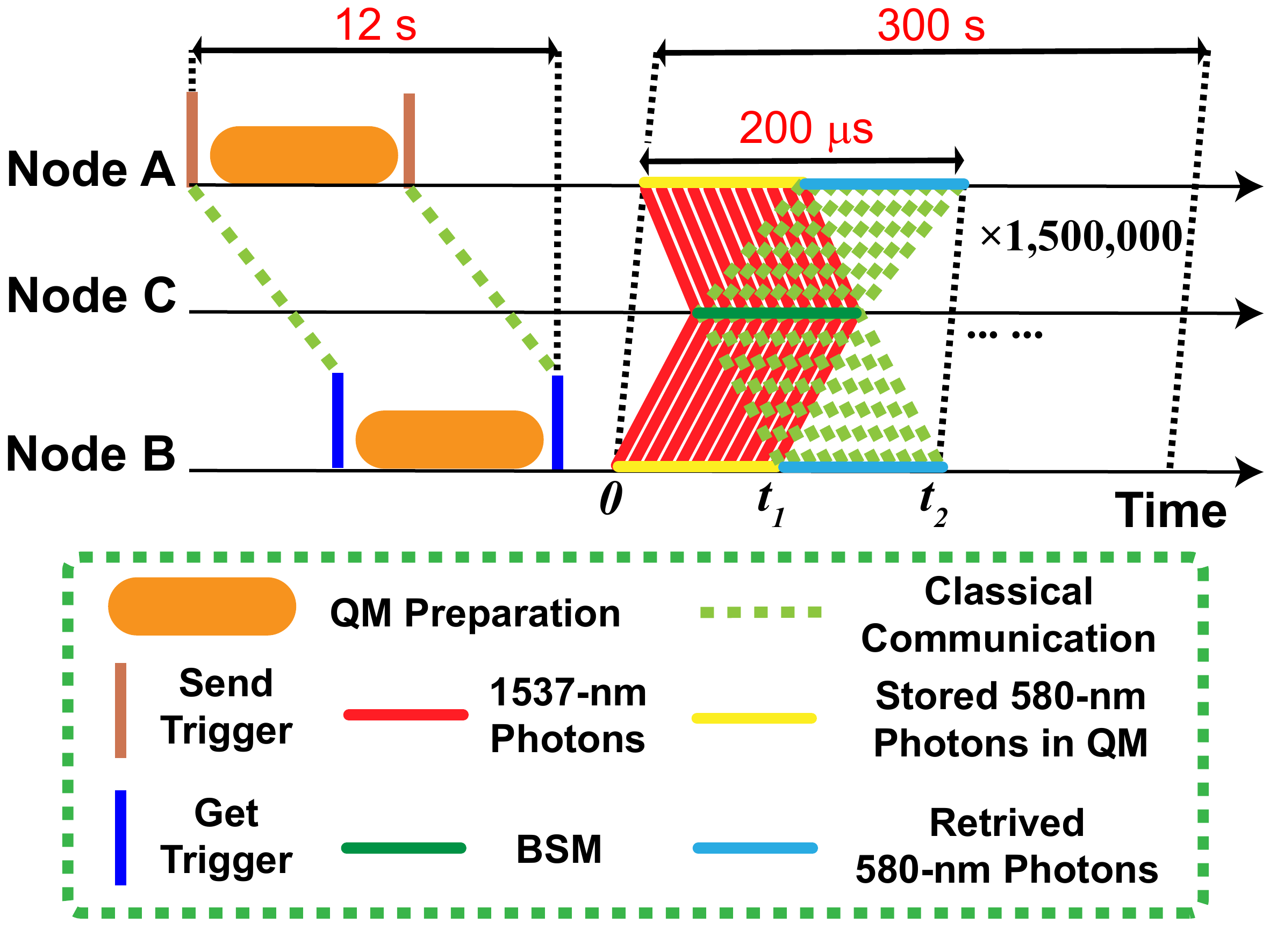}
\caption{The time sequence of our QR. The synchronization trigger transmitted from node A to node B serves as the start of the preparation of QMs at nodes A and B. The entangled photon-pair source at node B immediately generates photon pairs, while node A introduces a deliberate time delay to ensure simultaneous arrival of 1537-nm photons at node C for BSM. The BSM outcomes are then distributed to nodes A and B, enabling the preparation of desired entangled states before the final detection of 580-nm photons.}
\label{fig:total_sequence}
\end{figure*} 

Here we briefly introduce the overall experimental setup (as shown in Fig.~\ref{fig:totalsm} in a modular form). 
Node A serves as a QR node, located at the East Campus of the University of Science and Technology of China ($31^\circ 50^\prime 27.3^{\prime \prime}$ N, $117^\circ 16^\prime 20.7^{\prime \prime}$ E). Node B serves as another QR node, located at Hefei National Laboratory ($31^\circ 49^\prime 45.0^{\prime \prime}$ N, $117^\circ 07^\prime 05.4^{\prime \prime}$ E). Node C serves as the central Bell-state-measurement (BSM) node, located at a China Unicom office ($31^\circ 50^\prime 27.5^{\prime \prime}$ N, $117^\circ 11^\prime 55.5^{\prime \prime}$ E) approximately midway between nodes A and B. The direct inter-node distances are $14.5$ km (A–B), $7.0$ km (A–C), and $7.6$ km (B–C). The lengths of the deployed fibers are $7.9$ km (A–C) and $9.9$ km (B–C). 
     
QR nodes A and B are equipped with independent laser systems, cavity-enhanced spontaneous-parametric-down-conversion (cSPDC) sources, QMs, time-to-polarization conversion (TPC) modules, qubit analysis modules, and time-synchronization systems (TSSs). Node A additionally incorporates an Electro-Optic Modulator (EOM) module to perform the corresponding local operation that maps the heralded atomic entanglement into the desired state. Node C receives 1537-nm photons (telecom C-band) from nodes A and B, performing BSM to achieve entanglement swapping. The measurement outcomes from detectors $D1$ and $D2$ are copied and split into two paths: one routed independently to the TPC modules at nodes A and B through the transmission/reception terminals of the TSS (TSS-T/TSS-R), and the other is routed to the photon coincidence counter to await four-photon coincidence events. 

Fig.~\ref{fig:total_sequence} shows the complete timing of our QR system. After synchronized preparation of the QMs, the QR ``executes one emission-storage-BSM-and-retrieval" cycle every 200 \us~and repeats the sequence 1.5 million times.

Descriptions of the individual components are provided in the following sections: the laser systems are detailed in Sec.~\ref{Sec:Laser systems}; the photon-pair source is covered in Sec.~\ref{sec:cSPDC}, ~\ref{sec:Locking system and resonance of the two cSPDC sources}, and ~\ref{sec:Noise model}; the TPC module is explained in Sec.\ref{sec:TPC}; the QM is discussed in Sec.~\ref{Sec:Quantum memories}; optical losses are listed in Sec.~\ref{sec:Optical losses}; the synchronization and classical communication are described in Sec.~\ref{sec:Synchronization and classical communication} and the detailed experimental data is provided in Sec.~\ref{sec:Detailed experimental data}.

\section{Laser systems}
\label{Sec:Laser systems}
%（朱）

\begin{figure*}[tbph]
\includegraphics [width=0.9\textwidth]{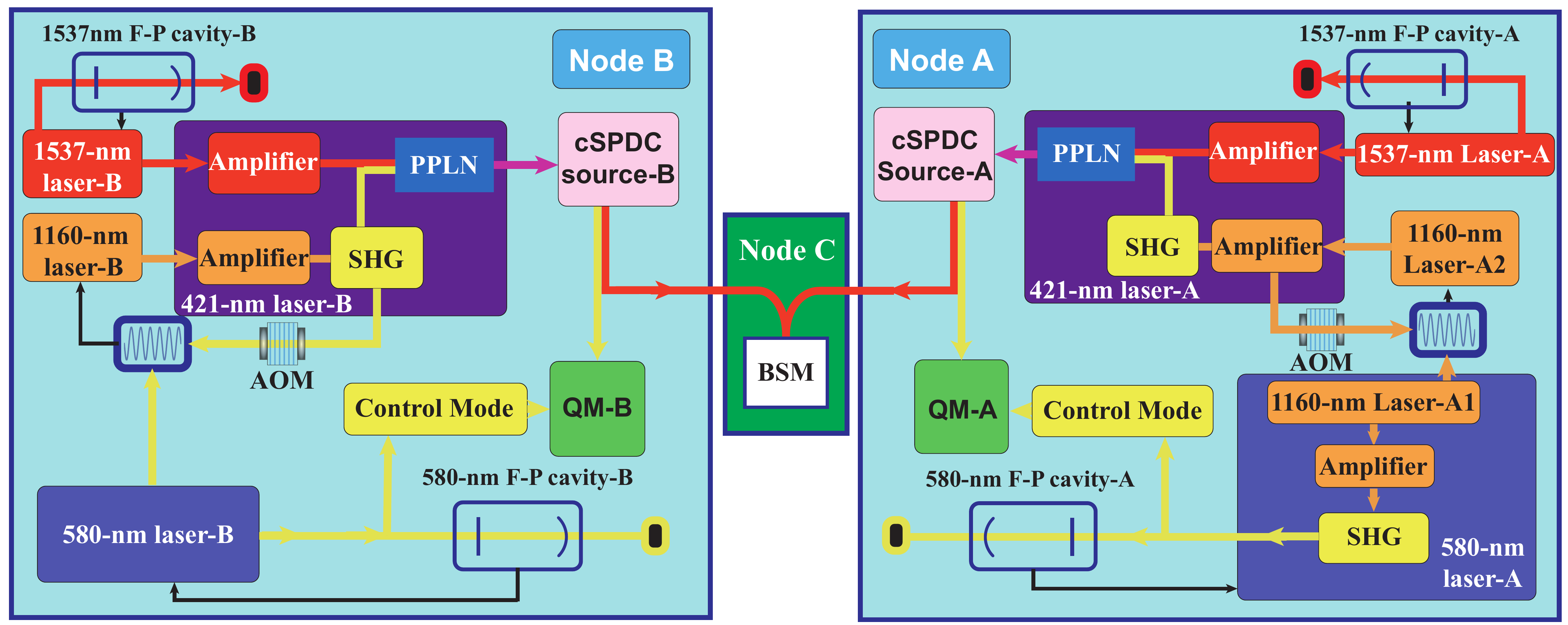}
\caption{Details of the laser system configuration. Lasers at nodes A and B are independently frequency-locked to local ultra-stable Fabry-Pérot (FP) cavities. The orange, yellow, red and purple lines correspond to the 1160-nm, 580-nm, 1537-nm and 421-nm beams, respectively. 
} 
\label{fig:laser_system}
\end{figure*}

\begin{figure*}[tbph]
\includegraphics [width=0.5\textwidth]{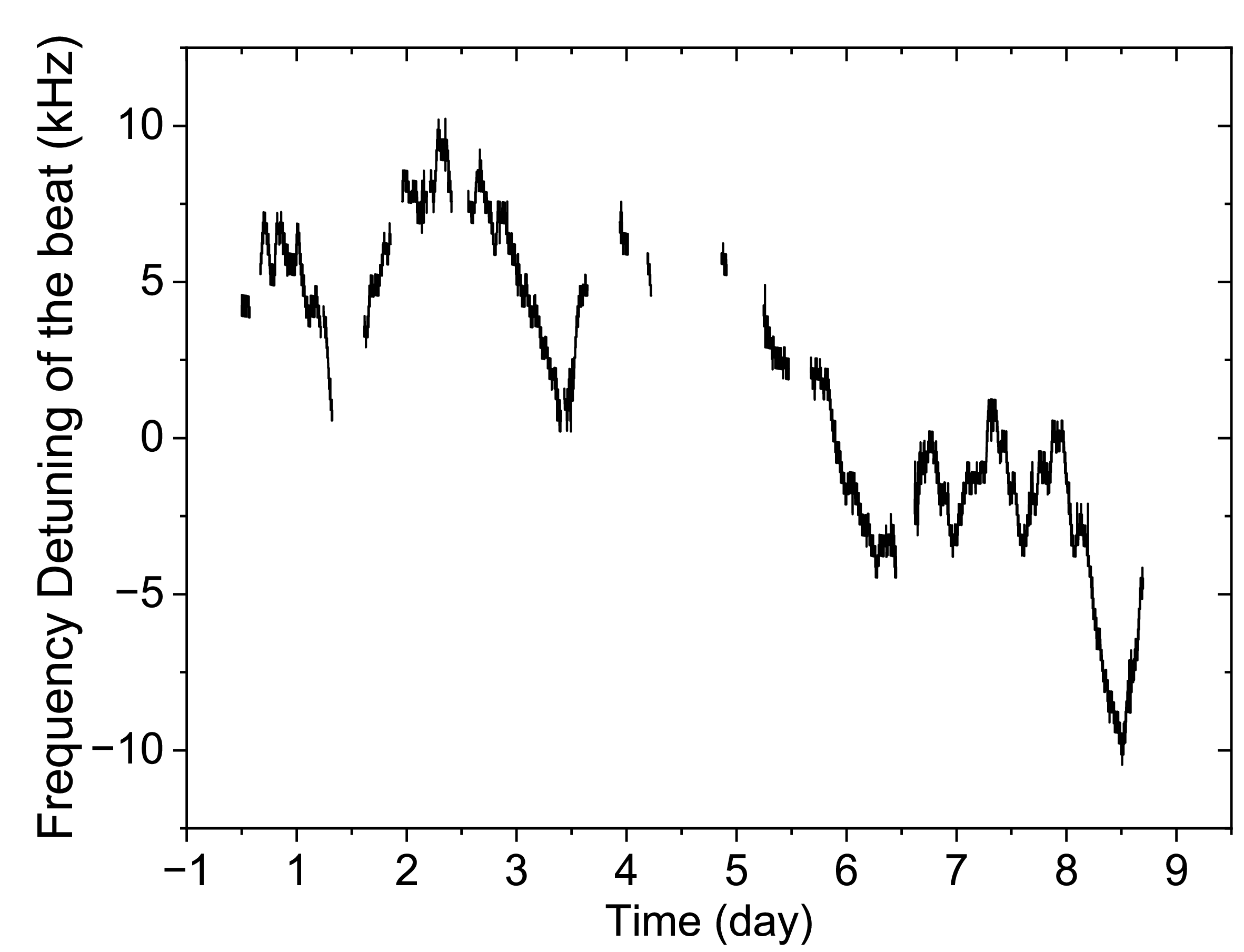}
\caption{Frequency detuning of the beat signal between the independently-locked 1537-nm lasers at nodes A and B. Data collected during lock-loss events caused by external disturbances are excluded; only locked-state data are retained.
} 
\label{fig:Beat_Frequency_detuning}
\end{figure*}

The detailed laser system for the QR is illustrated in Fig.~\ref{fig:laser_system}. Each node contains two independent laser systems dedicated to the QM and the cSPDC photon pair source, respectively. The Lasers within each node are frequency-stabilized independently, requiring no additional frequency locking between distant nodes. %To meet Bell-state measurement (BSM) requirements, the frequencies of heralded photons at both nodes must be closely matched. The frequency stability between nodes is characterized in Fig. ***.

Here we provide a detailed description of the laser system in Node A. The 580-nm laser-A (Toptica, TA-SHG) for QM-A is generated by second-harmonic generation (SHG) process pumped by amplified 1160-nm laser. The 580-nm laser-A is frequency locked to a 580-nm ultra-stable Fabry–Pérot (F-P) cavity using the Pound-Drever-Hall (PDH) scheme \cite{Black2001PDH}, achieving a linewidth of approximately $0.2$ kHz. The 421-nm Laser-A (Precilasers) serves as the pump laser for the cSPDC Source-A. It is generated via sum-frequency generation (SFG) of frequency-locked 580-nm and 1537-nm lasers within a PPLN crystal. The 580-nm laser originates from second-harmonic generation (SHG) of another amplified 1160-nm laser (1160-nm Laser-A2, Toptica). To ensure the down-converted 580-nm photons resonate with the QM, we lock the 1160-nm seed laser-A2 to the 1160-nm seed laser-A1 by beat-frequency locking. The 1537-nm laser is generated by a 1537-nm seed laser (1537-nm laser-A, Precilasers) amplified by an erbium-doped fiber amplifier (Precilasers), which is locked to an ultra-stable F-P cavity, yielding a linewidth of approximately 0.5 kHz. 

The laser system at node B closely resembles that of node A, with a distinction: the 580-nm Laser-B (Precilasers) for the QM is generated by SFG of a 931.5-nm laser and a 1537-nm laser. The 931.5-nm laser itself originates from SHG of a 1863-nm laser. 

To validate system stability, the frequency detuning between two 1537-nm lasers at node A and B is measured by beating the two lasers after transmission through deployed fiber to node C (Fig.~\ref{fig:Beat_Frequency_detuning}). Over a period of $8.2$ days, the frequency detuning of the beat remained within a $20$ kHz range, meeting the requirements for implementing BSM and enabling autonomous quantum node operations.

%\newpage
\section{cSPDC photon source}
\label{sec:cSPDC}
%（张）
%We build a widely nondegenerate cavity enhanced spontaneous parametric down conversion (cSPDC) source. The wavelength of the signal photon is close to 580 nm, compatible with the Eu$^{3+}$-doped solid state QM, while the wavelength of the idler photon is close to 1537 nm, falling within the telecom C-band. The spectral bandwidth of the down-coverted photons is about 10 MHz. Since the coherence length of the pump laser is much larger than the down-converted photons, the generated photon pairs are in a time-energy entangled state. By storing the signal photon as a collective excitation of the ions in the QM, we can produce entanglement between the QM and a telecom photon, which is critical in the long-distance quantum communication architectures. 

To construct a high-performance cSPDC source, we implement the following strategies: 1. A Type 0 phase-matching PPKTP crystal is adopted. Compared with PPLN crystals, it provides a higher optical damage threshold, reduced temperature sensitivity, and more stable phase-matching conditions. 2. The F-P cavity is designed to achieve triple resonance for the signal, idler, and locking wavelengths. The clustering effect resulting from the large wavelength separation between signal and idler helps improve the spectral brightness of the photon source. 3. The cavity length is stabilized using a continuous-wave reference beam at 1160 nm, enabling robust and continuous active locking.

\begin{figure*}[tbph]
\includegraphics [width=0.9\textwidth]{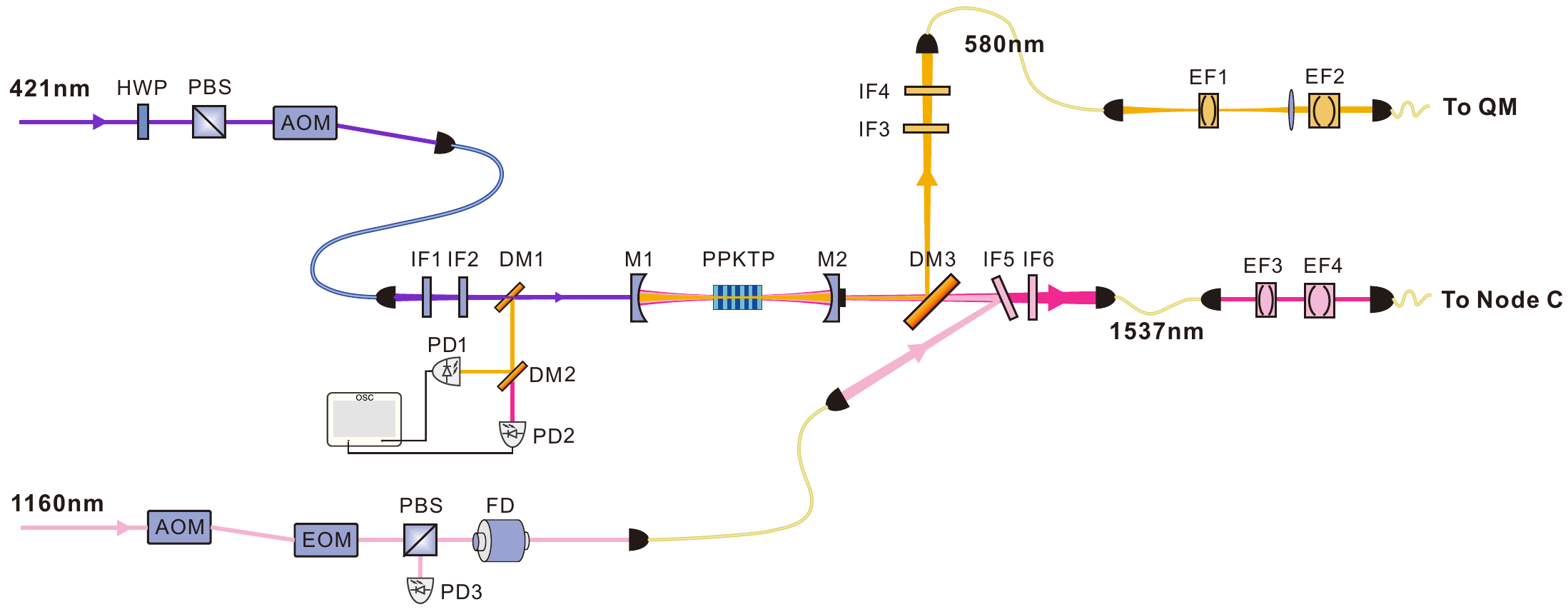}
\caption{The experimental setup for the cSPDC source. The AOM is used to gate the continuous 421-nm pump laser into pulses. The IF1 (Semrock, FF01-440/SP-25) and IF2 (Semrock, FF01-940/SP-25) are used to remove the residual 580 nm and 1537 nm beam in the SFG-based  421-nm laser respectively. The DM1 is used to transmit the pump laser and reflect wavelength at 580 nm and 1537 nm. DM2 and DM3 (Semrock, FF775-Di01-25*36) are used to reflect wavelength at 580 nm and transmit 1537 nm. The signal photon is filtered by a long-pass filter IF3 (Semrock, BLP01-442R-25) and a narrowband filter IF4 (centered at 580 nm, 1 nm bandwidth). The idler photon is filtered by a  narrowband filter IF5 (centered at 1550 nm, 12 nm bandwidth) and a long-pass filter IF6 (Thorlabs, FELH1500nm). IF5 is tilted so that its center wavelength shifts to 1537 nm, and it can also be used to reflect the reference beam at 1160 nm into the cavity. The reflected 1160 nm beam from the cavity is detected by PD3 and is used for locking the cavity length through PDH locking. The reflectivities of the plano-concave mirror M1 (M2) for signal, idler and locking wavelength are  designed to be $99.8\pm0.1\%$ ($92\pm2\%$), $99.8\pm0.1\%$ ($92\pm2\%$) and $99\pm0.3\%$ ($99\pm0.3\%$) respectively, and the transmittance for the pump wavelength exceeds $97\%$. The PPKTP crystal, with a poling period of 4.775 $\mu m$, is AR coated for the four wavelength (R$<$0.2$\%$). These result in a cavity finesse of about 50 and a FSR about 730 MHz. The FWHM of the phase matching envelope of the PPKTP crystal is 82.4 GHz. The signal photons pass through two cascaded etalons EF1 (bandwidth 3 GHz, FSR 89 GHz) and EF2 (bandwidth 250 MHz, FSR 9.4 GHz). The idler photons pass through two cascaded etalons EF3 (bandwidth 3 GHz, FSR 89 GHz) and EF4 (bandwidth 160 MHz, FSR 9.4 GHz). This ensures the photon pairs are filtered to a single spectral mode. FD, Faraday rotator; EF, etalon; IF, interference filter; DM, dichroic mirror; PD, photon detector.}
\label{fig:cspdc}
\end{figure*}

{\it Setup.--} The scheme of the photon-pair source is shown in Fig.~\ref{fig:cspdc}. The pump laser at 421 nm is gated into pulses according to the time sequence of the QM (Fig.~\ref{fig:time_sequence}) by an acousto-optic modulator (AOM) to reduce the background noise. The 2 cm long PPKTP crystal is placed in the middle of the two cavity mirrors M1 and M2 (R=100 mm). The cavity length is about 18.5 cm. The two cavity mirrors are mounted on a single stainless steel base to enhance passive stability. The PPKTP crystal is housed in a temperature-controlled enclosure with a temperature stability of 1 mK. The interference filters (IFs) and dichroic mirrors (DMs) are used to filter and separate photons of different wavelengths. The reference beam at 1160 nm first passes through a AOM for frequency tuning, then injects into the cavity by the reflection of IF5. The reflection beam from the cavity is used for locking the cavity length at resonance with both of the signal and idler frequencies. The generated signal (idler) photons are collected into single mode fibers and pass through two cascaded etalons and guided to the QMs (node C).

{\it Coincidence efficiency and counting rate.--} We first test the single-pass case without cavity (Source A). We choose a pump beam waist of 200 \um~and obtain a detected two-photon counting rate of 70 kHz/mW (bandwidth 82.4 GHz). The coincidence efficiency  is about $50\%$ with single-photon detector efficiency of $70\%$ ($90\%$) for the signal (idler) photon. The coincidence efficiency is defined as the geometric mean of the collection efficiencies of the signal photon and idler photon ($\sqrt{\eta_s\eta_D^s\eta_i\eta_D^i}$). After inserting the cavity mirrors, the collection collimator needs to be realigned to match the cavity mode. We find the coincidence efficiency and the two-photon counting rate dropped to $31\%$ and 60 kHz/mW respectively. This is due to the internal losses and the mismatch between the beam waist sizes of the signal and idler photons inside the cavity, which are 150 \um~and 250 \um~respectively in our case.  After passing through the etalons, both signal and idler photons are filtered into single spectral mode, resulting in a coincidence efficiency of $27\%$ and a two-photon counting rate of 4.5 kHz/mW (bandwidth 10 MHz). This corresponds to a 530-fold enhancement of the spectral brightness compared to the single-pass case. The peak transmittance for each etalon is about $90\%$. We estimate the heralding efficiency of the signal photon to be $34\%$ after correcting the detector efficiency. The results for source A and source B are summarized in Table~\ref{sourceparameter}. 

\begin{table}[htb]
\centering
\caption{The measured results of the Source parameters. The bandwidth for the signal and idler photons can be deduced from Fig.~\ref{fig:CorrelationofSource}. }
\begin{tabular}{|c|c|c|} \hline
Source parameters & Source A & Source B \\\hline
Coincidence efficiency & 27$\%$ & 31$\%$ \\\hline
Two-photon counting rate & 4.5 kHz/mW & 7.5 kHz/mW\\\hline
Generation rate & 61 kHz/mW & 78 kHz/mW\\\hline
Heralding efficiency of the signal & 34$\%$ & 36$\%$\\\hline
Bandwidth of the signal & 15.0 MHz & 13.3 MHz \\\hline
Bandwidth of the idler & 9.4 MHz &  10.1 MHz\\\hline
\end{tabular}
\label{sourceparameter}
\end{table}

{\it Estimation of the cavity parameters.--} We characterize the cavity parameters by injecting counter propagating 580 nm and 1537 nm reference beams and detecting the cavity transmission signals using PD1 and PD2. The measured linewidth and free spectral range (FSR) of the cavity for different wavelengths are shown in Table~\ref{cavityparameter}. The finesse of the cavity is equal to 
\begin{eqnarray}
F=\frac{\Delta \nu_{FSR}}{\Delta \nu}=\frac{\pi \rho^{1/4}}{1-\sqrt{\rho}},
\end{eqnarray}
$\rho$ is the power ratio recycled after one round trip
\begin{eqnarray}
\rho=R_1R_2(1-L_{RT}).
\end{eqnarray}
where $R_1$ ($R_2$) is the reflectivity of M1 (M2). Thus the internal round-trip loss of the cavity $L_{RT}$ can be inferred from the measured finesse. The escape efficiency can be deduced according to the internal losses
\begin{eqnarray}
\eta_{esc}=\frac{1-R_2}{1-R_2R_1(1-L_{RT})}.
\end{eqnarray}
The internal losses include the absorption in the crystal (absorption coefficient $\alpha<$ 50 ppm/cm at 1064 nm, $\alpha<$ 2000 ppm/cm at 532 nm), the reflectivity for the crystal facets (R$<$0.2$\%$) and the absorption of the cavity mirrors ($<$300 ppm). The reflectivity of M2 for 580 nm and 1537 nm is measured to be $90\%$ and $93\%$ respectively. We see the measured internal losses agree well with the above prediction ($2.6\%$ for signal, $1.1\%$ for idler).

\begin{table}[htb]
\centering
\caption{The measured results of the cavity parameters. $\Delta \nu_{FSR}$, free spectral range; $\Delta \nu$, linewidth; $F$, finesse; $L_{RT}$, the internal round-trip loss; $\eta_{esc}$, the escape efficiency. }
\begin{tabular}{|c|c|c|c|c|} \hline
 Cavity parameters & Source A, 580 nm & Source A, 1537 nm & Source B, 580 nm & Source B, 1537 nm  \\\hline
$\Delta \nu_{FSR}$ (MHz) & 713 & 737 & 728 & 731\\\hline
$\Delta \nu$ (MHz) & 15.2 & 9.5 & 16.6 & 9.9\\\hline
$F$ & 46.9 & 77.6 & 44.0 & 74.1\\\hline
$L_{RT}$ & 2.6$\%$ & 0.64$\%$ &3.5$\%$ &1.0$\%$ \\\hline
$\eta_{esc}$ & 79.8$\%$ & 90.0$\%$ &75.1$\%$ &86.2$\%$ \\\hline
\end{tabular}
\label{cavityparameter}
\end{table}

\begin{figure*}[tbph]
\includegraphics [width=0.9\textwidth]{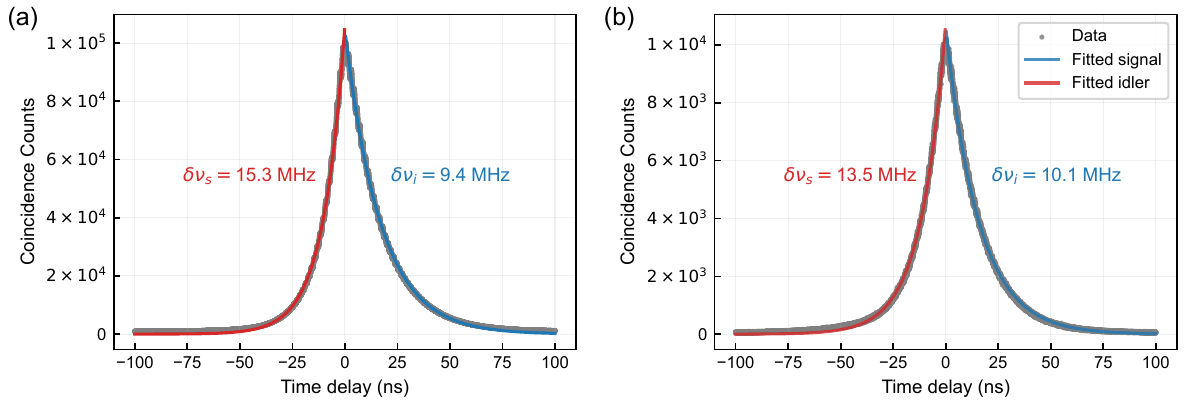}
\caption{Measured second-order cross-correlation functions for (a) source A and (b) source B.}
\label{fig:CorrelationofSource}
\end{figure*}

{\it Second-order cross-correlation function.--} To investigate the bandwidth of the down-converted photons we use the second-order cross-correlation function between the signal and idler photons $G_{s,i}^{(2)}$. The temporal $G_{s,i}^{(2)}(\tau)$ is defined as
\begin{eqnarray}
G_{si}^{\left(2 \right)}\left( \tau  \right) = \left\langle {\hat a_s^\dag \left( t \right)\hat a_i^\dag \left( {t + \tau } \right){{\hat a}_i}\left( {t + \tau } \right){{\hat a}_s}\left( t \right)} \right\rangle = |f_{R} (\tau)|^2
\end{eqnarray}
where $f_{R} (\tau)$ is the joint temporal function as the inverse Fourier transform of the joint spectral function (JSF). The JSF of the cavity enhanced down-converted photon pairs is the product of the conventional JSF and the Airy functions of the resonator ${f_R \left( {{\omega _s}{\rm{,}}{\omega _i}} \right)}=f \left( {\omega _s}{\rm{,}}{\omega _i} \right) A_s\left( {\omega _s} \right)A_i\left( {\omega _i} \right)$. The spectrum of the down-converted photons should be dominated by the Lorentzian shape of the cavity mode spectrum. And the temporal wave packet of the photons should feature an exponential decay. We record the detection times of both photons by using a fast coincidence unit (Swabin Time Tagger 20). We correlate the detected photon pairs as a function of the arrival time difference between the signal and idler photons $\tau=t_s-t_i$. The results are plotted in Fig.~\ref{fig:CorrelationofSource} (with signal photons as start events and idler photons as stop events). By using the function $e^{-2\pi\Delta\nu\tau}$ to fit the two sides of the curves separately, we obtain the bandwidth of the photons (shown in Table~\ref{sourceparameter}).

\section{Locking system and resonance of the two cSPDC sources}
\label{sec:Locking system and resonance of the two cSPDC sources}

The F-P cavity used in the entangled photon source must resonate simultaneously at both the signal and idler frequencies. Meanwhile, the frequency of the reference beam is tuned to achieve resonance with the cavity. Additionally, the signal frequency must resonate with the local \euyso~crystal, and the idler frequencies from the two sources need to be aligned to enable interference at node C. In this section, we present the locking system developed to meet these requirements.

As previously described, the 421-nm pump laser for the entangled photon source is generated by SFG of the 580-nm and 1537-nm lasers from the SPDC platform. First, we perform beat-frequency locking between this 580-nm laser and the 580-nm for the QMs to ensure that its frequency matching with the QM. The F-P cavity of the entangled photon source is equipped with a piezoelectric actuator that we can use it to scan the cavity length and to lock it. For node A, the F-P cavity length is first adjusted to resonate with the 580-nm laser. Then we tune the frequency of the 1537-nm laser and the frequency of the 1160-nm reference light to make them resonance with the F-P cavity, thereby realizing three-wavelength resonance at node A. Meanwhile we lock the 1537-nm laser to an ultra-stable cavity. For Node B, the 1537-nm laser is also locked to a local ultra-stable cavity. By detecting the beat frequency signal between the two 1537-nm lasers from node A and node B, we can ensure that the frequencies of the two lasers are identical. Since the 580-nm laser at node B is also frequency-locked to the QM's frequency, we need to perform a wide-range scan of the FP cavity length to achieve dual resonance at the fixed 580 nm and 1537 nm. We find that when scanning the cavity length with a period of 1537 nm, a minimum spacing between the dual-wavelength transmission peaks occurs every 20 cycles (corresponding to the least common multiple of the integer 580 and 1537). At this stage, slight adjustment of the PPKTP crystal’s temperature allows the two peaks to overlap. Finally, by tuning the frequency of the 1160 nm reference laser, three-wavelength resonance at node B is achieved.

\newpage
\section{Time-to-polarization conversion}
\label{sec:TPC}
%（张）

\begin{figure*}[tbph]
\includegraphics [width=0.8\textwidth]{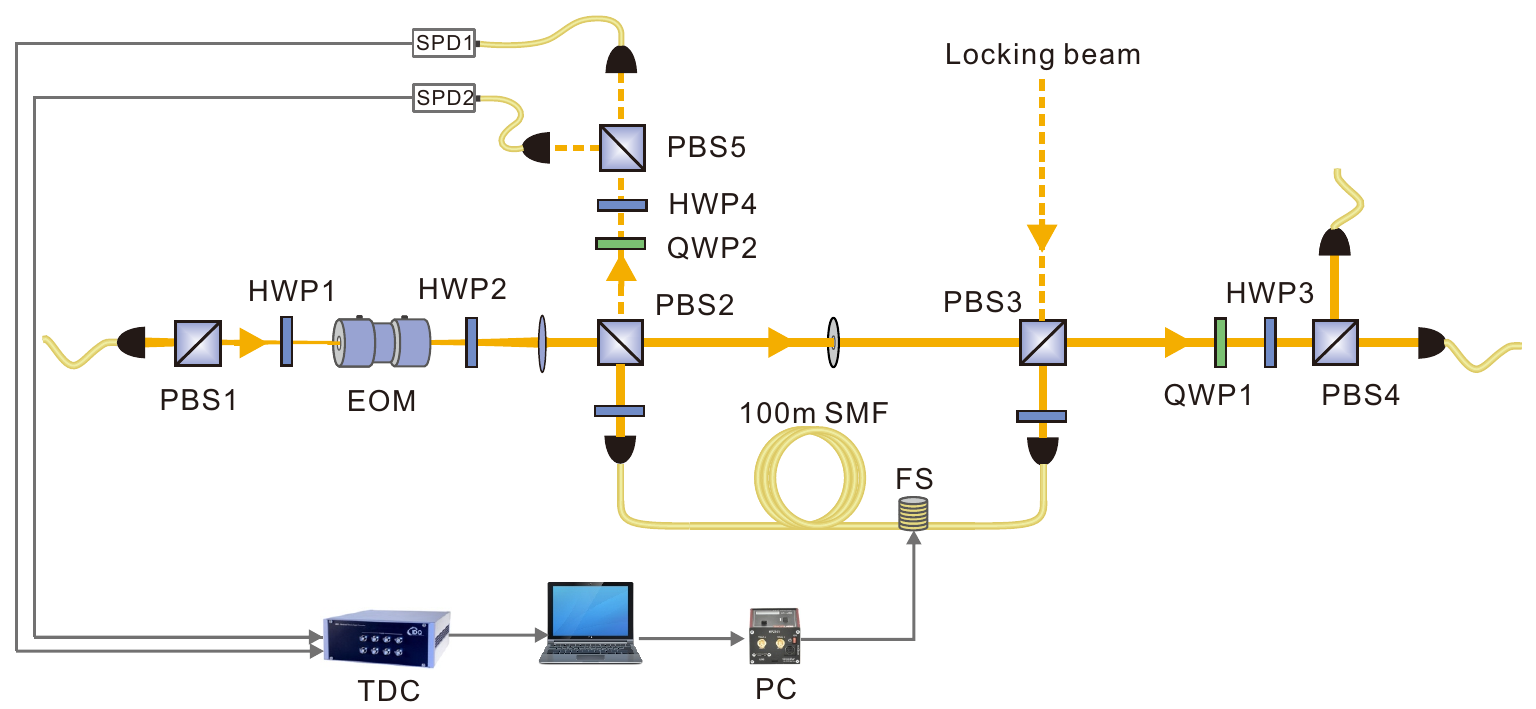}
\caption{Experimental setup for the TPC. The locking beam, prepared in the polarization state $H + \eta V$, is attenuated to the single-photon level (about 200 kHz). The parameter $\eta$ is adjusted such that the output polarization state becomes $H + e^{i\phi}V$. The long arm, which includes a 100-meter fiber and associated coupling components, introduces approximately 60\% optical loss. A tunable aperture is inserted in the short arm to balance the loss between the long and short arms, ensuring equal loss ratios at nodes A and B. QWP2 is set to $45^\circ$. By changing the angle of HWP4, we can change the projection basis and thus alter the locking point. The counting rate of the two single photon detectors is measured every 10 ms by a time-to-digital converter (TDC, IDQ ID800) and the error signal is equal to the counting rate difference between SPD1 and SPD2. The PID feedback signal produced by the computer is amplified by a piezo controller (PC, Thorlabs KPZ101) and applied to the fiber stretcher (FS) to actively compensate the relative phase between the two arms.}
\label{fig:interferometer}
\end{figure*}

The setup for time-to-polarization conversion (TPC) based on an uMZI is illustrated in Fig.~\ref{fig:interferometer}.  Signal photons enter from the left-side fiber with horizontal polarization. The EOM sandwiched by two $22.5^\circ$ HWPs functions as a high-speed polarization rotator. When a high voltage is applied to the EOM, it will flip the input photons from horizontal polarization to vertical polarization, thereby routing the early time-bin mode to the long arm and the late mode to the short arm. The long arm introduces a 500 ns delay to the early-mode photons, enabling the interferometer to convert time-bin encoded photons—separated by a 500 ns interval—into polarization-encoded photons. The output photons are then directed into a polarization analysis system composed of QWP1, HWP3, PBS4, and two fiber-coupled single-photon detectors.

\begin{figure*}[h]
\includegraphics [width=0.9\textwidth]{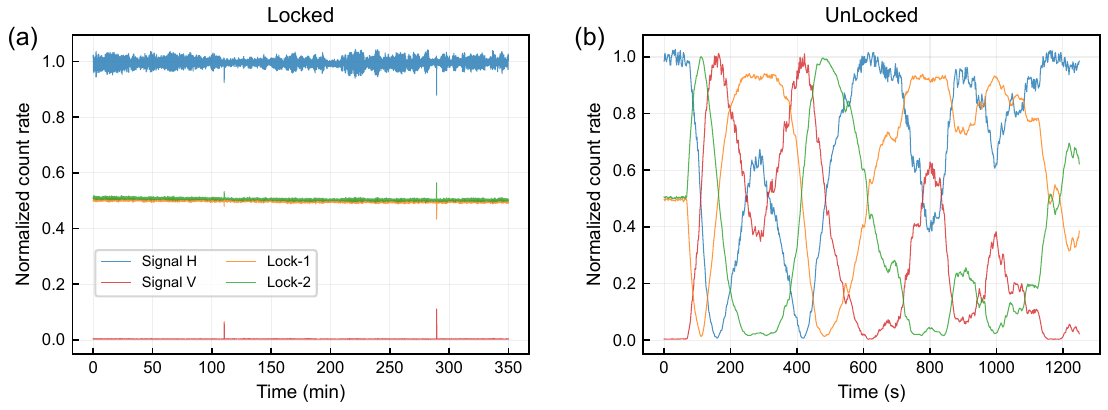}
\caption{(a) The phase locking performance of the uMZI over 6 hours. We reset the PC voltage to 25 V whenever it exceeds the operational range of 0–50 V (indicated by sharp resets in the trace), then the controller quickly locks to the nearest setpoint, enabling continuous phase stabilization. The average extinction ratio achieved is 276:1. (b) Without active feedback, the interferometer phase fluctuates significantly within seconds.}
\label{fig:Interference_stability}
\end{figure*}

We use a reference laser beam with the same frequency of the signal photons to lock the interferometer. This locking beam is attenuated to the single-photon level and counter-propagates relative to the signal photons, allowing continuous operation of the locking system without adding extraneous noise. The locking beam exits the interferometer through a separate port of the PBS2 and is detected by a polarization analysis system to extract phase information. The PID feedback signal is generated every 10 ms and is send to a piezo-electric fiber stretcher to actively compensate the relative phase between the two arms. To enhance the passive stability of the interferometer, we cover the 100 m fiber and fiber stretcher with thermal insulated material and implement active temperature control with a temperature stability of 1 mK. The locking performance was evaluated by injecting an additional attenuated laser beam from the left-side fiber, as shown in Fig.~\ref{fig:Interference_stability}.

\newpage
\section{Noise model for the swapped photonic entanglement}
\label{sec:Noise model}

The noise for the swapped photonic entanglement mainly includes white noise and dephasing noise. Suppose the targe state is $|\psi^+\rangle$, then we can model the noise state as \cite{nielsen2010quantum}
\begin{eqnarray}
\rho_{q,r}=(1-q)(1-r)|\psi^+\rangle\langle\psi^+|+(1-q)r|\psi^-\rangle\langle\psi^-|+q\frac{\mathbbm{1}^2}{2^2}
\end{eqnarray}
where $q$ represents the proportion of the white noise and $r$ represents the proportion of the dephasing noise. The white noise mainly originates from the higher-order emission noise of the SPDC process. The probability that both sources emit one photon pair is proportional to $p^2$, if one source emits two photon pairs, it will introduce higher-order noise with a probability proportional to $p^3$. The proportion of the white noise can be estimated from the result of the ZZ measurement. The dephasing noise arises mainly from the imperfection of the Hong-Ou-Mandel interference, imperfection of the TPCs, and phase instability of lasers. It can be estimated from the result of the XX measurement. For example, for pump power of $\{3,9,18\}$ mW and coincidence window of 20 ns, the measured state fidelity is $\{77.42\%,73.45\%,62.13\%\}$ for $|\psi^+\rangle$ respectively, we can estimate the parameters $(q,r)$ to be $\{(0.161, 0.125),(0.233,0.118),(0.404,0.127)\}$ based on the measurement results of the entanglement witness. By using more stable lasers and TPC interferometers, the parameter $r$ could be reduced to 0.04, limited only by the imperfection of the HOM interference. By further improving the coincidence efficiency of the photon source to $50\%$, we can employ a lower pump power to reduce the parameter $q$ below 0.08, thus a entanglement fidelity exceeding $90\%$ could be achieved.

\newpage
\section{Quantum memories} 
\label{Sec:Quantum memories}

\begin{figure*}[tbph]
\includegraphics [width=0.8\textwidth]{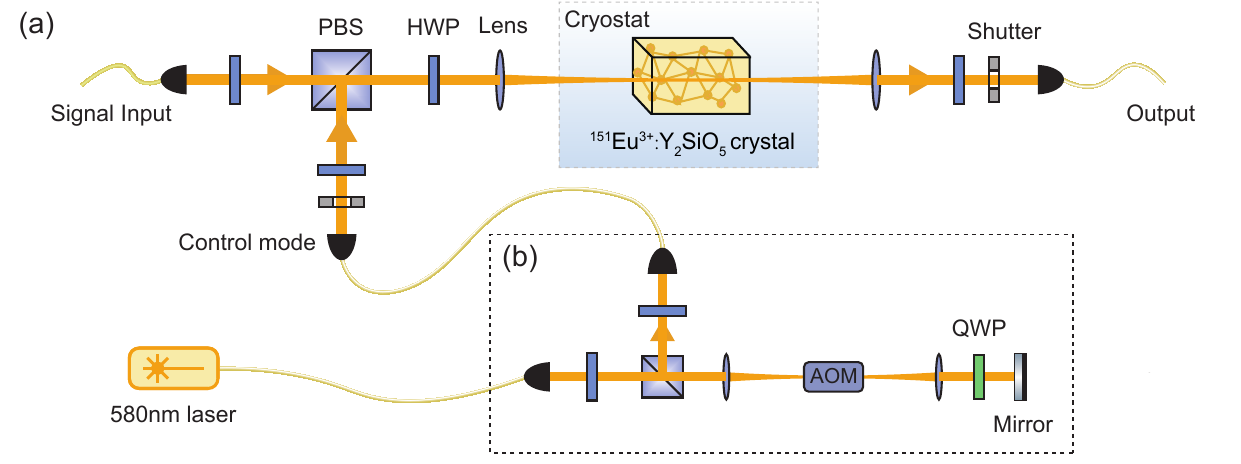}
\caption{Schematic diagram of quantum memories in node A and B. (a) \euyso~crystals are cooled to approximately 3 K within closed-cycle cryostats (Montana Instruments). Signal and pump beams are combined using a polarization beam splitter (PBS). A $4f$ imaging system (two lenses with focal length 300 mm) focuses the combined beam into the crystal. Optical shutters are employed to protect single-photon detectors (SPDs). (b) A double-pass acousto-optic modulator (AOM) generates the control mode light. QWP: quarter-wave plate.}
\label{fig:qm_light}
\end{figure*}

As shown in Fig.~\ref{fig:totalsm}, two QM systems with similar designs are equipped at nodes A and B. The optical configurations for both systems are presented in detail in Fig.~\ref{fig:qm_light}(a) and (b). In this configuration, \euyso~ crystals are maintained at a temperature of $\sim 3$ K using closed-cycle cryostats (Montana Instruments). To minimize mechanical disturbances, the designed vibration-isolated sample holders (VI-SH) are supported by six stainless-steel springs, which effectively attenuate vibrations originating from the cryostat cold heads. Through measuring the amplitude decay of the two-pulse photon echoes as a function of the temporal spacing of the two pulses, we obtain the optical coherence times are $T_2=927\pm 17$ \us~ for QM-A and $T_2=491\pm 10$ \us~ for QM-B. These exceptionally long coherence times facilitate the preparation of high-finesse atomic frequency combs (AFC), a critical requirement for achieving multiplexed quantum storage with extended storage times.

To fully utilize the sample absorption, the QMs are operated at the central frequency of optical inhomogeneous broadening of the specific crystal sample \cite{Afzelius2009Multimode}. After characterizing the optical inhomogeneous broadening, we set the center frequencies of signal photons generated from cSPDC to 516.84711~THz for node A and 516.84713~THz for node B.
The crystals' nature absorption depths are $4.1$ for QM-A and $4.2$ for QM-B, with inhomogeneous broadening of 580 MHz for QM-A and 470 MHz for QM-B, respectively.
The 580-nm pump light is generated using a double-pass acousto-optic modulator (AOM), as shown in Fig.~\ref{fig:qm_light}(b). The pump light is combined with the signal mode via a polarization beam splitter (PBS), with the signal mode polarized along the $D_1$ axis of the crystal and the pump light along the $D_2$ axis. The beam waist diameters at the crystal focus are 90~\textmu m for the signal mode and 210~\textmu m for the pump beam.

\begin{figure*}[tbph]
\includegraphics [width=1\textwidth]{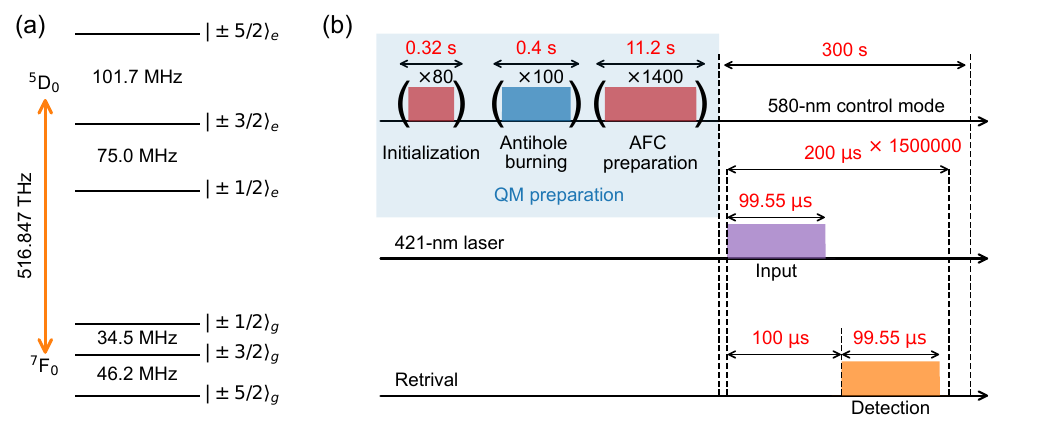}
\caption{(a) Energy levels for the $\mathrm{{^7}F{_0}\rightarrow{^5}D{_0}}$ transition for site-1 \euyso~ions at zero magnetic field. (b) Experimental timing sequence for the 580-nm pump laser, 421-nm laser, and measurement. The total experimental cycle lasts 312~s, including AFC initialization (12~s) and $1.5 \times 10^6$ input-detection cycles (300~s). Each input-detection cycle gates the 421-nm laser into 99.95-\textmu s pulses to generate input photons, with 1205 temporal modes stored simultaneously within a 99.55-\textmu s input window. After a 100-\textmu s storage period, photons are retrieved and detected within a 99.55-\textmu s window.}
\label{fig:time_sequence}
\end{figure*}

\begin{figure}[tbph]
\includegraphics [width=1\textwidth]{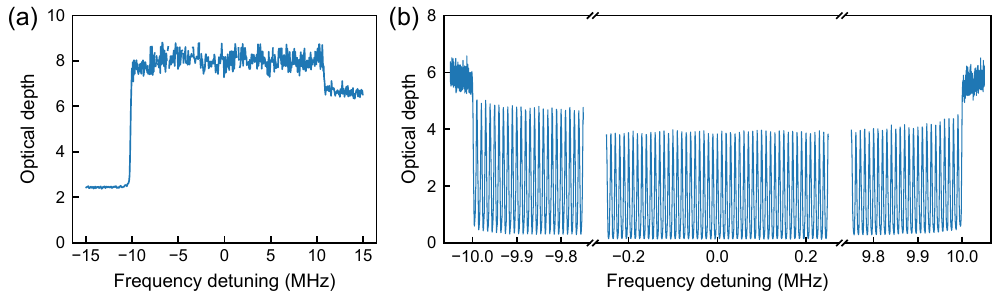}
\caption{(a) Enhanced absorption profile for QM-A with a 20-MHz bandwidth. (b) Measured AFC structure for QM-B, with a total bandwidth of 20 MHz and a comb spacing, comprising 2000 spectral teeth (three subsets shown for clarity).}
\label{fig:al_AFC}
\end{figure}

We employ site-1 $^{151}\mathrm{Eu}$ ions for AFC storage, leveraging the optical transition $\mathrm{{^7}F_0~\rightarrow~^5D_0}$, as shown in Fig.~\ref{fig:time_sequence}(a). The operational time sequence for the QMs is depicted in Fig.~\ref{fig:time_sequence}(b). As illustrated in Fig.~\ref{fig:total_sequence}, a synchronization trigger transmitted from node A to node B initiates QM preparation at both nodes, which consists of three stages:

\begin{enumerate}
    \item \textit{Population Initialization}: Chirped pulses with a frequency range of $[f_0 - 10~\mathrm{MHz}, f_0 + 10~\mathrm{MHz}]$, a duration of 4~ms, and 80 repetitions are applied to hole-burn a 20-MHz frequency band aligned with the signal photon frequency, where $f_0$ is the center frequency of signal photons for nodes A (516.84711~THz) and B (516.84713~THz).
    \item \textit{Enhanced Absorption Band Preparation}: Chirped pulses spanning $[f_0 - 90.85~\mathrm{MHz}, f_0 - 10.15~\mathrm{MHz}]$, with a 4-ms duration and 100 repetitions, pump the population back into the 20-MHz band. Inspired by Ref.~\cite{Zhu2022On-Demand}, this sequence creates an enhanced absorption band, as shown in Fig.~\ref{fig:al_AFC}(a) for QM-A, with a bandwidth of 20~MHz and an absorption depth of $\sim$8.0 for QM-A (1.95-fold enhancement over the natural absorption depth of 4.1) and $\sim$7.9 for QM-B.
    \item \textit{AFC Structure Preparation}: For a 20-MHz, 100-\textmu s AFC, 2000 comb teeth are prepared simultaneously. We optimized the long-term frequency stability of the 580-nm lasers (580-nm laser-A and 580-nm laser-B) and adopted a high-efficiency AFC-preparation waveform from Ref.~\cite{Liu2024Nonlocal}. The waveform parameters are: temporal duration $T_{\text{prep}} = 8.018~\mathrm{ms}$, complex-hyperbolic-secant parameter $\beta = 17.627/T_{\text{prep}}$, and spectral hole-burning comb width $\Delta_f = \Delta - \gamma$ of 4.1~kHz for node A and 4.4~kHz for node B, where $\Delta$ is the AFC frequency period and $\gamma$ is the comb tooth width. The AFC-preparation pulse is repeated 1400 times at each node, resulting in a total preparation time of 11.2~s. The final 20-MHz AFC structure for QM-B is shown in Fig.~\ref{fig:al_AFC}(b).
\end{enumerate}

\begin{figure*}[tbph]
\includegraphics [width=0.9\textwidth]{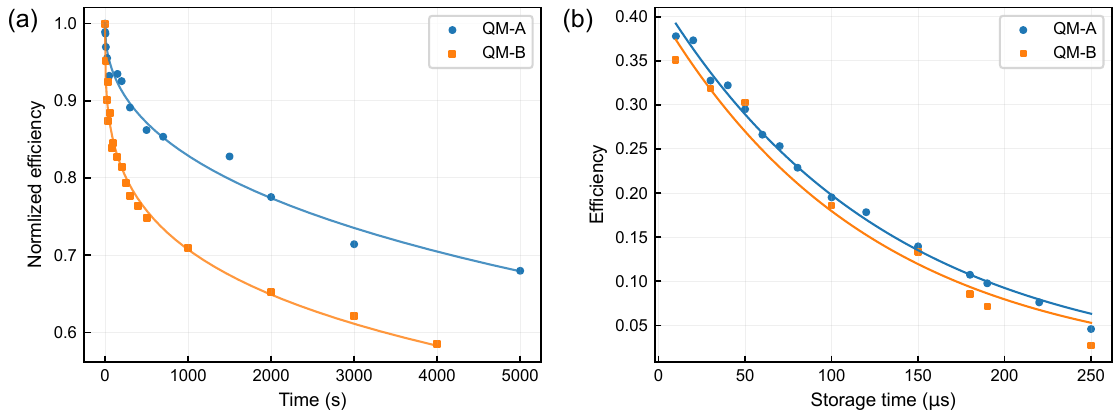}       
\caption{(a) Normalized storage efficiency versus time interval between AFC preparation and signal input. Blue circles and orange squares correspond to QM-A and QM-B, respectively. Both datasets are fitted to $\exp[-(t/T_1^{\mathrm{AFC}})^x]$, where $x$ is a fitting parameter, giving $T_1^{\mathrm{AFC}} = (4.12\pm 0.74) \times 10^{4}~\mathrm{s}$ for QM-A and $T_1^{\mathrm{AFC}} = (2.73 \pm 0.38) \times 10^{4}~\mathrm{s}$ for QM-B. 
(b) Storage efficiency as a function of storage time with 20-MHz storage bandwidth. Blue circles (QM-A) and orange squares (QM-B) are fitted with $\eta_0 \exp(-4t/T_2^{\text{AFC}})$, yielding $T_2^{\text{AFC}} = 526 \pm 15$ \us~ (QM-A, blue line) and $T_2^{\text{AFC}} = 491 \pm 44$ \us~(QM-B, orange line). Input pulses are Gaussian with a 100-ns FWHM.}
\label{fig:AFC_T1_T2}
\end{figure*}

To evaluate QM performance, we measure storage efficiency as a function of time after the AFC preparation, as shown in Fig.~\ref{fig:AFC_T1_T2}(a). The AFC structure lifetimes are $T_1^{\text{AFC}} = (4.12 \pm 0.74) \times 10^4~\mathrm{s}$ for QM-A and $T_1^{\text{AFC}} = (2.73 \pm 0.38) \times 10^4~\mathrm{s}$ for QM-B, representing a $\sim$10,000-fold improvement over the 1.6~s lifetime of 0.2\% $\mathrm{^{153}Eu^{3+}}$:$\mathrm{Y_2SiO_5}$ crystals~\cite{Liu2024Nonlocal}. This extended lifetime enables multiple storage/detection cycles, achieving a duty cycle of 48.1\%, a significant improvement over the 3.7\% reported previously~\cite{Liu2024Nonlocal}, thereby  greatly enhancing the overall EDR.

We further measure the AFC storage efficiency versus storage time (Fig.~\ref{fig:AFC_T1_T2}(b)) and extract AFC coherence times of $T_2^{\text{AFC}} = 526 \pm 15~\textmu s$ for QM-A and $T_2^{\text{AFC}} = 491 \pm 44~\textmu s$ for QM-B—more than twice the values reported in previous studies~\cite{Liu2024Nonlocal,Ortu2022Storage}. To the best of our knowledge, these are the longest AFC-coherence times reported to date and approach the fundamental limit imposed by the optical-coherence lifetime. The improvement is attributed to the use of a lower-concentration (0.01\%) $\mathrm{^{151}Eu^{3+}}$:$\mathrm{Y_2SiO_5}$ crystal, which reduces Eu-Eu interactions and spectral broadening, thereby enhancing both the spectral precision and hole-burning lifetime.

The 100-\textmu s storage efficiency is $\eta_A = 19.5 \pm 0.9\%$ for node A and $\eta_B = 18.6 \pm 0.4\%$ for node B, measured with 100-ns FWHM Gaussian input photons. For bandwidth-matched 580-nm signal photons, as shown in Fig.~1(c) of the main text and Fig.~\ref{fig:qm_trace_205}, the 100~\us~quantum storage average efficiencies reach $16.6 \pm 0.1\%$ for node A and $15.7 \pm 0.1\%$ for node B (Fig. \ref{fig:qm_trace_205}), integrated over the duration of $300$ s after the AFC preparation.
 
{\begin{figure*}[tbph]
\includegraphics [width=0.6\textwidth]{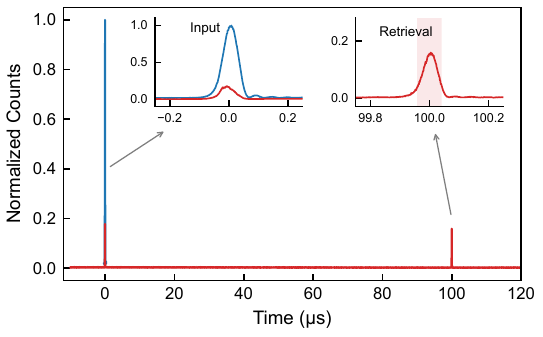}
\caption{ Heralded storage of 580-nm photons for 100~\us~at node B. The blue trace shows transmitted photons after a 20-MHz transparency window while the red trace shows the storage process. The inset provides a close-up of the retrieved echo, revealing a single-mode duration of 83 ns and an efficiency of $15.7\pm0.1\%$.}
\label{fig:qm_trace_205}
\end{figure*}

% \subsection{8. The characteristics of quantum channels and classical channels}
%（朱，欧）1

\newpage
\section{Optical losses}
\label{sec:Optical losses}
We have characterized the component efficiencies in detail, as summarized in Table \ref{tab:component efficiencies}. The transmittance of the cascaded etalons (580 nm and 1537 nm), the optical transmission efficiency of QMs, the TPC, and the field-deployed optical fiber (1537 nm) are measured with classical light. The heralding efficiency of the cSPDC sources, the bandwidth matching efficiency, and the internal storage efficiency were evaluated using heralded single-photon measurements. $\eta_{T_w}$ varies between approximately 44\% and 90\% for coincidence windows ranging from 15 ns to 40 ns. The bandwidth matching efficiencies between the 580-nm signal photons and the QMs are determined by measuring the ratio of the signal transmission through a 20 MHz spectral pit to that through a 34 MHz spectral pit. These efficiencies are found to be 90\% at node A and 91\% at node B, demonstrating excellent frequency mode matching between the photons and the memories. The internal storage efficiency was specifically determined by storing 580-nm signal photons generated via cSPDC, with a storage duty cycle of 48.1\% as depicted in the timing sequence in Fig. \ref{fig:time_sequence}(b).

%This efficient photon–memory interface, combined with multiplexing capability, enables heralding rates between signal and idler photons of 3.7 kHz (Node: A–C) and 3.1 kHz (Node: B–C) at a pump power of  $P\approx 18$ mW and a 40-ns coincidence window.

\begin{table}[h]
\centering
\caption{The success probabilities or efficiencies of each step and the losses of each component.}
\begin{tabular}{|c|c|c|} \hline
 Steps and components& node A& node B\\\hline
 Loss of the field-deployed optical fiber (1537) &1.54 dB (7.9 km)& 3.24 dB (9.9km) \\\hline
 Transmittance of the cascaded etalons (580-nm)& 83.5\% & 81\%\\\hline
 Heralding efficiency of the cSPDC sources & 34\% &36\%\\\hline
 Optical transmission efficiency for QMs & 82\% & 85\%\\\hline
Photon–quantum memory bandwidth-matching efficiency & 90\% & 91\% \\\hline
Internal storage efficiency of quantum memory & 16.6\% & 15.7\% \\\hline
Average transmission of TPC &58\% &60\% \\\hline
Detection efficiency (580-nm) & 80\% & 70\% \\\hline
Transmittance of the cascaded etalons (1537-nm)& \multicolumn{2}{|c|}{81\% (Node C)} \\\hline
Detection efficiency (1537-nm) & \multicolumn{2}{|c|}{90\% (Node C)} \\\hline
\end{tabular}
\label{tab:component efficiencies}
\end{table}

\newpage
%（欧）
\section{Synchronization and classical communication}
\label{sec:Synchronization and classical communication}
%（朱）

As shown in Fig.~\ref{fig:total_sequence}, after the preparation of the QMs, the 421-nm pump laser is turned on every 100 \us~to generate photon pairs, followed by a 100 \us~signal detection period during which the detector gate is open. This clock cycle is controlled by the arbitrary waveform generator (AWG) associated with the QM. At each node, the AWG signal is sent to the cSPDC module to regulate the gating AOM for the 421-nm pump laser, thereby synchronizing photon-pair generation with the QM operation. Additionally, the AWG signal from node A is used to trigger the AWG at node B. By counter-propagating a 580-nm reference beam in each source, a 1537-nm difference-frequency signal is generated. Synchronization between the two nodes is achieved by adjusting the relative delay between the AWG signals until the two 1537-nm difference-frequency signals overlap at node C.

%We test the polarization extinction ratio of the EOM under high voltage to be approximately 200:1, which is significantly higher than the fidelity of the heralded entangled state.
% The optical to electrical conversion devices at Nodes A and B are equipped with adjustable delay functionality, the synchronization precision is 5 ns.
Node C comprises two TDCs, with TDC1 dedicated to analyzing two-photon detection events at node C. The success signal is converted optical pulses and transmitted to nodes A and B to trigger the EOM inside TPC modules, enabling deterministic TPC. When the two detection events in node C originates from different detectors, a signal is further sent to node A to trigger the EOM (with an extinction ratio of 200:1), applying a $\sigma_z$ operation for the state to always prepare the heralded entangled state into $|\psi^+\rangle$. Then the electrical signals of the 580 nm single-photon detectors at nodes A and B are converted to optical signals and transmitted to node C. TDC2 is used to analyze the four-fold coincidence events. The synchronization precision for the single-photon detection signals is 220 ps.

Our quantum signal and classical signal are transmitted through the same multi-core fiber optic cable (24 cores). We measure the loss of the two fiber cables, with an average insertion loss of 1.54 dB for the 7.9 km cable and 3.24 dB for the 9.9 km cable. Although both fibers are of the low-loss type (G654E, $\sim 0.17$ dB/km), the higher loss in the 9.9 km cable is due to additional splicing points. For the experiment, we selected the fiber core with the lowest insertion loss as the quantum channel. Furthermore, we evaluated the polarization stability of both fiber cables. As shown in Fig. \ref{fig:Polarization_stability}, significant polarization drift is observed primarily during the transition between day and night. Consequently, only two polarization adjustments per day are required to maintain stable operation.

\begin{figure*}[htbp]
\includegraphics [width=0.5\textwidth]{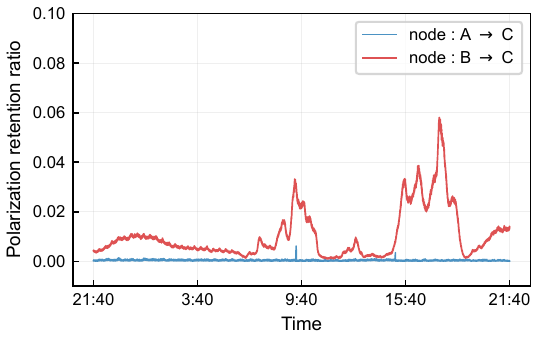}
\caption{The polarization extinction ratio stability of fiber links A–C and B–C, measured using attenuated single photons. The polarization extinction ratio is defined as the ratio of the number of photons in one polarization state (e.g., horizontal) to the number of photons in the orthogonal polarization state (e.g., vertical).}
\label{fig:Polarization_stability}
\end{figure*}

\newpage
\section{Detailed experimental data}
\label{sec:Detailed experimental data}
%欧，张，朱）

Here we provide supplementary data corresponding to Fig. 2 in the main text. Fig.~\ref{fig:SM_Witness_and_CHSH_without_QMs}(a) and (b) show the entanglement witness measurement for the swapped photonic entanglement with different pump power and coincidence window. Fig.~\ref{fig:SM_Witness_and_CHSH_without_QMs}(c) and (d) show the CHSH test with $3$-mW pump power and variable coincidence window. 
Fig.~\ref{fig:SM_tomography} and Fig.~\ref{fig:SM_tomography_fidelity} show the tomographic measurements with $3$-mW pump power and variable coincidence windows. 

Raw data corresponding to Fig. 3 and Fig. 4 in the main text are provided in Tab.~\ref{data_entanglement_witness}–\ref{data_CHSH_with_QM}.

\begin{figure*}[htbp]
\includegraphics [width=0.95\textwidth]{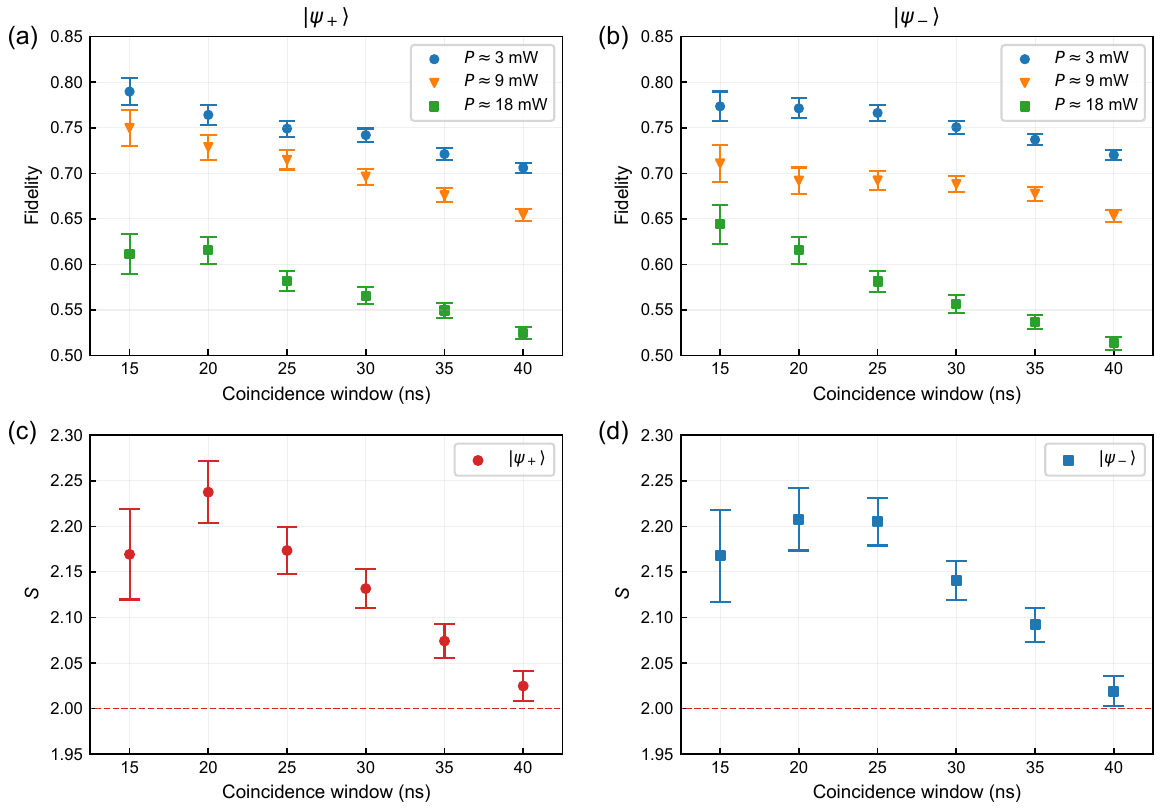}
\caption{Entanglement witness and CHSH-Bell tests of swapped photonic entanglement, measured in a delayed-choice configuration. (a,b) Entanglement witness fidelity for $|\psi_+\rangle$ (a) and $|\psi_-\rangle$ (b) as a function of the coincidence time window, measured at pump powers of $P \approx 3~\text{mW}$ (blue circles), $P \approx 9~\text{mW}$ (orange triangles), and $P \approx 18~\text{mW}$ (green squares). The measurement time for the data is $9$~h, $0.9$~h, and $0.2$~h for $P\approx 3$~mW, 9~mW, and 18~mW.
(c,d) CHSH-Bell values for $|\psi_+\rangle$ (c) and $|\psi_-\rangle$ (d) as a function of the coincidence window at $P \approx 3\text{mW}$. Maximum CHSH values of $2.24 \pm 0.03$ (for $|\psi_+\rangle$) and $2.21 \pm 0.03$ (for $|\psi_-\rangle$) are achieved at a 20-ns coincidence window, clearly violating the classical bound (red dashed lines). The measurement time for the data is $33$~h.}
\label{fig:SM_Witness_and_CHSH_without_QMs}
\end{figure*}

\begin{figure*}[htbp]
\includegraphics [width=0.95\textwidth]{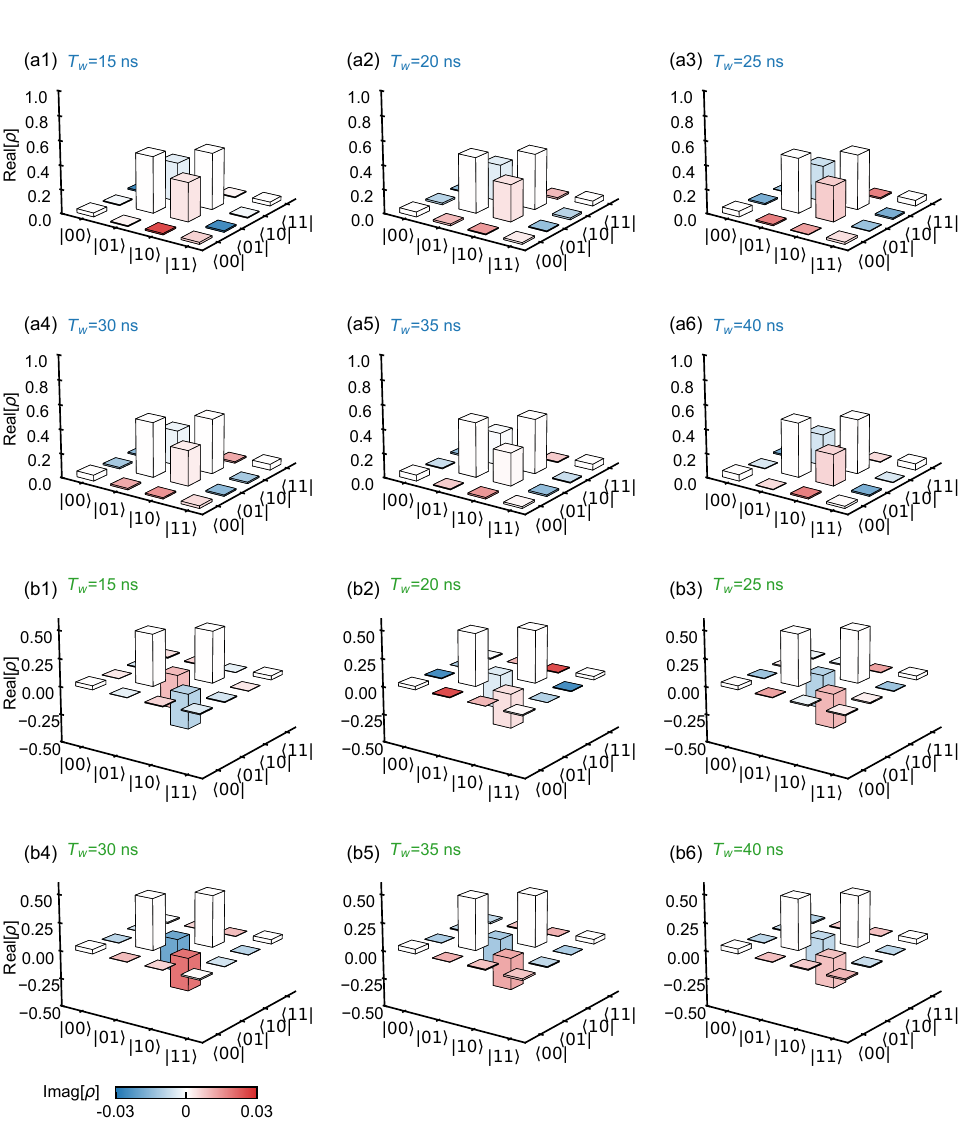}
\caption{Detailed tomography of heralded photonic entanglement with a pump power of $3$ mW. (a1-a6) Reconstructed density matrices of the $|\psi_+\rangle$ state using coincidence windows ($T_\text{w}$) ranging from 15 ns to 40 ns. (b1-b6) Corresponding reconstructed density matrices of $|\psi_-\rangle$ under the same $T_\text{w}$ conditions. Bar height and color represent the real and imaginary parts of the matrix elements, respectively.}
\label{fig:SM_tomography}
\end{figure*}

\begin{figure*}[htbp]
\includegraphics [width=0.95\textwidth]{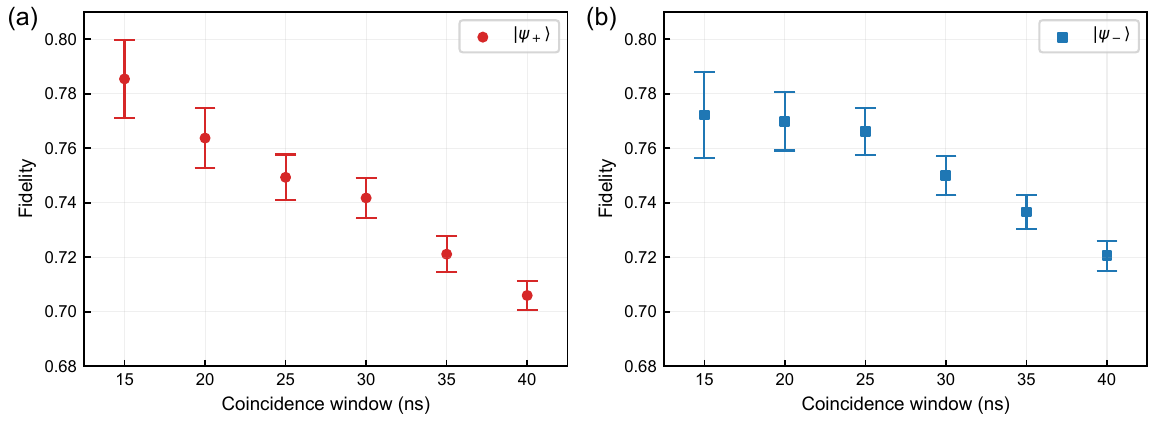}
\caption{Fidelity of swapped photonic entanglement as a function of the coincidence window for the entangled states (a) $|\psi_+\rangle$ and (b) $|\psi_-\rangle$, obtained by tomographic measurements. The measurement time for the data is $23$~h.}
\label{fig:SM_tomography_fidelity}
\end{figure*}

\newpage
\begin{table}[H]
\centering
\caption{Raw data for the entanglement witness measurement with a 20-ns coincidence window and a 3-mW pump power. X, Y and Z are Pauli operators. We record all the combinations of the four-fold coincidence events by using the coincidence unit (Swabian Time Tagger 20). Here we use H(V) to denote the H or V port detector firing for the two 580 nm photons, and use 1(2) to denote the two central detectors for 1537 nm photons firing, 1'(2') denotes the two central detector with 500 ns delay.   }
\begin{tabular}{|c|c|c|c|c|c|c|c|c|c|c|c|c|c|c|c|c|} \hline
Setting & HH11' & HH22' & HH12' & HH21' & HV11' & HV22' & HV12' & HV21' & VH11' & VH22' & VH12' & VH21' & VV11' & VV22' & VV12' & VV21' \\\hline
XX  & 25 & 24 & 18 & 24 & 5 & 5 & 4 & 3 & 3 & 3 & 6 & 3 & 33 & 28 & 15 & 17 \\\hline
YY & 24 & 15 & 23 & 29 & 3 & 3 & 9 & 6 & 5 & 4 & 9 & 5 & 18 & 17 & 16 & 11\\\hline 
ZZ & 1 & 4 & 2 & 1 & 45 & 28 & 45 & 31 & 46 & 26 & 26 & 33 & 2 & 3 & 1 & 3 \\\hline 
\end{tabular}
\label{data_entanglement_witness}
\end{table}

\begin{table}[htbp]
\centering
\caption{Raw data for the entanglement witness measurement with a 20-ns coincidence window and a 18-mW pump power. }
\begin{tabular}{|c|c|c|c|c|c|c|c|c|c|c|c|c|c|c|c|c|} \hline
Setting & HH11' & HH22' & HH12' & HH21' & HV11' & HV22' & HV12' & HV21' & VH11' & VH22' & VH12' & VH21' & VV11' & VV22' & VV12' & VV21' \\\hline
XX  & 49 & 46 & 42 & 49 & 19 & 17 & 16 & 16 & 13 & 13 & 12 & 20 & 36 & 40 & 47 & 41 \\\hline
YY & 49 & 56 & 36 & 40 & 19 & 21 & 17 & 20 & 11 & 12 & 20 & 13 & 35 & 32 & 45 & 48 \\\hline
ZZ & 12 & 23 & 15 & 21 & 63 & 74 & 80 & 87 & 89 & 87 & 88 & 73 & 17 & 19 & 20 & 14 \\\hline
\end{tabular}
\label{data_witness_with_QM}
\end{table}

%We collected a total of 3290 detection events in 215 hours.
\begin{table}[htbp]
\centering
\caption{Raw data for the CHSH inequality measurement with a 20-ns coincidence window and a 3-mW pump power.}
\begin{tabular}{|c|c|c|c|c|c|c|c|c|c|c|c|c|c|c|c|c|} \hline
Setting & HH11' & HH22' & HH12' & HH21' & HV11' & HV22' & HV12' & HV21' & VH11' & VH22' & VH12' & VH21' & VV11' & VV22' & VV12' & VV21' \\\hline
$Z\frac{(Z+X)}{\sqrt{2}}$  & 26 & 21 & 22 & 24 & 86 & 104 & 86 & 88 & 98 & 97 & 99 & 81 & 18 & 16 & 20 & 14  \\\hline
$Z\frac{(Z-X)}{\sqrt{2}}$  &21 & 23 & 17 & 14 & 82 & 65 & 63 & 81 & 82 & 87 & 73 & 90 & 20 & 9 & 15 & 14 \\\hline
$X\frac{(Z+X)}{\sqrt{2}}$  & 73 & 64 & 80 & 79 & 34 & 33 & 34 & 27 & 22 & 24 & 25 & 27 & 80 & 73 & 79 & 75 \\\hline
$X\frac{(Z-X)}{\sqrt{2}}$  & 22 & 40 & 35 & 27 & 78 & 83 & 78 & 72 & 55 & 83 & 69 & 77 & 16 & 21 & 25 & 24 \\\hline
\end{tabular}
\label{data_CHSH_with_QM}
\end{table}

% \clearpage
% \newpage
% \bibliographystyle{naturemag}
% \bibliography{bibliography}

\begin{thebibliography}{53}%
\makeatletter
\providecommand \@ifxundefined [1]{%
 \@ifx{#1\undefined}
}%
\providecommand \@ifnum [1]{%
 \ifnum #1\expandafter \@firstoftwo
 \else \expandafter \@secondoftwo
 \fi
}%
\providecommand \@ifx [1]{%
 \ifx #1\expandafter \@firstoftwo
 \else \expandafter \@secondoftwo
 \fi
}%
\providecommand \natexlab [1]{#1}%
\providecommand \enquote  [1]{``#1''}%
\providecommand \bibnamefont  [1]{#1}%
\providecommand \bibfnamefont [1]{#1}%
\providecommand \citenamefont [1]{#1}%
\providecommand \href@noop [0]{\@secondoftwo}%
\providecommand \href [0]{\begingroup \@sanitize@url \@href}%
\providecommand \@href[1]{\@@startlink{#1}\@@href}%
\providecommand \@@href[1]{\endgroup#1\@@endlink}%
\providecommand \@sanitize@url [0]{\catcode `\\12\catcode `\$12\catcode
  `\&12\catcode `\#12\catcode `\^12\catcode `\_12\catcode `\%12\relax}%
\providecommand \@@startlink[1]{}%
\providecommand \@@endlink[0]{}%
\providecommand \url  [0]{\begingroup\@sanitize@url \@url }%
\providecommand \@url [1]{\endgroup\@href {#1}{\urlprefix }}%
\providecommand \urlprefix  [0]{URL }%
\providecommand \Eprint [0]{\href }%
\providecommand \doibase [0]{http://dx.doi.org/}%
\providecommand \selectlanguage [0]{\@gobble}%
\providecommand \bibinfo  [0]{\@secondoftwo}%
\providecommand \bibfield  [0]{\@secondoftwo}%
\providecommand \translation [1]{[#1]}%
\providecommand \BibitemOpen [0]{}%
\providecommand \bibitemStop [0]{}%
\providecommand \bibitemNoStop [0]{.\EOS\space}%
\providecommand \EOS [0]{\spacefactor3000\relax}%
\providecommand \BibitemShut  [1]{\csname bibitem#1\endcsname}%
\let\auto@bib@innerbib\@empty
%</preamble>
\bibitem [{\citenamefont {Briegel}\ \emph {et~al.}(1998)\citenamefont
  {Briegel}, \citenamefont {Dür}, \citenamefont {Cirac},\ and\ \citenamefont
  {Zoller}}]{Briegel1998Quantum}%
  \BibitemOpen
  \bibfield  {author} {\bibinfo {author} {\bibfnamefont {H.~J.}\ \bibnamefont
  {Briegel}}, \bibinfo {author} {\bibfnamefont {W.}~\bibnamefont {Dür}},
  \bibinfo {author} {\bibfnamefont {J.~I.}\ \bibnamefont {Cirac}}, \ and\
  \bibinfo {author} {\bibfnamefont {P.}~\bibnamefont {Zoller}},\ }\href
  {\doibase 10.1103/PhysRevLett.81.5932} {\bibfield  {journal} {\bibinfo
  {journal} {Physical Review Letters}\ }\textbf {\bibinfo {volume} {81}},\
  \bibinfo {pages} {5932} (\bibinfo {year} {1998})}\BibitemShut {NoStop}%
\bibitem [{\citenamefont {Kimble}(2008)}]{Kimble2008The}%
  \BibitemOpen
  \bibfield  {author} {\bibinfo {author} {\bibfnamefont {H.~J.}\ \bibnamefont
  {Kimble}},\ }\href {\doibase 10.1038/nature07127} {\bibfield  {journal}
  {\bibinfo  {journal} {Nature}\ }\textbf {\bibinfo {volume} {453}},\ \bibinfo
  {pages} {1023} (\bibinfo {year} {2008})}\BibitemShut {NoStop}%
\bibitem [{\citenamefont {Sangouard}\ \emph {et~al.}(2011)\citenamefont
  {Sangouard}, \citenamefont {Simon}, \citenamefont {de~Riedmatten},\ and\
  \citenamefont {Gisin}}]{Sangouard2011Quantum}%
  \BibitemOpen
  \bibfield  {author} {\bibinfo {author} {\bibfnamefont {N.}~\bibnamefont
  {Sangouard}}, \bibinfo {author} {\bibfnamefont {C.}~\bibnamefont {Simon}},
  \bibinfo {author} {\bibfnamefont {H.}~\bibnamefont {de~Riedmatten}}, \ and\
  \bibinfo {author} {\bibfnamefont {N.}~\bibnamefont {Gisin}},\ }\href
  {\doibase 10.1103/RevModPhys.83.33} {\bibfield  {journal} {\bibinfo
  {journal} {Rev. Mod. Phys.}\ }\textbf {\bibinfo {volume} {83}},\ \bibinfo
  {pages} {33} (\bibinfo {year} {2011})}\BibitemShut {NoStop}%
\bibitem [{\citenamefont {Wehner}\ \emph {et~al.}(2018)\citenamefont {Wehner},
  \citenamefont {Elkouss},\ and\ \citenamefont {Hanson}}]{Wehner2018Quantum}%
  \BibitemOpen
  \bibfield  {author} {\bibinfo {author} {\bibfnamefont {S.}~\bibnamefont
  {Wehner}}, \bibinfo {author} {\bibfnamefont {D.}~\bibnamefont {Elkouss}}, \
  and\ \bibinfo {author} {\bibfnamefont {R.}~\bibnamefont {Hanson}},\ }\href
  {\doibase doi:10.1126/science.aam9288} {\bibfield  {journal} {\bibinfo
  {journal} {Science}\ }\textbf {\bibinfo {volume} {362}},\ \bibinfo {pages}
  {eaam9288} (\bibinfo {year} {2018})}\BibitemShut {NoStop}%
\bibitem [{\citenamefont {Chou}\ \emph {et~al.}(2007)\citenamefont {Chou},
  \citenamefont {Laurat}, \citenamefont {Deng}, \citenamefont {Choi},
  \citenamefont {de~Riedmatten}, \citenamefont {Felinto},\ and\ \citenamefont
  {Kimble}}]{Chou2007functional}%
  \BibitemOpen
  \bibfield  {author} {\bibinfo {author} {\bibfnamefont {C.-W.}\ \bibnamefont
  {Chou}}, \bibinfo {author} {\bibfnamefont {J.}~\bibnamefont {Laurat}},
  \bibinfo {author} {\bibfnamefont {H.}~\bibnamefont {Deng}}, \bibinfo {author}
  {\bibfnamefont {K.~S.}\ \bibnamefont {Choi}}, \bibinfo {author}
  {\bibfnamefont {H.}~\bibnamefont {de~Riedmatten}}, \bibinfo {author}
  {\bibfnamefont {D.}~\bibnamefont {Felinto}}, \ and\ \bibinfo {author}
  {\bibfnamefont {H.~J.}\ \bibnamefont {Kimble}},\ }\href {\doibase
  doi:10.1126/science.1140300} {\bibfield  {journal} {\bibinfo  {journal}
  {Science}\ }\textbf {\bibinfo {volume} {316}},\ \bibinfo {pages} {1316}
  (\bibinfo {year} {2007})}\BibitemShut {NoStop}%
\bibitem [{\citenamefont {Yu}\ \emph {et~al.}(2020)\citenamefont {Yu},
  \citenamefont {Ma}, \citenamefont {Luo}, \citenamefont {Jing}, \citenamefont
  {Sun}, \citenamefont {Fang}, \citenamefont {Yang}, \citenamefont {Liu},
  \citenamefont {Zheng}, \citenamefont {Xie}, \citenamefont {Zhang},
  \citenamefont {You}, \citenamefont {Wang}, \citenamefont {Chen},
  \citenamefont {Zhang}, \citenamefont {Bao},\ and\ \citenamefont
  {Pan}}]{Yu2020Entanglement}%
  \BibitemOpen
  \bibfield  {author} {\bibinfo {author} {\bibfnamefont {Y.}~\bibnamefont
  {Yu}}, \bibinfo {author} {\bibfnamefont {F.}~\bibnamefont {Ma}}, \bibinfo
  {author} {\bibfnamefont {X.-Y.}\ \bibnamefont {Luo}}, \bibinfo {author}
  {\bibfnamefont {B.}~\bibnamefont {Jing}}, \bibinfo {author} {\bibfnamefont
  {P.-F.}\ \bibnamefont {Sun}}, \bibinfo {author} {\bibfnamefont {R.-Z.}\
  \bibnamefont {Fang}}, \bibinfo {author} {\bibfnamefont {C.-W.}\ \bibnamefont
  {Yang}}, \bibinfo {author} {\bibfnamefont {H.}~\bibnamefont {Liu}}, \bibinfo
  {author} {\bibfnamefont {M.-Y.}\ \bibnamefont {Zheng}}, \bibinfo {author}
  {\bibfnamefont {X.-P.}\ \bibnamefont {Xie}}, \bibinfo {author} {\bibfnamefont
  {W.-J.}\ \bibnamefont {Zhang}}, \bibinfo {author} {\bibfnamefont {L.-X.}\
  \bibnamefont {You}}, \bibinfo {author} {\bibfnamefont {Z.}~\bibnamefont
  {Wang}}, \bibinfo {author} {\bibfnamefont {T.-Y.}\ \bibnamefont {Chen}},
  \bibinfo {author} {\bibfnamefont {Q.}~\bibnamefont {Zhang}}, \bibinfo
  {author} {\bibfnamefont {X.-H.}\ \bibnamefont {Bao}}, \ and\ \bibinfo
  {author} {\bibfnamefont {J.-W.}\ \bibnamefont {Pan}},\ }\href {\doibase
  10.1038/s41586-020-1976-7} {\bibfield  {journal} {\bibinfo  {journal}
  {Nature}\ }\textbf {\bibinfo {volume} {578}},\ \bibinfo {pages} {240}
  (\bibinfo {year} {2020})}\BibitemShut {NoStop}%
\bibitem [{\citenamefont {Luo}\ \emph {et~al.}(2025)\citenamefont {Luo},
  \citenamefont {Wang}, \citenamefont {Zheng}, \citenamefont {Wang},
  \citenamefont {Liu}, \citenamefont {Gao}, \citenamefont {Li}, \citenamefont
  {Yan}, \citenamefont {Ke}, \citenamefont {Teng}, \citenamefont {Wang},
  \citenamefont {Wu}, \citenamefont {Huang}, \citenamefont {Li}, \citenamefont
  {You}, \citenamefont {Xie}, \citenamefont {Xu}, \citenamefont {Zhang},
  \citenamefont {Bao},\ and\ \citenamefont {Pan}}]{Luo2025Entangling}%
  \BibitemOpen
  \bibfield  {author} {\bibinfo {author} {\bibfnamefont {X.-Y.}\ \bibnamefont
  {Luo}}, \bibinfo {author} {\bibfnamefont {C.-Y.}\ \bibnamefont {Wang}},
  \bibinfo {author} {\bibfnamefont {M.-Y.}\ \bibnamefont {Zheng}}, \bibinfo
  {author} {\bibfnamefont {B.}~\bibnamefont {Wang}}, \bibinfo {author}
  {\bibfnamefont {J.-L.}\ \bibnamefont {Liu}}, \bibinfo {author} {\bibfnamefont
  {B.-F.}\ \bibnamefont {Gao}}, \bibinfo {author} {\bibfnamefont
  {J.}~\bibnamefont {Li}}, \bibinfo {author} {\bibfnamefont {Z.}~\bibnamefont
  {Yan}}, \bibinfo {author} {\bibfnamefont {Q.-M.}\ \bibnamefont {Ke}},
  \bibinfo {author} {\bibfnamefont {D.}~\bibnamefont {Teng}}, \bibinfo {author}
  {\bibfnamefont {R.-C.}\ \bibnamefont {Wang}}, \bibinfo {author}
  {\bibfnamefont {J.}~\bibnamefont {Wu}}, \bibinfo {author} {\bibfnamefont
  {J.}~\bibnamefont {Huang}}, \bibinfo {author} {\bibfnamefont
  {H.}~\bibnamefont {Li}}, \bibinfo {author} {\bibfnamefont {L.-X.}\
  \bibnamefont {You}}, \bibinfo {author} {\bibfnamefont {X.-P.}\ \bibnamefont
  {Xie}}, \bibinfo {author} {\bibfnamefont {F.}~\bibnamefont {Xu}}, \bibinfo
  {author} {\bibfnamefont {Q.}~\bibnamefont {Zhang}}, \bibinfo {author}
  {\bibfnamefont {X.-H.}\ \bibnamefont {Bao}}, \ and\ \bibinfo {author}
  {\bibfnamefont {J.-W.}\ \bibnamefont {Pan}},\ }\href
  {https://arxiv.org/abs/2504.05660} {\  (\bibinfo {year} {2025})},\ \Eprint
  {http://arxiv.org/abs/2504.05660} {arXiv:2504.05660 [quant-ph]} \BibitemShut
  {NoStop}%
\bibitem [{\citenamefont {Hofmann}\ \emph {et~al.}(2012)\citenamefont
  {Hofmann}, \citenamefont {Krug}, \citenamefont {Ortegel}, \citenamefont
  {Gérard}, \citenamefont {Weber}, \citenamefont {Rosenfeld},\ and\
  \citenamefont {Weinfurter}}]{Hofmann2012Heralded}%
  \BibitemOpen
  \bibfield  {author} {\bibinfo {author} {\bibfnamefont {J.}~\bibnamefont
  {Hofmann}}, \bibinfo {author} {\bibfnamefont {M.}~\bibnamefont {Krug}},
  \bibinfo {author} {\bibfnamefont {N.}~\bibnamefont {Ortegel}}, \bibinfo
  {author} {\bibfnamefont {L.}~\bibnamefont {Gérard}}, \bibinfo {author}
  {\bibfnamefont {M.}~\bibnamefont {Weber}}, \bibinfo {author} {\bibfnamefont
  {W.}~\bibnamefont {Rosenfeld}}, \ and\ \bibinfo {author} {\bibfnamefont
  {H.}~\bibnamefont {Weinfurter}},\ }\href {\doibase
  doi:10.1126/science.1221856} {\bibfield  {journal} {\bibinfo  {journal}
  {Science}\ }\textbf {\bibinfo {volume} {337}},\ \bibinfo {pages} {72}
  (\bibinfo {year} {2012})}\BibitemShut {NoStop}%
\bibitem [{\citenamefont {van Leent}\ \emph {et~al.}(2022)\citenamefont {van
  Leent}, \citenamefont {Bock}, \citenamefont {Fertig}, \citenamefont
  {Garthoff}, \citenamefont {Eppelt}, \citenamefont {Zhou}, \citenamefont
  {Malik}, \citenamefont {Seubert}, \citenamefont {Bauer}, \citenamefont
  {Rosenfeld}, \citenamefont {Zhang}, \citenamefont {Becher},\ and\
  \citenamefont {Weinfurter}}]{van2022entangling}%
  \BibitemOpen
  \bibfield  {author} {\bibinfo {author} {\bibfnamefont {T.}~\bibnamefont {van
  Leent}}, \bibinfo {author} {\bibfnamefont {M.}~\bibnamefont {Bock}}, \bibinfo
  {author} {\bibfnamefont {F.}~\bibnamefont {Fertig}}, \bibinfo {author}
  {\bibfnamefont {R.}~\bibnamefont {Garthoff}}, \bibinfo {author}
  {\bibfnamefont {S.}~\bibnamefont {Eppelt}}, \bibinfo {author} {\bibfnamefont
  {Y.}~\bibnamefont {Zhou}}, \bibinfo {author} {\bibfnamefont {P.}~\bibnamefont
  {Malik}}, \bibinfo {author} {\bibfnamefont {M.}~\bibnamefont {Seubert}},
  \bibinfo {author} {\bibfnamefont {T.}~\bibnamefont {Bauer}}, \bibinfo
  {author} {\bibfnamefont {W.}~\bibnamefont {Rosenfeld}}, \bibinfo {author}
  {\bibfnamefont {W.}~\bibnamefont {Zhang}}, \bibinfo {author} {\bibfnamefont
  {C.}~\bibnamefont {Becher}}, \ and\ \bibinfo {author} {\bibfnamefont
  {H.}~\bibnamefont {Weinfurter}},\ }\href {\doibase
  10.1038/s41586-022-04764-4} {\bibfield  {journal} {\bibinfo  {journal}
  {Nature}\ }\textbf {\bibinfo {volume} {607}},\ \bibinfo {pages} {69}
  (\bibinfo {year} {2022})}\BibitemShut {NoStop}%
\bibitem [{\citenamefont {Zhang}\ \emph {et~al.}(2022)\citenamefont {Zhang},
  \citenamefont {van Leent}, \citenamefont {Redeker}, \citenamefont {Garthoff},
  \citenamefont {Schwonnek}, \citenamefont {Fertig}, \citenamefont {Eppelt},
  \citenamefont {Rosenfeld}, \citenamefont {Scarani}, \citenamefont {Lim},\
  and\ \citenamefont {Weinfurter}}]{Zhang2022device-independent}%
  \BibitemOpen
  \bibfield  {author} {\bibinfo {author} {\bibfnamefont {W.}~\bibnamefont
  {Zhang}}, \bibinfo {author} {\bibfnamefont {T.}~\bibnamefont {van Leent}},
  \bibinfo {author} {\bibfnamefont {K.}~\bibnamefont {Redeker}}, \bibinfo
  {author} {\bibfnamefont {R.}~\bibnamefont {Garthoff}}, \bibinfo {author}
  {\bibfnamefont {R.}~\bibnamefont {Schwonnek}}, \bibinfo {author}
  {\bibfnamefont {F.}~\bibnamefont {Fertig}}, \bibinfo {author} {\bibfnamefont
  {S.}~\bibnamefont {Eppelt}}, \bibinfo {author} {\bibfnamefont
  {W.}~\bibnamefont {Rosenfeld}}, \bibinfo {author} {\bibfnamefont
  {V.}~\bibnamefont {Scarani}}, \bibinfo {author} {\bibfnamefont {C.~C.~W.}\
  \bibnamefont {Lim}}, \ and\ \bibinfo {author} {\bibfnamefont
  {H.}~\bibnamefont {Weinfurter}},\ }\href {\doibase
  10.1038/s41586-022-04891-y} {\bibfield  {journal} {\bibinfo  {journal}
  {Nature}\ }\textbf {\bibinfo {volume} {607}},\ \bibinfo {pages} {687}
  (\bibinfo {year} {2022})}\BibitemShut {NoStop}%
\bibitem [{\citenamefont {Nadlinger}\ \emph {et~al.}(2022)\citenamefont
  {Nadlinger}, \citenamefont {Drmota}, \citenamefont {Nichol}, \citenamefont
  {Araneda}, \citenamefont {Main}, \citenamefont {Srinivas}, \citenamefont
  {Lucas}, \citenamefont {Ballance}, \citenamefont {Ivanov}, \citenamefont
  {Tan}, \citenamefont {Sekatski}, \citenamefont {Urbanke}, \citenamefont
  {Renner}, \citenamefont {Sangouard},\ and\ \citenamefont
  {Bancal}}]{Nadlinger2022Experimental}%
  \BibitemOpen
  \bibfield  {author} {\bibinfo {author} {\bibfnamefont {D.~P.}\ \bibnamefont
  {Nadlinger}}, \bibinfo {author} {\bibfnamefont {P.}~\bibnamefont {Drmota}},
  \bibinfo {author} {\bibfnamefont {B.~C.}\ \bibnamefont {Nichol}}, \bibinfo
  {author} {\bibfnamefont {G.}~\bibnamefont {Araneda}}, \bibinfo {author}
  {\bibfnamefont {D.}~\bibnamefont {Main}}, \bibinfo {author} {\bibfnamefont
  {R.}~\bibnamefont {Srinivas}}, \bibinfo {author} {\bibfnamefont {D.~M.}\
  \bibnamefont {Lucas}}, \bibinfo {author} {\bibfnamefont {C.~J.}\ \bibnamefont
  {Ballance}}, \bibinfo {author} {\bibfnamefont {K.}~\bibnamefont {Ivanov}},
  \bibinfo {author} {\bibfnamefont {E.~Y.~Z.}\ \bibnamefont {Tan}}, \bibinfo
  {author} {\bibfnamefont {P.}~\bibnamefont {Sekatski}}, \bibinfo {author}
  {\bibfnamefont {R.~L.}\ \bibnamefont {Urbanke}}, \bibinfo {author}
  {\bibfnamefont {R.}~\bibnamefont {Renner}}, \bibinfo {author} {\bibfnamefont
  {N.}~\bibnamefont {Sangouard}}, \ and\ \bibinfo {author} {\bibfnamefont
  {J.~D.}\ \bibnamefont {Bancal}},\ }\href {\doibase
  10.1038/s41586-022-04941-5} {\bibfield  {journal} {\bibinfo  {journal}
  {Nature}\ }\textbf {\bibinfo {volume} {607}},\ \bibinfo {pages} {682}
  (\bibinfo {year} {2022})}\BibitemShut {NoStop}%
\bibitem [{\citenamefont {Krutyanskiy}\ \emph {et~al.}(2023)\citenamefont
  {Krutyanskiy}, \citenamefont {Galli}, \citenamefont {Krcmarsky},
  \citenamefont {Baier}, \citenamefont {Fioretto}, \citenamefont {Pu},
  \citenamefont {Mazloom}, \citenamefont {Sekatski}, \citenamefont {Canteri},
  \citenamefont {Teller}, \citenamefont {Schupp}, \citenamefont {Bate},
  \citenamefont {Meraner}, \citenamefont {Sangouard}, \citenamefont {Lanyon},\
  and\ \citenamefont {Northup}}]{Krutyanskiy2023Entanglement}%
  \BibitemOpen
  \bibfield  {author} {\bibinfo {author} {\bibfnamefont {V.}~\bibnamefont
  {Krutyanskiy}}, \bibinfo {author} {\bibfnamefont {M.}~\bibnamefont {Galli}},
  \bibinfo {author} {\bibfnamefont {V.}~\bibnamefont {Krcmarsky}}, \bibinfo
  {author} {\bibfnamefont {S.}~\bibnamefont {Baier}}, \bibinfo {author}
  {\bibfnamefont {D.~â.}\ \bibnamefont {Fioretto}}, \bibinfo {author}
  {\bibfnamefont {Y.}~\bibnamefont {Pu}}, \bibinfo {author} {\bibfnamefont
  {A.}~\bibnamefont {Mazloom}}, \bibinfo {author} {\bibfnamefont
  {P.}~\bibnamefont {Sekatski}}, \bibinfo {author} {\bibfnamefont
  {M.}~\bibnamefont {Canteri}}, \bibinfo {author} {\bibfnamefont
  {M.}~\bibnamefont {Teller}}, \bibinfo {author} {\bibfnamefont
  {J.}~\bibnamefont {Schupp}}, \bibinfo {author} {\bibfnamefont
  {J.}~\bibnamefont {Bate}}, \bibinfo {author} {\bibfnamefont {M.}~\bibnamefont
  {Meraner}}, \bibinfo {author} {\bibfnamefont {N.}~\bibnamefont {Sangouard}},
  \bibinfo {author} {\bibfnamefont {B.~â.}\ \bibnamefont {Lanyon}}, \ and\
  \bibinfo {author} {\bibfnamefont {T.~â.}\ \bibnamefont {Northup}},\ }\href
  {\doibase 10.1103/PhysRevLett.130.050803} {\bibfield  {journal} {\bibinfo
  {journal} {Physical Review Letters}\ }\textbf {\bibinfo {volume} {130}},\
  \bibinfo {pages} {050803} (\bibinfo {year} {2023})}\BibitemShut {NoStop}%
\bibitem [{\citenamefont {Hensen}\ \emph {et~al.}(2015)\citenamefont {Hensen},
  \citenamefont {Bernien}, \citenamefont {Dréau}, \citenamefont {Reiserer},
  \citenamefont {Kalb}, \citenamefont {Blok}, \citenamefont {Ruitenberg},
  \citenamefont {Vermeulen}, \citenamefont {Schouten}, \citenamefont
  {Abellán}, \citenamefont {Amaya}, \citenamefont {Pruneri}, \citenamefont
  {Mitchell}, \citenamefont {Markham}, \citenamefont {Twitchen}, \citenamefont
  {Elkouss}, \citenamefont {Wehner}, \citenamefont {Taminiau},\ and\
  \citenamefont {Hanson}}]{Hensen2015Loophole}%
  \BibitemOpen
  \bibfield  {author} {\bibinfo {author} {\bibfnamefont {B.}~\bibnamefont
  {Hensen}}, \bibinfo {author} {\bibfnamefont {H.}~\bibnamefont {Bernien}},
  \bibinfo {author} {\bibfnamefont {A.~E.}\ \bibnamefont {Dréau}}, \bibinfo
  {author} {\bibfnamefont {A.}~\bibnamefont {Reiserer}}, \bibinfo {author}
  {\bibfnamefont {N.}~\bibnamefont {Kalb}}, \bibinfo {author} {\bibfnamefont
  {M.~S.}\ \bibnamefont {Blok}}, \bibinfo {author} {\bibfnamefont
  {J.}~\bibnamefont {Ruitenberg}}, \bibinfo {author} {\bibfnamefont {R.~F.~L.}\
  \bibnamefont {Vermeulen}}, \bibinfo {author} {\bibfnamefont {R.~N.}\
  \bibnamefont {Schouten}}, \bibinfo {author} {\bibfnamefont {C.}~\bibnamefont
  {Abellán}}, \bibinfo {author} {\bibfnamefont {W.}~\bibnamefont {Amaya}},
  \bibinfo {author} {\bibfnamefont {V.}~\bibnamefont {Pruneri}}, \bibinfo
  {author} {\bibfnamefont {M.~W.}\ \bibnamefont {Mitchell}}, \bibinfo {author}
  {\bibfnamefont {M.}~\bibnamefont {Markham}}, \bibinfo {author} {\bibfnamefont
  {D.~J.}\ \bibnamefont {Twitchen}}, \bibinfo {author} {\bibfnamefont
  {D.}~\bibnamefont {Elkouss}}, \bibinfo {author} {\bibfnamefont
  {S.}~\bibnamefont {Wehner}}, \bibinfo {author} {\bibfnamefont {T.~H.}\
  \bibnamefont {Taminiau}}, \ and\ \bibinfo {author} {\bibfnamefont
  {R.}~\bibnamefont {Hanson}},\ }\href {\doibase 10.1038/nature15759}
  {\bibfield  {journal} {\bibinfo  {journal} {Nature}\ }\textbf {\bibinfo
  {volume} {526}},\ \bibinfo {pages} {682} (\bibinfo {year}
  {2015})}\BibitemShut {NoStop}%
\bibitem [{\citenamefont {Delteil}\ \emph {et~al.}(2016)\citenamefont
  {Delteil}, \citenamefont {Sun}, \citenamefont {Gao}, \citenamefont {Togan},
  \citenamefont {Faelt},\ and\ \citenamefont
  {Imamoğlu}}]{Delteil2016Generation}%
  \BibitemOpen
  \bibfield  {author} {\bibinfo {author} {\bibfnamefont {A.}~\bibnamefont
  {Delteil}}, \bibinfo {author} {\bibfnamefont {Z.}~\bibnamefont {Sun}},
  \bibinfo {author} {\bibfnamefont {W.-b.}\ \bibnamefont {Gao}}, \bibinfo
  {author} {\bibfnamefont {E.}~\bibnamefont {Togan}}, \bibinfo {author}
  {\bibfnamefont {S.}~\bibnamefont {Faelt}}, \ and\ \bibinfo {author}
  {\bibfnamefont {A.}~\bibnamefont {Imamoğlu}},\ }\href {\doibase
  10.1038/nphys3605} {\bibfield  {journal} {\bibinfo  {journal} {Nature
  Physics}\ }\textbf {\bibinfo {volume} {12}},\ \bibinfo {pages} {218}
  (\bibinfo {year} {2016})}\BibitemShut {NoStop}%
\bibitem [{\citenamefont {Stolk}\ \emph
  {et~al.}(2024{\natexlab{a}})\citenamefont {Stolk}, \citenamefont {van~der
  Enden}, \citenamefont {Slater}, \citenamefont {te~Raa-Derckx}, \citenamefont
  {Botma}, \citenamefont {van Rantwijk}, \citenamefont {Biemond}, \citenamefont
  {Hagen}, \citenamefont {Herfst}, \citenamefont {Koek}, \citenamefont
  {Meskers}, \citenamefont {Vollmer}, \citenamefont {van Zwet}, \citenamefont
  {Markham}, \citenamefont {Edmonds}, \citenamefont {Geus}, \citenamefont
  {Elsen}, \citenamefont {Jungbluth}, \citenamefont {Haefner}, \citenamefont
  {Tresp}, \citenamefont {Stuhler}, \citenamefont {Ritter},\ and\ \citenamefont
  {Hanson}}]{Stolk2024Metropolitan-scale}%
  \BibitemOpen
  \bibfield  {author} {\bibinfo {author} {\bibfnamefont {A.~J.}\ \bibnamefont
  {Stolk}}, \bibinfo {author} {\bibfnamefont {K.~L.}\ \bibnamefont {van~der
  Enden}}, \bibinfo {author} {\bibfnamefont {M.-C.}\ \bibnamefont {Slater}},
  \bibinfo {author} {\bibfnamefont {I.}~\bibnamefont {te~Raa-Derckx}}, \bibinfo
  {author} {\bibfnamefont {P.}~\bibnamefont {Botma}}, \bibinfo {author}
  {\bibfnamefont {J.}~\bibnamefont {van Rantwijk}}, \bibinfo {author}
  {\bibfnamefont {J.~J.~B.}\ \bibnamefont {Biemond}}, \bibinfo {author}
  {\bibfnamefont {R.~A.~J.}\ \bibnamefont {Hagen}}, \bibinfo {author}
  {\bibfnamefont {R.~W.}\ \bibnamefont {Herfst}}, \bibinfo {author}
  {\bibfnamefont {W.~D.}\ \bibnamefont {Koek}}, \bibinfo {author}
  {\bibfnamefont {A.~J.~H.}\ \bibnamefont {Meskers}}, \bibinfo {author}
  {\bibfnamefont {R.}~\bibnamefont {Vollmer}}, \bibinfo {author} {\bibfnamefont
  {E.~J.}\ \bibnamefont {van Zwet}}, \bibinfo {author} {\bibfnamefont
  {M.}~\bibnamefont {Markham}}, \bibinfo {author} {\bibfnamefont {A.~M.}\
  \bibnamefont {Edmonds}}, \bibinfo {author} {\bibfnamefont {J.~F.}\
  \bibnamefont {Geus}}, \bibinfo {author} {\bibfnamefont {F.}~\bibnamefont
  {Elsen}}, \bibinfo {author} {\bibfnamefont {B.}~\bibnamefont {Jungbluth}},
  \bibinfo {author} {\bibfnamefont {C.}~\bibnamefont {Haefner}}, \bibinfo
  {author} {\bibfnamefont {C.}~\bibnamefont {Tresp}}, \bibinfo {author}
  {\bibfnamefont {J.}~\bibnamefont {Stuhler}}, \bibinfo {author} {\bibfnamefont
  {S.}~\bibnamefont {Ritter}}, \ and\ \bibinfo {author} {\bibfnamefont
  {R.}~\bibnamefont {Hanson}},\ }\href {\doibase doi:10.1126/sciadv.adp6442}
  {\bibfield  {journal} {\bibinfo  {journal} {Science Advances}\ }\textbf
  {\bibinfo {volume} {10}},\ \bibinfo {pages} {eadp6442} (\bibinfo {year}
  {2024}{\natexlab{a}})}\BibitemShut {NoStop}%
\bibitem [{\citenamefont {Knaut}\ \emph {et~al.}(2024)\citenamefont {Knaut},
  \citenamefont {Suleymanzade}, \citenamefont {Wei}, \citenamefont {Assumpcao},
  \citenamefont {Stas}, \citenamefont {Huan}, \citenamefont {Machielse},
  \citenamefont {Knall}, \citenamefont {Sutula}, \citenamefont {Baranes},
  \citenamefont {Sinclair}, \citenamefont {De-Eknamkul}, \citenamefont
  {Levonian}, \citenamefont {Bhaskar}, \citenamefont {Park}, \citenamefont
  {Lončar},\ and\ \citenamefont {Lukin}}]{Knaut2024Entanglement}%
  \BibitemOpen
  \bibfield  {author} {\bibinfo {author} {\bibfnamefont {C.~M.}\ \bibnamefont
  {Knaut}}, \bibinfo {author} {\bibfnamefont {A.}~\bibnamefont {Suleymanzade}},
  \bibinfo {author} {\bibfnamefont {Y.~C.}\ \bibnamefont {Wei}}, \bibinfo
  {author} {\bibfnamefont {D.~R.}\ \bibnamefont {Assumpcao}}, \bibinfo {author}
  {\bibfnamefont {P.~J.}\ \bibnamefont {Stas}}, \bibinfo {author}
  {\bibfnamefont {Y.~Q.}\ \bibnamefont {Huan}}, \bibinfo {author}
  {\bibfnamefont {B.}~\bibnamefont {Machielse}}, \bibinfo {author}
  {\bibfnamefont {E.~N.}\ \bibnamefont {Knall}}, \bibinfo {author}
  {\bibfnamefont {M.}~\bibnamefont {Sutula}}, \bibinfo {author} {\bibfnamefont
  {G.}~\bibnamefont {Baranes}}, \bibinfo {author} {\bibfnamefont
  {N.}~\bibnamefont {Sinclair}}, \bibinfo {author} {\bibfnamefont
  {C.}~\bibnamefont {De-Eknamkul}}, \bibinfo {author} {\bibfnamefont {D.~S.}\
  \bibnamefont {Levonian}}, \bibinfo {author} {\bibfnamefont {M.~K.}\
  \bibnamefont {Bhaskar}}, \bibinfo {author} {\bibfnamefont {H.}~\bibnamefont
  {Park}}, \bibinfo {author} {\bibfnamefont {M.}~\bibnamefont {Lončar}}, \
  and\ \bibinfo {author} {\bibfnamefont {M.~D.}\ \bibnamefont {Lukin}},\ }\href
  {\doibase 10.1038/s41586-024-07252-z} {\bibfield  {journal} {\bibinfo
  {journal} {Nature}\ }\textbf {\bibinfo {volume} {629}},\ \bibinfo {pages}
  {573} (\bibinfo {year} {2024})}\BibitemShut {NoStop}%
\bibitem [{\citenamefont {Inc}\ \emph {et~al.}(2024)\citenamefont {Inc},
  \citenamefont {:}, \citenamefont {Afzal}, \citenamefont {Akhlaghi},
  \citenamefont {Beale}, \citenamefont {Bedroya}, \citenamefont {Bell},
  \citenamefont {Bergeron}, \citenamefont {Bonsma-Fisher}, \citenamefont
  {Bychkova}, \citenamefont {Chaisson}, \citenamefont {Chartrand},
  \citenamefont {Clear}, \citenamefont {Darcie}, \citenamefont {DeAbreu},
  \citenamefont {DeLisle}, \citenamefont {Duncan}, \citenamefont {Smith},
  \citenamefont {Dunn}, \citenamefont {Ebrahimi}, \citenamefont {Evetts},
  \citenamefont {Pinheiro}, \citenamefont {Fuentes}, \citenamefont {Georgiou},
  \citenamefont {Guha}, \citenamefont {Haenel}, \citenamefont {Higginbottom},
  \citenamefont {Jackson}, \citenamefont {Jahed}, \citenamefont
  {Khorshidahmad}, \citenamefont {Shandilya}, \citenamefont {Kurkjian},
  \citenamefont {Lauk}, \citenamefont {Lee-Hone}, \citenamefont {Lin},
  \citenamefont {Litynskyy}, \citenamefont {Lock}, \citenamefont {Ma},
  \citenamefont {MacGilp}, \citenamefont {MacQuarrie}, \citenamefont {Mar},
  \citenamefont {Khah}, \citenamefont {Matiash}, \citenamefont {Meyer-Scott},
  \citenamefont {Michaels}, \citenamefont {Motira}, \citenamefont {Noori},
  \citenamefont {Ospadov}, \citenamefont {Patel}, \citenamefont {Patscheider},
  \citenamefont {Paulson}, \citenamefont {Petruk}, \citenamefont
  {Ravindranath}, \citenamefont {Reznychenko}, \citenamefont {Ruether},
  \citenamefont {Ruscica}, \citenamefont {Saxena}, \citenamefont {Schaller},
  \citenamefont {Seidlitz}, \citenamefont {Senger}, \citenamefont {Lee},
  \citenamefont {Sevoyan}, \citenamefont {Simmons}, \citenamefont {Soykal},
  \citenamefont {Stott}, \citenamefont {Tran}, \citenamefont {Tserkis},
  \citenamefont {Ulhaq}, \citenamefont {Vine}, \citenamefont {Weeks},
  \citenamefont {Wolfowicz},\ and\ \citenamefont
  {Yoneda}}]{inc2024distributed}%
  \BibitemOpen
  \bibfield  {author} {\bibinfo {author} {\bibfnamefont {P.}~\bibnamefont
  {Inc}}, \bibinfo {author} {\bibnamefont {:}}, \bibinfo {author}
  {\bibfnamefont {F.}~\bibnamefont {Afzal}}, \bibinfo {author} {\bibfnamefont
  {M.}~\bibnamefont {Akhlaghi}}, \bibinfo {author} {\bibfnamefont {S.~J.}\
  \bibnamefont {Beale}}, \bibinfo {author} {\bibfnamefont {O.}~\bibnamefont
  {Bedroya}}, \bibinfo {author} {\bibfnamefont {K.}~\bibnamefont {Bell}},
  \bibinfo {author} {\bibfnamefont {L.}~\bibnamefont {Bergeron}}, \bibinfo
  {author} {\bibfnamefont {K.}~\bibnamefont {Bonsma-Fisher}}, \bibinfo {author}
  {\bibfnamefont {P.}~\bibnamefont {Bychkova}}, \bibinfo {author}
  {\bibfnamefont {Z.~M.~E.}\ \bibnamefont {Chaisson}}, \bibinfo {author}
  {\bibfnamefont {C.}~\bibnamefont {Chartrand}}, \bibinfo {author}
  {\bibfnamefont {C.}~\bibnamefont {Clear}}, \bibinfo {author} {\bibfnamefont
  {A.}~\bibnamefont {Darcie}}, \bibinfo {author} {\bibfnamefont
  {A.}~\bibnamefont {DeAbreu}}, \bibinfo {author} {\bibfnamefont
  {C.}~\bibnamefont {DeLisle}}, \bibinfo {author} {\bibfnamefont {L.~A.}\
  \bibnamefont {Duncan}}, \bibinfo {author} {\bibfnamefont {C.~D.}\
  \bibnamefont {Smith}}, \bibinfo {author} {\bibfnamefont {J.}~\bibnamefont
  {Dunn}}, \bibinfo {author} {\bibfnamefont {A.}~\bibnamefont {Ebrahimi}},
  \bibinfo {author} {\bibfnamefont {N.}~\bibnamefont {Evetts}}, \bibinfo
  {author} {\bibfnamefont {D.~F.}\ \bibnamefont {Pinheiro}}, \bibinfo {author}
  {\bibfnamefont {P.}~\bibnamefont {Fuentes}}, \bibinfo {author} {\bibfnamefont
  {T.}~\bibnamefont {Georgiou}}, \bibinfo {author} {\bibfnamefont
  {B.}~\bibnamefont {Guha}}, \bibinfo {author} {\bibfnamefont {R.}~\bibnamefont
  {Haenel}}, \bibinfo {author} {\bibfnamefont {D.}~\bibnamefont
  {Higginbottom}}, \bibinfo {author} {\bibfnamefont {D.~M.}\ \bibnamefont
  {Jackson}}, \bibinfo {author} {\bibfnamefont {N.}~\bibnamefont {Jahed}},
  \bibinfo {author} {\bibfnamefont {A.}~\bibnamefont {Khorshidahmad}}, \bibinfo
  {author} {\bibfnamefont {P.~K.}\ \bibnamefont {Shandilya}}, \bibinfo {author}
  {\bibfnamefont {A.~T.~K.}\ \bibnamefont {Kurkjian}}, \bibinfo {author}
  {\bibfnamefont {N.}~\bibnamefont {Lauk}}, \bibinfo {author} {\bibfnamefont
  {N.~R.}\ \bibnamefont {Lee-Hone}}, \bibinfo {author} {\bibfnamefont
  {E.}~\bibnamefont {Lin}}, \bibinfo {author} {\bibfnamefont {R.}~\bibnamefont
  {Litynskyy}}, \bibinfo {author} {\bibfnamefont {D.}~\bibnamefont {Lock}},
  \bibinfo {author} {\bibfnamefont {L.}~\bibnamefont {Ma}}, \bibinfo {author}
  {\bibfnamefont {I.}~\bibnamefont {MacGilp}}, \bibinfo {author} {\bibfnamefont
  {E.~R.}\ \bibnamefont {MacQuarrie}}, \bibinfo {author} {\bibfnamefont
  {A.}~\bibnamefont {Mar}}, \bibinfo {author} {\bibfnamefont {A.~M.}\
  \bibnamefont {Khah}}, \bibinfo {author} {\bibfnamefont {A.}~\bibnamefont
  {Matiash}}, \bibinfo {author} {\bibfnamefont {E.}~\bibnamefont
  {Meyer-Scott}}, \bibinfo {author} {\bibfnamefont {C.~P.}\ \bibnamefont
  {Michaels}}, \bibinfo {author} {\bibfnamefont {J.}~\bibnamefont {Motira}},
  \bibinfo {author} {\bibfnamefont {N.~K.}\ \bibnamefont {Noori}}, \bibinfo
  {author} {\bibfnamefont {E.}~\bibnamefont {Ospadov}}, \bibinfo {author}
  {\bibfnamefont {E.}~\bibnamefont {Patel}}, \bibinfo {author} {\bibfnamefont
  {A.}~\bibnamefont {Patscheider}}, \bibinfo {author} {\bibfnamefont
  {D.}~\bibnamefont {Paulson}}, \bibinfo {author} {\bibfnamefont
  {A.}~\bibnamefont {Petruk}}, \bibinfo {author} {\bibfnamefont {A.~L.}\
  \bibnamefont {Ravindranath}}, \bibinfo {author} {\bibfnamefont
  {B.}~\bibnamefont {Reznychenko}}, \bibinfo {author} {\bibfnamefont
  {M.}~\bibnamefont {Ruether}}, \bibinfo {author} {\bibfnamefont
  {J.}~\bibnamefont {Ruscica}}, \bibinfo {author} {\bibfnamefont
  {K.}~\bibnamefont {Saxena}}, \bibinfo {author} {\bibfnamefont
  {Z.}~\bibnamefont {Schaller}}, \bibinfo {author} {\bibfnamefont
  {A.}~\bibnamefont {Seidlitz}}, \bibinfo {author} {\bibfnamefont
  {J.}~\bibnamefont {Senger}}, \bibinfo {author} {\bibfnamefont {Y.~S.}\
  \bibnamefont {Lee}}, \bibinfo {author} {\bibfnamefont {O.}~\bibnamefont
  {Sevoyan}}, \bibinfo {author} {\bibfnamefont {S.}~\bibnamefont {Simmons}},
  \bibinfo {author} {\bibfnamefont {O.}~\bibnamefont {Soykal}}, \bibinfo
  {author} {\bibfnamefont {L.}~\bibnamefont {Stott}}, \bibinfo {author}
  {\bibfnamefont {Q.}~\bibnamefont {Tran}}, \bibinfo {author} {\bibfnamefont
  {S.}~\bibnamefont {Tserkis}}, \bibinfo {author} {\bibfnamefont
  {A.}~\bibnamefont {Ulhaq}}, \bibinfo {author} {\bibfnamefont
  {W.}~\bibnamefont {Vine}}, \bibinfo {author} {\bibfnamefont {R.}~\bibnamefont
  {Weeks}}, \bibinfo {author} {\bibfnamefont {G.}~\bibnamefont {Wolfowicz}}, \
  and\ \bibinfo {author} {\bibfnamefont {I.}~\bibnamefont {Yoneda}},\ }\href
  {https://arxiv.org/abs/2406.01704} {\enquote {\bibinfo {title} {Distributed
  quantum computing in silicon},}\ } (\bibinfo {year} {2024}),\ \Eprint
  {http://arxiv.org/abs/2406.01704} {arXiv:2406.01704 [quant-ph]} \BibitemShut
  {NoStop}%
\bibitem [{\citenamefont {Riedinger}\ \emph {et~al.}(2018)\citenamefont
  {Riedinger}, \citenamefont {Wallucks}, \citenamefont {Marinković},
  \citenamefont {Löschnauer}, \citenamefont {Aspelmeyer}, \citenamefont
  {Hong},\ and\ \citenamefont {Gröblacher}}]{Riedinger2018Remote}%
  \BibitemOpen
  \bibfield  {author} {\bibinfo {author} {\bibfnamefont {R.}~\bibnamefont
  {Riedinger}}, \bibinfo {author} {\bibfnamefont {A.}~\bibnamefont {Wallucks}},
  \bibinfo {author} {\bibfnamefont {I.}~\bibnamefont {Marinković}}, \bibinfo
  {author} {\bibfnamefont {C.}~\bibnamefont {Löschnauer}}, \bibinfo {author}
  {\bibfnamefont {M.}~\bibnamefont {Aspelmeyer}}, \bibinfo {author}
  {\bibfnamefont {S.}~\bibnamefont {Hong}}, \ and\ \bibinfo {author}
  {\bibfnamefont {S.}~\bibnamefont {Gröblacher}},\ }\href {\doibase
  10.1038/s41586-018-0036-z} {\bibfield  {journal} {\bibinfo  {journal}
  {Nature}\ }\textbf {\bibinfo {volume} {556}},\ \bibinfo {pages} {473}
  (\bibinfo {year} {2018})}\BibitemShut {NoStop}%
\bibitem [{\citenamefont {Liu}\ \emph {et~al.}(2021)\citenamefont {Liu},
  \citenamefont {Hu}, \citenamefont {Li}, \citenamefont {Li}, \citenamefont
  {Li}, \citenamefont {Liang}, \citenamefont {Zhou}, \citenamefont {Li},\ and\
  \citenamefont {Guo}}]{Liu2021Heralded}%
  \BibitemOpen
  \bibfield  {author} {\bibinfo {author} {\bibfnamefont {X.}~\bibnamefont
  {Liu}}, \bibinfo {author} {\bibfnamefont {J.}~\bibnamefont {Hu}}, \bibinfo
  {author} {\bibfnamefont {Z.-F.}\ \bibnamefont {Li}}, \bibinfo {author}
  {\bibfnamefont {X.}~\bibnamefont {Li}}, \bibinfo {author} {\bibfnamefont
  {P.-Y.}\ \bibnamefont {Li}}, \bibinfo {author} {\bibfnamefont {P.-J.}\
  \bibnamefont {Liang}}, \bibinfo {author} {\bibfnamefont {Z.-Q.}\ \bibnamefont
  {Zhou}}, \bibinfo {author} {\bibfnamefont {C.-F.}\ \bibnamefont {Li}}, \ and\
  \bibinfo {author} {\bibfnamefont {G.-C.}\ \bibnamefont {Guo}},\ }\href
  {\doibase 10.1038/s41586-021-03505-3} {\bibfield  {journal} {\bibinfo
  {journal} {Nature}\ }\textbf {\bibinfo {volume} {594}},\ \bibinfo {pages}
  {41} (\bibinfo {year} {2021})}\BibitemShut {NoStop}%
\bibitem [{\citenamefont {Lago-Rivera}\ \emph {et~al.}(2021)\citenamefont
  {Lago-Rivera}, \citenamefont {Grandi}, \citenamefont {Rakonjac},
  \citenamefont {Seri},\ and\ \citenamefont
  {de~Riedmatten}}]{Lago-Rivera2021telecom}%
  \BibitemOpen
  \bibfield  {author} {\bibinfo {author} {\bibfnamefont {D.}~\bibnamefont
  {Lago-Rivera}}, \bibinfo {author} {\bibfnamefont {S.}~\bibnamefont {Grandi}},
  \bibinfo {author} {\bibfnamefont {J.~V.}\ \bibnamefont {Rakonjac}}, \bibinfo
  {author} {\bibfnamefont {A.}~\bibnamefont {Seri}}, \ and\ \bibinfo {author}
  {\bibfnamefont {H.}~\bibnamefont {de~Riedmatten}},\ }\href {\doibase
  10.1038/s41586-021-03481-8} {\bibfield  {journal} {\bibinfo  {journal}
  {Nature}\ }\textbf {\bibinfo {volume} {594}},\ \bibinfo {pages} {37}
  (\bibinfo {year} {2021})}\BibitemShut {NoStop}%
\bibitem [{\citenamefont {Ruskuc}\ \emph {et~al.}(2025)\citenamefont {Ruskuc},
  \citenamefont {Wu}, \citenamefont {Green}, \citenamefont {Hermans},
  \citenamefont {Pajak}, \citenamefont {Choi},\ and\ \citenamefont
  {Faraon}}]{Ruskuc2025Multiplexed}%
  \BibitemOpen
  \bibfield  {author} {\bibinfo {author} {\bibfnamefont {A.}~\bibnamefont
  {Ruskuc}}, \bibinfo {author} {\bibfnamefont {C.~J.}\ \bibnamefont {Wu}},
  \bibinfo {author} {\bibfnamefont {E.}~\bibnamefont {Green}}, \bibinfo
  {author} {\bibfnamefont {S.~L.~N.}\ \bibnamefont {Hermans}}, \bibinfo
  {author} {\bibfnamefont {W.}~\bibnamefont {Pajak}}, \bibinfo {author}
  {\bibfnamefont {J.}~\bibnamefont {Choi}}, \ and\ \bibinfo {author}
  {\bibfnamefont {A.}~\bibnamefont {Faraon}},\ }\href {\doibase
  10.1038/s41586-024-08537-z} {\bibfield  {journal} {\bibinfo  {journal}
  {Nature}\ }\textbf {\bibinfo {volume} {639}},\ \bibinfo {pages} {54}
  (\bibinfo {year} {2025})}\BibitemShut {NoStop}%
\bibitem [{\citenamefont {Zhu}\ \emph {et~al.}(2025)\citenamefont {Zhu},
  \citenamefont {Liu}, \citenamefont {Zhou},\ and\ \citenamefont
  {Li}}]{Zhu2025Remote}%
  \BibitemOpen
  \bibfield  {author} {\bibinfo {author} {\bibfnamefont {T.-X.}\ \bibnamefont
  {Zhu}}, \bibinfo {author} {\bibfnamefont {X.}~\bibnamefont {Liu}}, \bibinfo
  {author} {\bibfnamefont {Z.-Q.}\ \bibnamefont {Zhou}}, \ and\ \bibinfo
  {author} {\bibfnamefont {C.-F.}\ \bibnamefont {Li}},\ }\href {\doibase
  doi:10.1515/nanoph-2024-0487} {\bibfield  {journal} {\bibinfo  {journal}
  {Nanophotonics}\ }\textbf {\bibinfo {volume} {14}},\ \bibinfo {pages} {1975}
  (\bibinfo {year} {2025})}\BibitemShut {NoStop}%
\bibitem [{\citenamefont {Liu}\ \emph {et~al.}(2024{\natexlab{a}})\citenamefont
  {Liu}, \citenamefont {Luo}, \citenamefont {Yu}, \citenamefont {Wang},
  \citenamefont {Wang}, \citenamefont {Hu}, \citenamefont {Li}, \citenamefont
  {Zheng}, \citenamefont {Yao}, \citenamefont {Yan}, \citenamefont {Teng},
  \citenamefont {Jiang}, \citenamefont {Liu}, \citenamefont {Xie},
  \citenamefont {Zhang}, \citenamefont {Mao}, \citenamefont {Jiang},
  \citenamefont {Zhang}, \citenamefont {Bao},\ and\ \citenamefont
  {Pan}}]{Liu2024Creation}%
  \BibitemOpen
  \bibfield  {author} {\bibinfo {author} {\bibfnamefont {J.-L.}\ \bibnamefont
  {Liu}}, \bibinfo {author} {\bibfnamefont {X.-Y.}\ \bibnamefont {Luo}},
  \bibinfo {author} {\bibfnamefont {Y.}~\bibnamefont {Yu}}, \bibinfo {author}
  {\bibfnamefont {C.-Y.}\ \bibnamefont {Wang}}, \bibinfo {author}
  {\bibfnamefont {B.}~\bibnamefont {Wang}}, \bibinfo {author} {\bibfnamefont
  {Y.}~\bibnamefont {Hu}}, \bibinfo {author} {\bibfnamefont {J.}~\bibnamefont
  {Li}}, \bibinfo {author} {\bibfnamefont {M.-Y.}\ \bibnamefont {Zheng}},
  \bibinfo {author} {\bibfnamefont {B.}~\bibnamefont {Yao}}, \bibinfo {author}
  {\bibfnamefont {Z.}~\bibnamefont {Yan}}, \bibinfo {author} {\bibfnamefont
  {D.}~\bibnamefont {Teng}}, \bibinfo {author} {\bibfnamefont {J.-W.}\
  \bibnamefont {Jiang}}, \bibinfo {author} {\bibfnamefont {X.-B.}\ \bibnamefont
  {Liu}}, \bibinfo {author} {\bibfnamefont {X.-P.}\ \bibnamefont {Xie}},
  \bibinfo {author} {\bibfnamefont {J.}~\bibnamefont {Zhang}}, \bibinfo
  {author} {\bibfnamefont {Q.-H.}\ \bibnamefont {Mao}}, \bibinfo {author}
  {\bibfnamefont {X.}~\bibnamefont {Jiang}}, \bibinfo {author} {\bibfnamefont
  {Q.}~\bibnamefont {Zhang}}, \bibinfo {author} {\bibfnamefont {X.-H.}\
  \bibnamefont {Bao}}, \ and\ \bibinfo {author} {\bibfnamefont {J.-W.}\
  \bibnamefont {Pan}},\ }\href {\doibase 10.1038/s41586-024-07308-0} {\bibfield
   {journal} {\bibinfo  {journal} {Nature}\ }\textbf {\bibinfo {volume}
  {629}},\ \bibinfo {pages} {579} (\bibinfo {year}
  {2024}{\natexlab{a}})}\BibitemShut {NoStop}%
\bibitem [{\citenamefont {Acín}\ \emph {et~al.}(2007)\citenamefont {Acín},
  \citenamefont {Brunner}, \citenamefont {Gisin}, \citenamefont {Massar},
  \citenamefont {Pironio},\ and\ \citenamefont
  {Scarani}}]{Acin2007Device-Independent}%
  \BibitemOpen
  \bibfield  {author} {\bibinfo {author} {\bibfnamefont {A.}~\bibnamefont
  {Acín}}, \bibinfo {author} {\bibfnamefont {N.}~\bibnamefont {Brunner}},
  \bibinfo {author} {\bibfnamefont {N.}~\bibnamefont {Gisin}}, \bibinfo
  {author} {\bibfnamefont {S.}~\bibnamefont {Massar}}, \bibinfo {author}
  {\bibfnamefont {S.}~\bibnamefont {Pironio}}, \ and\ \bibinfo {author}
  {\bibfnamefont {V.}~\bibnamefont {Scarani}},\ }\href {\doibase
  10.1103/PhysRevLett.98.230501} {\bibfield  {journal} {\bibinfo  {journal}
  {Physical Review Letters}\ }\textbf {\bibinfo {volume} {98}},\ \bibinfo
  {pages} {230501} (\bibinfo {year} {2007})}\BibitemShut {NoStop}%
\bibitem [{\citenamefont {Pironio}\ \emph {et~al.}(2010)\citenamefont
  {Pironio}, \citenamefont {Acín}, \citenamefont {Massar}, \citenamefont
  {de~la Giroday}, \citenamefont {Matsukevich}, \citenamefont {Maunz},
  \citenamefont {Olmschenk}, \citenamefont {Hayes}, \citenamefont {Luo},
  \citenamefont {Manning},\ and\ \citenamefont {Monroe}}]{Pironio2010Random}%
  \BibitemOpen
  \bibfield  {author} {\bibinfo {author} {\bibfnamefont {S.}~\bibnamefont
  {Pironio}}, \bibinfo {author} {\bibfnamefont {A.}~\bibnamefont {Acín}},
  \bibinfo {author} {\bibfnamefont {S.}~\bibnamefont {Massar}}, \bibinfo
  {author} {\bibfnamefont {A.~B.}\ \bibnamefont {de~la Giroday}}, \bibinfo
  {author} {\bibfnamefont {D.~N.}\ \bibnamefont {Matsukevich}}, \bibinfo
  {author} {\bibfnamefont {P.}~\bibnamefont {Maunz}}, \bibinfo {author}
  {\bibfnamefont {S.}~\bibnamefont {Olmschenk}}, \bibinfo {author}
  {\bibfnamefont {D.}~\bibnamefont {Hayes}}, \bibinfo {author} {\bibfnamefont
  {L.}~\bibnamefont {Luo}}, \bibinfo {author} {\bibfnamefont {T.~A.}\
  \bibnamefont {Manning}}, \ and\ \bibinfo {author} {\bibfnamefont
  {C.}~\bibnamefont {Monroe}},\ }\href {\doibase 10.1038/nature09008}
  {\bibfield  {journal} {\bibinfo  {journal} {Nature}\ }\textbf {\bibinfo
  {volume} {464}},\ \bibinfo {pages} {1021} (\bibinfo {year}
  {2010})}\BibitemShut {NoStop}%
\bibitem [{\citenamefont {Adamson}(2025)}]{Adamson2025Parallel}%
  \BibitemOpen
  \bibfield  {author} {\bibinfo {author} {\bibfnamefont {S.~A.}\ \bibnamefont
  {Adamson}},\ }\href {\doibase 10.1103/PhysRevResearch.7.013069} {\bibfield
  {journal} {\bibinfo  {journal} {Physical Review Research}\ }\textbf {\bibinfo
  {volume} {7}},\ \bibinfo {pages} {013069} (\bibinfo {year}
  {2025})}\BibitemShut {NoStop}%
\bibitem [{\citenamefont {{\v{S}}upi{\'{c}}}\ and\ \citenamefont
  {Bowles}(2020)}]{Supic2020self-testing}%
  \BibitemOpen
  \bibfield  {author} {\bibinfo {author} {\bibfnamefont {I.}~\bibnamefont
  {{\v{S}}upi{\'{c}}}}\ and\ \bibinfo {author} {\bibfnamefont {J.}~\bibnamefont
  {Bowles}},\ }\href {\doibase 10.22331/q-2020-09-30-337} {\bibfield  {journal}
  {\bibinfo  {journal} {{Quantum}}\ }\textbf {\bibinfo {volume} {4}},\ \bibinfo
  {pages} {337} (\bibinfo {year} {2020})}\BibitemShut {NoStop}%
\bibitem [{\citenamefont {Sekatski}\ \emph {et~al.}(2023)\citenamefont
  {Sekatski}, \citenamefont {Bancal}, \citenamefont {Ioannou}, \citenamefont
  {Afzelius},\ and\ \citenamefont {Brunner}}]{Sekatski2023Toward}%
  \BibitemOpen
  \bibfield  {author} {\bibinfo {author} {\bibfnamefont {P.}~\bibnamefont
  {Sekatski}}, \bibinfo {author} {\bibfnamefont {J.-D.}\ \bibnamefont
  {Bancal}}, \bibinfo {author} {\bibfnamefont {M.}~\bibnamefont {Ioannou}},
  \bibinfo {author} {\bibfnamefont {M.}~\bibnamefont {Afzelius}}, \ and\
  \bibinfo {author} {\bibfnamefont {N.}~\bibnamefont {Brunner}},\ }\href
  {\doibase 10.1103/PhysRevLett.131.170802} {\bibfield  {journal} {\bibinfo
  {journal} {Physical Review Letters}\ }\textbf {\bibinfo {volume} {131}},\
  \bibinfo {pages} {170802} (\bibinfo {year} {2023})}\BibitemShut {NoStop}%
\bibitem [{\citenamefont {Barrett}\ and\ \citenamefont
  {Kok}(2005)}]{Barrett2005Efficient}%
  \BibitemOpen
  \bibfield  {author} {\bibinfo {author} {\bibfnamefont {S.~D.}\ \bibnamefont
  {Barrett}}\ and\ \bibinfo {author} {\bibfnamefont {P.}~\bibnamefont {Kok}},\
  }\href {\doibase 10.1103/PhysRevA.71.060310} {\bibfield  {journal} {\bibinfo
  {journal} {Phys. Rev. A}\ }\textbf {\bibinfo {volume} {71}},\ \bibinfo
  {pages} {060310} (\bibinfo {year} {2005})}\BibitemShut {NoStop}%
\bibitem [{\citenamefont {Halder}\ \emph {et~al.}(2007)\citenamefont {Halder},
  \citenamefont {Beveratos}, \citenamefont {Gisin}, \citenamefont {Scarani},
  \citenamefont {Simon},\ and\ \citenamefont {Zbinden}}]{Halder2007Entangling}%
  \BibitemOpen
  \bibfield  {author} {\bibinfo {author} {\bibfnamefont {M.}~\bibnamefont
  {Halder}}, \bibinfo {author} {\bibfnamefont {A.}~\bibnamefont {Beveratos}},
  \bibinfo {author} {\bibfnamefont {N.}~\bibnamefont {Gisin}}, \bibinfo
  {author} {\bibfnamefont {V.}~\bibnamefont {Scarani}}, \bibinfo {author}
  {\bibfnamefont {C.}~\bibnamefont {Simon}}, \ and\ \bibinfo {author}
  {\bibfnamefont {H.}~\bibnamefont {Zbinden}},\ }\href {\doibase
  10.1038/nphys700} {\bibfield  {journal} {\bibinfo  {journal} {Nature
  Physics}\ }\textbf {\bibinfo {volume} {3}},\ \bibinfo {pages} {692} (\bibinfo
  {year} {2007})}\BibitemShut {NoStop}%
\bibitem [{\citenamefont {Clauser}\ \emph {et~al.}(1970)\citenamefont
  {Clauser}, \citenamefont {Horne}, \citenamefont {Shimony},\ and\
  \citenamefont {Holt}}]{clauser1970proposed}%
  \BibitemOpen
  \bibfield  {author} {\bibinfo {author} {\bibfnamefont {J.~F.}\ \bibnamefont
  {Clauser}}, \bibinfo {author} {\bibfnamefont {M.~A.}\ \bibnamefont {Horne}},
  \bibinfo {author} {\bibfnamefont {A.}~\bibnamefont {Shimony}}, \ and\
  \bibinfo {author} {\bibfnamefont {R.~A.}\ \bibnamefont {Holt}},\ }\href
  {\doibase 10.1103/PhysRevLett.24.549} {\bibfield  {journal} {\bibinfo
  {journal} {Phys. Rev. Lett.}\ }\textbf {\bibinfo {volume} {24}},\ \bibinfo
  {pages} {549} (\bibinfo {year} {1970})}\BibitemShut {NoStop}%
\bibitem [{\citenamefont {Ortu}\ \emph {et~al.}(2022)\citenamefont {Ortu},
  \citenamefont {Holzäpfel}, \citenamefont {Etesse},\ and\ \citenamefont
  {Afzelius}}]{Ortu2022Storage}%
  \BibitemOpen
  \bibfield  {author} {\bibinfo {author} {\bibfnamefont {A.}~\bibnamefont
  {Ortu}}, \bibinfo {author} {\bibfnamefont {A.}~\bibnamefont {Holzäpfel}},
  \bibinfo {author} {\bibfnamefont {J.}~\bibnamefont {Etesse}}, \ and\ \bibinfo
  {author} {\bibfnamefont {M.}~\bibnamefont {Afzelius}},\ }\href {\doibase
  10.1038/s41534-022-00541-3} {\bibfield  {journal} {\bibinfo  {journal} {npj
  Quantum Information}\ }\textbf {\bibinfo {volume} {8}},\ \bibinfo {pages}
  {29} (\bibinfo {year} {2022})}\BibitemShut {NoStop}%
\bibitem [{\citenamefont {Ma}\ \emph {et~al.}(2021{\natexlab{a}})\citenamefont
  {Ma}, \citenamefont {Jin}, \citenamefont {Chen}, \citenamefont {Zhou},
  \citenamefont {Li},\ and\ \citenamefont {Guo}}]{Ma2021Elimination}%
  \BibitemOpen
  \bibfield  {author} {\bibinfo {author} {\bibfnamefont {Y.-Z.}\ \bibnamefont
  {Ma}}, \bibinfo {author} {\bibfnamefont {M.}~\bibnamefont {Jin}}, \bibinfo
  {author} {\bibfnamefont {D.-L.}\ \bibnamefont {Chen}}, \bibinfo {author}
  {\bibfnamefont {Z.-Q.}\ \bibnamefont {Zhou}}, \bibinfo {author}
  {\bibfnamefont {C.-F.}\ \bibnamefont {Li}}, \ and\ \bibinfo {author}
  {\bibfnamefont {G.-C.}\ \bibnamefont {Guo}},\ }\href {\doibase
  10.1038/s41467-021-24679-4} {\bibfield  {journal} {\bibinfo  {journal}
  {Nature Communications}\ }\textbf {\bibinfo {volume} {12}},\ \bibinfo {pages}
  {4378} (\bibinfo {year} {2021}{\natexlab{a}})}\BibitemShut {NoStop}%
\bibitem [{\citenamefont {Liu}\ \emph {et~al.}(2025)\citenamefont {Liu},
  \citenamefont {Ou}, \citenamefont {Zhu}, \citenamefont {Su}, \citenamefont
  {Liu}, \citenamefont {Han}, \citenamefont {Zhou}, \citenamefont {Li},\ and\
  \citenamefont {Guo}}]{Liu2025millisecond}%
  \BibitemOpen
  \bibfield  {author} {\bibinfo {author} {\bibfnamefont {Y.-P.}\ \bibnamefont
  {Liu}}, \bibinfo {author} {\bibfnamefont {Z.-W.}\ \bibnamefont {Ou}},
  \bibinfo {author} {\bibfnamefont {T.-X.}\ \bibnamefont {Zhu}}, \bibinfo
  {author} {\bibfnamefont {M.-X.}\ \bibnamefont {Su}}, \bibinfo {author}
  {\bibfnamefont {C.}~\bibnamefont {Liu}}, \bibinfo {author} {\bibfnamefont
  {Y.-J.}\ \bibnamefont {Han}}, \bibinfo {author} {\bibfnamefont {Z.-Q.}\
  \bibnamefont {Zhou}}, \bibinfo {author} {\bibfnamefont {C.-F.}\ \bibnamefont
  {Li}}, \ and\ \bibinfo {author} {\bibfnamefont {G.-C.}\ \bibnamefont {Guo}},\
  }\href {\doibase doi:10.1126/sciadv.adu5264} {\bibfield  {journal} {\bibinfo
  {journal} {Science Advances}\ }\textbf {\bibinfo {volume} {11}},\ \bibinfo
  {pages} {eadu5264} (\bibinfo {year} {2025})}\BibitemShut {NoStop}%
\bibitem [{\citenamefont {Simon}\ \emph {et~al.}(2007)\citenamefont {Simon},
  \citenamefont {de~Riedmatten}, \citenamefont {Afzelius}, \citenamefont
  {Sangouard}, \citenamefont {Zbinden},\ and\ \citenamefont
  {Gisin}}]{Simon2007Quantum}%
  \BibitemOpen
  \bibfield  {author} {\bibinfo {author} {\bibfnamefont {C.}~\bibnamefont
  {Simon}}, \bibinfo {author} {\bibfnamefont {H.}~\bibnamefont
  {de~Riedmatten}}, \bibinfo {author} {\bibfnamefont {M.}~\bibnamefont
  {Afzelius}}, \bibinfo {author} {\bibfnamefont {N.}~\bibnamefont {Sangouard}},
  \bibinfo {author} {\bibfnamefont {H.}~\bibnamefont {Zbinden}}, \ and\
  \bibinfo {author} {\bibfnamefont {N.}~\bibnamefont {Gisin}},\ }\href
  {\doibase 10.1103/PhysRevLett.98.190503} {\bibfield  {journal} {\bibinfo
  {journal} {Physical Review Letters}\ }\textbf {\bibinfo {volume} {98}},\
  \bibinfo {pages} {190503} (\bibinfo {year} {2007})}\BibitemShut {NoStop}%
\bibitem [{\citenamefont {Jiang}\ \emph {et~al.}(2007)\citenamefont {Jiang},
  \citenamefont {Taylor},\ and\ \citenamefont {Lukin}}]{Jiang2007Fast}%
  \BibitemOpen
  \bibfield  {author} {\bibinfo {author} {\bibfnamefont {L.}~\bibnamefont
  {Jiang}}, \bibinfo {author} {\bibfnamefont {J.~M.}\ \bibnamefont {Taylor}}, \
  and\ \bibinfo {author} {\bibfnamefont {M.~D.}\ \bibnamefont {Lukin}},\ }\href
  {\doibase 10.1103/PhysRevA.76.012301} {\bibfield  {journal} {\bibinfo
  {journal} {Physical Review A}\ }\textbf {\bibinfo {volume} {76}},\ \bibinfo
  {pages} {012301} (\bibinfo {year} {2007})}\BibitemShut {NoStop}%
\bibitem [{\citenamefont {Zhao}\ \emph {et~al.}(2007)\citenamefont {Zhao},
  \citenamefont {Chen}, \citenamefont {Chen}, \citenamefont {Schmiedmayer},\
  and\ \citenamefont {Pan}}]{Zhao2007Robust}%
  \BibitemOpen
  \bibfield  {author} {\bibinfo {author} {\bibfnamefont {B.}~\bibnamefont
  {Zhao}}, \bibinfo {author} {\bibfnamefont {Z.-B.}\ \bibnamefont {Chen}},
  \bibinfo {author} {\bibfnamefont {Y.-A.}\ \bibnamefont {Chen}}, \bibinfo
  {author} {\bibfnamefont {J.}~\bibnamefont {Schmiedmayer}}, \ and\ \bibinfo
  {author} {\bibfnamefont {J.-W.}\ \bibnamefont {Pan}},\ }\href {\doibase
  10.1103/PhysRevLett.98.240502} {\bibfield  {journal} {\bibinfo  {journal}
  {Physical Review Letters}\ }\textbf {\bibinfo {volume} {98}},\ \bibinfo
  {pages} {240502} (\bibinfo {year} {2007})}\BibitemShut {NoStop}%
\bibitem [{\citenamefont {Simon}\ and\ \citenamefont
  {Irvine}(2003)}]{PhysRevLett.91.110405}%
  \BibitemOpen
  \bibfield  {author} {\bibinfo {author} {\bibfnamefont {C.}~\bibnamefont
  {Simon}}\ and\ \bibinfo {author} {\bibfnamefont {W.~T.~M.}\ \bibnamefont
  {Irvine}},\ }\href {\doibase 10.1103/PhysRevLett.91.110405} {\bibfield
  {journal} {\bibinfo  {journal} {Phys. Rev. Lett.}\ }\textbf {\bibinfo
  {volume} {91}},\ \bibinfo {pages} {110405} (\bibinfo {year}
  {2003})}\BibitemShut {NoStop}%
\bibitem [{\citenamefont {Li}\ \emph {et~al.}(2024)\citenamefont {Li},
  \citenamefont {Yin},\ and\ \citenamefont {Chen}}]{Li_2024}%
  \BibitemOpen
  \bibfield  {author} {\bibinfo {author} {\bibfnamefont {C.-L.}\ \bibnamefont
  {Li}}, \bibinfo {author} {\bibfnamefont {H.-L.}\ \bibnamefont {Yin}}, \ and\
  \bibinfo {author} {\bibfnamefont {Z.-B.}\ \bibnamefont {Chen}},\ }\href
  {\doibase 10.1088/1361-6633/ad91de} {\bibfield  {journal} {\bibinfo
  {journal} {Reports on Progress in Physics}\ }\textbf {\bibinfo {volume}
  {87}},\ \bibinfo {pages} {127901} (\bibinfo {year} {2024})}\BibitemShut
  {NoStop}%
\bibitem [{\citenamefont {Fekete}\ \emph {et~al.}(2013)\citenamefont {Fekete},
  \citenamefont {Rieländer}, \citenamefont {Cristiani},\ and\ \citenamefont
  {de~Riedmatten}}]{Fekete2013Ultranarrow-Band}%
  \BibitemOpen
  \bibfield  {author} {\bibinfo {author} {\bibfnamefont {J.}~\bibnamefont
  {Fekete}}, \bibinfo {author} {\bibfnamefont {D.}~\bibnamefont {Rieländer}},
  \bibinfo {author} {\bibfnamefont {M.}~\bibnamefont {Cristiani}}, \ and\
  \bibinfo {author} {\bibfnamefont {H.}~\bibnamefont {de~Riedmatten}},\ }\href
  {\doibase 10.1103/PhysRevLett.110.220502} {\bibfield  {journal} {\bibinfo
  {journal} {Physical Review Letters}\ }\textbf {\bibinfo {volume} {110}},\
  \bibinfo {pages} {220502} (\bibinfo {year} {2013})}\BibitemShut {NoStop}%
\bibitem [{\citenamefont {de~Riedmatten}\ \emph {et~al.}(2008)\citenamefont
  {de~Riedmatten}, \citenamefont {Afzelius}, \citenamefont {Staudt},
  \citenamefont {Simon},\ and\ \citenamefont {Gisin}}]{de2008solid-state}%
  \BibitemOpen
  \bibfield  {author} {\bibinfo {author} {\bibfnamefont {H.}~\bibnamefont
  {de~Riedmatten}}, \bibinfo {author} {\bibfnamefont {M.}~\bibnamefont
  {Afzelius}}, \bibinfo {author} {\bibfnamefont {M.~U.}\ \bibnamefont
  {Staudt}}, \bibinfo {author} {\bibfnamefont {C.}~\bibnamefont {Simon}}, \
  and\ \bibinfo {author} {\bibfnamefont {N.}~\bibnamefont {Gisin}},\ }\href
  {\doibase 10.1038/nature07607} {\bibfield  {journal} {\bibinfo  {journal}
  {Nature}\ }\textbf {\bibinfo {volume} {456}},\ \bibinfo {pages} {773}
  (\bibinfo {year} {2008})}\BibitemShut {NoStop}%
\bibitem [{\citenamefont {Liu}\ \emph {et~al.}(2024{\natexlab{b}})\citenamefont
  {Liu}, \citenamefont {Hu}, \citenamefont {Zhu}, \citenamefont {Zhang},
  \citenamefont {Xiao}, \citenamefont {Miao}, \citenamefont {Ou}, \citenamefont
  {Li}, \citenamefont {Liu}, \citenamefont {Zhou}, \citenamefont {Li},\ and\
  \citenamefont {Guo}}]{Liu2024Nonlocal}%
  \BibitemOpen
  \bibfield  {author} {\bibinfo {author} {\bibfnamefont {X.}~\bibnamefont
  {Liu}}, \bibinfo {author} {\bibfnamefont {X.-M.}\ \bibnamefont {Hu}},
  \bibinfo {author} {\bibfnamefont {T.-X.}\ \bibnamefont {Zhu}}, \bibinfo
  {author} {\bibfnamefont {C.}~\bibnamefont {Zhang}}, \bibinfo {author}
  {\bibfnamefont {Y.-X.}\ \bibnamefont {Xiao}}, \bibinfo {author}
  {\bibfnamefont {J.-L.}\ \bibnamefont {Miao}}, \bibinfo {author}
  {\bibfnamefont {Z.-W.}\ \bibnamefont {Ou}}, \bibinfo {author} {\bibfnamefont
  {P.-Y.}\ \bibnamefont {Li}}, \bibinfo {author} {\bibfnamefont {B.-H.}\
  \bibnamefont {Liu}}, \bibinfo {author} {\bibfnamefont {Z.-Q.}\ \bibnamefont
  {Zhou}}, \bibinfo {author} {\bibfnamefont {C.-F.}\ \bibnamefont {Li}}, \ and\
  \bibinfo {author} {\bibfnamefont {G.-C.}\ \bibnamefont {Guo}},\ }\href
  {\doibase 10.1038/s41467-024-52912-3} {\bibfield  {journal} {\bibinfo
  {journal} {Nature Communications}\ }\textbf {\bibinfo {volume} {15}},\
  \bibinfo {pages} {8529} (\bibinfo {year} {2024}{\natexlab{b}})}\BibitemShut
  {NoStop}%
\bibitem [{\citenamefont {Gühne}\ and\ \citenamefont
  {Tóth}(2009)}]{Guhne2009Entanglement}%
  \BibitemOpen
  \bibfield  {author} {\bibinfo {author} {\bibfnamefont {O.}~\bibnamefont
  {Gühne}}\ and\ \bibinfo {author} {\bibfnamefont {G.}~\bibnamefont {Tóth}},\
  }\href {\doibase https://doi.org/10.1016/j.physrep.2009.02.004} {\bibfield
  {journal} {\bibinfo  {journal} {Physics Reports}\ }\textbf {\bibinfo {volume}
  {474}},\ \bibinfo {pages} {1} (\bibinfo {year} {2009})}\BibitemShut {NoStop}%
\bibitem [{\citenamefont {Ma}\ \emph {et~al.}(2021{\natexlab{b}})\citenamefont
  {Ma}, \citenamefont {Ma}, \citenamefont {Zhou}, \citenamefont {Li},\ and\
  \citenamefont {Guo}}]{Ma2021One-hour}%
  \BibitemOpen
  \bibfield  {author} {\bibinfo {author} {\bibfnamefont {Y.}~\bibnamefont
  {Ma}}, \bibinfo {author} {\bibfnamefont {Y.-Z.}\ \bibnamefont {Ma}}, \bibinfo
  {author} {\bibfnamefont {Z.-Q.}\ \bibnamefont {Zhou}}, \bibinfo {author}
  {\bibfnamefont {C.-F.}\ \bibnamefont {Li}}, \ and\ \bibinfo {author}
  {\bibfnamefont {G.-C.}\ \bibnamefont {Guo}},\ }\href {\doibase
  10.1038/s41467-021-22706-y} {\bibfield  {journal} {\bibinfo  {journal}
  {Nature Communications}\ }\textbf {\bibinfo {volume} {12}},\ \bibinfo {pages}
  {2381} (\bibinfo {year} {2021}{\natexlab{b}})}\BibitemShut {NoStop}%
\bibitem [{\citenamefont {Zhu}\ \emph {et~al.}(2026)\citenamefont {Zhu},
  \citenamefont {Zhang}, \citenamefont {Ou}, \citenamefont {Liu}, \citenamefont
  {Liang}, \citenamefont {Hu}, \citenamefont {Huang}, \citenamefont {Zhou},
  \citenamefont {Li},\ and\ \citenamefont {Guo}}]{Zhu2026Data}%
  \BibitemOpen
  \bibfield  {author} {\bibinfo {author} {\bibfnamefont {T.-X.}\ \bibnamefont
  {Zhu}}, \bibinfo {author} {\bibfnamefont {C.}~\bibnamefont {Zhang}}, \bibinfo
  {author} {\bibfnamefont {Z.-W.}\ \bibnamefont {Ou}}, \bibinfo {author}
  {\bibfnamefont {X.}~\bibnamefont {Liu}}, \bibinfo {author} {\bibfnamefont
  {P.-J.}\ \bibnamefont {Liang}}, \bibinfo {author} {\bibfnamefont {X.-M.}\
  \bibnamefont {Hu}}, \bibinfo {author} {\bibfnamefont {Y.-F.}\ \bibnamefont
  {Huang}}, \bibinfo {author} {\bibfnamefont {Z.-Q.}\ \bibnamefont {Zhou}},
  \bibinfo {author} {\bibfnamefont {C.-F.}\ \bibnamefont {Li}}, \ and\ \bibinfo
  {author} {\bibfnamefont {G.-C.}\ \bibnamefont {Guo}},\ }\href@noop {}
  {\bibfield  {journal} {\bibinfo  {journal} {Figshare
  https://doi.org/10.6084/m9.figshare.31778782}\ } (\bibinfo {year}
  {2026})}\BibitemShut {NoStop}%
\bibitem [{\citenamefont {Ou}\ \emph {et~al.}(2025)\citenamefont {Ou},
  \citenamefont {Zhu}, \citenamefont {Liang}, \citenamefont {Hu}, \citenamefont
  {Zhou}, \citenamefont {Li},\ and\ \citenamefont {Guo}}]{ou2025multichannel}%
  \BibitemOpen
  \bibfield  {author} {\bibinfo {author} {\bibfnamefont {Z.-W.}\ \bibnamefont
  {Ou}}, \bibinfo {author} {\bibfnamefont {T.-X.}\ \bibnamefont {Zhu}},
  \bibinfo {author} {\bibfnamefont {P.-J.}\ \bibnamefont {Liang}}, \bibinfo
  {author} {\bibfnamefont {X.-M.}\ \bibnamefont {Hu}}, \bibinfo {author}
  {\bibfnamefont {Z.-Q.}\ \bibnamefont {Zhou}}, \bibinfo {author}
  {\bibfnamefont {C.-F.}\ \bibnamefont {Li}}, \ and\ \bibinfo {author}
  {\bibfnamefont {G.-C.}\ \bibnamefont {Guo}},\ }\href@noop {} {\bibfield
  {journal} {\bibinfo  {journal} {arXiv preprint arXiv:2508.19605}\ } (\bibinfo
  {year} {2025})}\BibitemShut {NoStop}%
\bibitem [{\citenamefont {Zhang}\ \emph {et~al.}(2024)\citenamefont {Zhang},
  \citenamefont {Shi}, \citenamefont {Cui}, \citenamefont {Wang}, \citenamefont
  {Wu}, \citenamefont {Duan},\ and\ \citenamefont {Pu}}]{zhang2024realization}%
  \BibitemOpen
  \bibfield  {author} {\bibinfo {author} {\bibfnamefont {S.}~\bibnamefont
  {Zhang}}, \bibinfo {author} {\bibfnamefont {J.}~\bibnamefont {Shi}}, \bibinfo
  {author} {\bibfnamefont {Z.}~\bibnamefont {Cui}}, \bibinfo {author}
  {\bibfnamefont {Y.}~\bibnamefont {Wang}}, \bibinfo {author} {\bibfnamefont
  {Y.}~\bibnamefont {Wu}}, \bibinfo {author} {\bibfnamefont {L.}~\bibnamefont
  {Duan}}, \ and\ \bibinfo {author} {\bibfnamefont {Y.}~\bibnamefont {Pu}},\
  }\href {\doibase 10.1103/PhysRevX.14.021018} {\bibfield  {journal} {\bibinfo
  {journal} {Phys. Rev. X}\ }\textbf {\bibinfo {volume} {14}},\ \bibinfo
  {pages} {021018} (\bibinfo {year} {2024})}\BibitemShut {NoStop}%
\bibitem [{\citenamefont {Stolk}\ \emph
  {et~al.}(2024{\natexlab{b}})\citenamefont {Stolk}, \citenamefont {van~der
  Enden}, \citenamefont {Slater}, \citenamefont {te~Raa-Derckx}, \citenamefont
  {Botma}, \citenamefont {Van~Rantwijk}, \citenamefont {Biemond}, \citenamefont
  {Hagen}, \citenamefont {Herfst}, \citenamefont {Koek} \emph
  {et~al.}}]{stolk2024metropolitan}%
  \BibitemOpen
  \bibfield  {author} {\bibinfo {author} {\bibfnamefont {A.~J.}\ \bibnamefont
  {Stolk}}, \bibinfo {author} {\bibfnamefont {K.~L.}\ \bibnamefont {van~der
  Enden}}, \bibinfo {author} {\bibfnamefont {M.-C.}\ \bibnamefont {Slater}},
  \bibinfo {author} {\bibfnamefont {I.}~\bibnamefont {te~Raa-Derckx}}, \bibinfo
  {author} {\bibfnamefont {P.}~\bibnamefont {Botma}}, \bibinfo {author}
  {\bibfnamefont {J.}~\bibnamefont {Van~Rantwijk}}, \bibinfo {author}
  {\bibfnamefont {J.~B.}\ \bibnamefont {Biemond}}, \bibinfo {author}
  {\bibfnamefont {R.~A.}\ \bibnamefont {Hagen}}, \bibinfo {author}
  {\bibfnamefont {R.~W.}\ \bibnamefont {Herfst}}, \bibinfo {author}
  {\bibfnamefont {W.~D.}\ \bibnamefont {Koek}},  \emph {et~al.},\ }\href@noop
  {} {\bibfield  {journal} {\bibinfo  {journal} {Science advances}\ }\textbf
  {\bibinfo {volume} {10}},\ \bibinfo {pages} {eadp6442} (\bibinfo {year}
  {2024}{\natexlab{b}})}\BibitemShut {NoStop}%
\bibitem [{\citenamefont {Liu}\ \emph {et~al.}(2026)\citenamefont {Liu},
  \citenamefont {Lin}, \citenamefont {Zhou}, \citenamefont {Li},\ and\
  \citenamefont {Guo}}]{Liu2026}%
  \BibitemOpen
  \bibfield  {author} {\bibinfo {author} {\bibfnamefont {P.-X.}\ \bibnamefont
  {Liu}}, \bibinfo {author} {\bibfnamefont {Y.-P.}\ \bibnamefont {Lin}},
  \bibinfo {author} {\bibfnamefont {Z.-Q.}\ \bibnamefont {Zhou}}, \bibinfo
  {author} {\bibfnamefont {C.-F.}\ \bibnamefont {Li}}, \ and\ \bibinfo {author}
  {\bibfnamefont {G.-C.}\ \bibnamefont {Guo}},\ }\href@noop {} {\  (\bibinfo
  {year} {2026})}\BibitemShut {NoStop}%
\bibitem [{\citenamefont {Black}(2001)}]{Black2001PDH}%
  \BibitemOpen
  \bibfield  {author} {\bibinfo {author} {\bibfnamefont {E.~D.}\ \bibnamefont
  {Black}},\ }\href {\doibase 10.1119/1.1286663} {\bibfield  {journal}
  {\bibinfo  {journal} {American Journal of Physics}\ }\textbf {\bibinfo
  {volume} {69}},\ \bibinfo {pages} {79} (\bibinfo {year} {2001})}\BibitemShut
  {NoStop}%
\bibitem [{\citenamefont {Nielsen}\ and\ \citenamefont
  {Chuang}(2010)}]{nielsen2010quantum}%
  \BibitemOpen
  \bibfield  {author} {\bibinfo {author} {\bibfnamefont {M.~A.}\ \bibnamefont
  {Nielsen}}\ and\ \bibinfo {author} {\bibfnamefont {I.~L.}\ \bibnamefont
  {Chuang}},\ }\href@noop {} {\emph {\bibinfo {title} {Quantum computation and
  quantum information}}}\ (\bibinfo  {publisher} {Cambridge university press},\
  \bibinfo {year} {2010})\BibitemShut {NoStop}%
\bibitem [{\citenamefont {Afzelius}\ \emph {et~al.}(2009)\citenamefont
  {Afzelius}, \citenamefont {Simon}, \citenamefont {de~Riedmatten},\ and\
  \citenamefont {Gisin}}]{Afzelius2009Multimode}%
  \BibitemOpen
  \bibfield  {author} {\bibinfo {author} {\bibfnamefont {M.}~\bibnamefont
  {Afzelius}}, \bibinfo {author} {\bibfnamefont {C.}~\bibnamefont {Simon}},
  \bibinfo {author} {\bibfnamefont {H.}~\bibnamefont {de~Riedmatten}}, \ and\
  \bibinfo {author} {\bibfnamefont {N.}~\bibnamefont {Gisin}},\ }\href
  {\doibase 10.1103/PhysRevA.79.052329} {\bibfield  {journal} {\bibinfo
  {journal} {Physical Review A}\ }\textbf {\bibinfo {volume} {79}},\ \bibinfo
  {pages} {052329} (\bibinfo {year} {2009})}\BibitemShut {NoStop}%
\bibitem [{\citenamefont {Zhu}\ \emph {et~al.}(2022)\citenamefont {Zhu},
  \citenamefont {Liu}, \citenamefont {Jin}, \citenamefont {Su}, \citenamefont
  {Liu}, \citenamefont {Li}, \citenamefont {Ye}, \citenamefont {Zhou},
  \citenamefont {Li},\ and\ \citenamefont {Guo}}]{Zhu2022On-Demand}%
  \BibitemOpen
  \bibfield  {author} {\bibinfo {author} {\bibfnamefont {T.-X.}\ \bibnamefont
  {Zhu}}, \bibinfo {author} {\bibfnamefont {C.}~\bibnamefont {Liu}}, \bibinfo
  {author} {\bibfnamefont {M.}~\bibnamefont {Jin}}, \bibinfo {author}
  {\bibfnamefont {M.-X.}\ \bibnamefont {Su}}, \bibinfo {author} {\bibfnamefont
  {Y.-P.}\ \bibnamefont {Liu}}, \bibinfo {author} {\bibfnamefont {W.-J.}\
  \bibnamefont {Li}}, \bibinfo {author} {\bibfnamefont {Y.}~\bibnamefont {Ye}},
  \bibinfo {author} {\bibfnamefont {Z.-Q.}\ \bibnamefont {Zhou}}, \bibinfo
  {author} {\bibfnamefont {C.-F.}\ \bibnamefont {Li}}, \ and\ \bibinfo {author}
  {\bibfnamefont {G.-C.}\ \bibnamefont {Guo}},\ }\href {\doibase
  10.1103/PhysRevLett.128.180501} {\bibfield  {journal} {\bibinfo  {journal}
  {Physical Review Letters}\ }\textbf {\bibinfo {volume} {128}},\ \bibinfo
  {pages} {180501} (\bibinfo {year} {2022})}\BibitemShut {NoStop}%
\end{thebibliography}

% \end{document}

% \printbibliography[title={reference}]
% \end{refsection}

\end{document}